\documentclass[11pt]{iopart}
\usepackage{color}
\usepackage{epsfig}
\usepackage{amssymb}
\usepackage{mathrsfs}
\newcommand{\be}{\begin{equation}}
\newcommand{\ee}{\end{equation}}
\newcommand{\bea}{\begin{eqnarray}}
\newcommand{\eea}{\end{eqnarray}}

\def\nn{\nonumber\\}

\def\fr#1{(\ref{#1})}

\def\eps{\epsilon}
\def\alb{\overline{\alpha}}
\def\veps{\varepsilon}
\def\bs{\bar{\sigma}}
\def\sth{\{\th_j\}}
\def\ot{\Omega(\theta,\omega)}
\def\qt{Q(\theta,q)}

\def\t{\theta}
\def\ks{K_{\rm SM}}
\newcommand{\ket}[1]{\left|#1\right\rangle}
\newcommand{\bra}[1]{\left\langle#1\right|}
\def\th{\theta}
\def\thh{a}
\def\Xint#1{\mathchoice
   {\XXint\displaystyle\textstyle{#1}}%
   {\XXint\textstyle\scriptstyle{#1}}%
   {\XXint\scriptstyle\scriptscriptstyle{#1}}%
   {\XXint\scriptscriptstyle\scriptscriptstyle{#1}}%
   \!\int}
\def\XXint#1#2#3{{\setbox0=\hbox{$#1{#2#3}{\int}$}
     \vcenter{\hbox{$#2#3$}}\kern-.5\wd0}}

\def\dashint{\Xint-}
\begin{document}
\title[Finite Temperature Dynamical Correlations in Massive Integrable QFTs]{Finite Temperature Dynamical Correlations in Massive
  Integrable Quantum Field Theories} 
\author{Fabian H.L. Essler$^{(a)}$\footnote{email:fab@thphys.ox.ac.uk}
  and Robert M. Konik$^{(b)}$} 
\address{
$^{(a)}$ The Rudolf Peierls Centre for Theoretical Physics, Oxford
University, Oxford OX1 3NP, UK\\ 
$^{(b)}$ Department of Physics, Brookhaven National Laboratory, Upton
NY 11973, USA }
\begin{abstract}
We consider the finite-temperature frequency and momentum dependent
two-point functions of local operators in integrable quantum field
theories. We focus on the case where the zero temperature correlation
function is dominated by a delta-function line arising from the
coherent propagation of single particle modes. Our specific examples
are the two-point function of spin fields in the disordered phase
of the quantum Ising and the O(3) nonlinear sigma models. We employ a Lehmann
representation in terms of the known exact zero-temperature form
factors to carry out a low-temperature expansion of two-point functions.
We present two different but equivalent  methods of regularizing the
divergences present in the Lehmann expansion: one directly regulates
the integral expressions of the squares of matrix elements in the
infinite volume whereas the other  operates through subtracting
divergences in a large, finite volume. Our central results are that the
temperature broadening of the line shape exhibits a pronounced
asymmetry and a shift of the maximum upwards in energy (``temperature
dependent gap''). The field theory results presented here describe the
scaling limits of the dynamical structure factor in the quantum Ising
and integer spin Heisenberg chains. We discuss the 
relevance of our results for the analysis of inelastic neutron
scattering experiments on gapped spin chain systems such as ${\rm
  CsNiCl_3}$ and ${\rm YBaNiO_5}$.  
\end{abstract}

\maketitle
\section{Introduction}
Progress over the last two decades 
\cite{Smirnov92book,Lukyanov95,DelMuss,FF,Delfino04,BH,BN,ks,mussardobook,karowski}
has made it possible to determine zero temperature dynamical response
functions in massive integrable models of quantum field theory
(QFT) by means of the ``form factor bootstrap approach''. More
precisely, response functions can be calculated exactly at low
frequencies (several times the mass gap) and to high accuracy at
intermediate frequencies. The $T=0$ dynamics described by integrable
QFTs exhibits a number of interesting phenomena such as dynamical mass 
generation, spin-charge separation, and other kinds of quantum number
fractionalization. The results obtained by the form factor bootstrap
approach have important applications in condensed matter systems \cite{review} such
as quantum magnets \cite{QM}, Mott insulators \cite{MI}, doped ladder
materials \cite{2leg}, carbon nanotubes \cite{nano} and ultra-cold
atomic gases \cite{gas}.  

At the heart of the form factor bootstrap approach is the notion that
in an integrable model the scattering of elementary excitations is
purely elastic by virtue of the existence of an infinite number of
local conservation laws. There is no particle production and the
individual particle momenta are conserved in scattering events.
Correspondingly, the scattering of n particles in an integrable model
can always be reduced to a sum of two-body scattering events. The
resulting simplified nature of the exact Hamiltonian eigenstates in a
massive integrable field theory permits the computation of zero
temperature correlation functions as follows. One first employs a
Lehmann representation in terms of n-particle Hamiltonian eigenstates
$|n,\{s_n\}\rangle$ with energy $E[\{s_n\}]$, where $\{s_n\}$ labels
the corresponding sets of good quantum numbers
\begin{eqnarray}
\fl
\chi_{\cal O}(\tau,x)\Big|_{T=0} &=& \langle 0|T_\tau {\cal O}(\tau,x){\cal
  O}^\dagger(0,0)|0\rangle\cr\cr 
&=&\sum^\infty_{n=0} e^{-\tau E[\{{s_n}\}]}\langle 0|{\cal O}(x,0)|n;\{s_n\}\rangle 
\langle n;\{s_n\}|{\cal O}^\dagger(0,0)|0\rangle ~~~(\tau > 0).
\label{intro1}
\end{eqnarray}
As the energies are simply given as sums over the single particle
energies of the elementary excitations, we have reduced the
computation of the response function to computing a set of matrix
elements or ``form factors'',  $\langle 0|{\cal
  O}(\tau,x)|n;\{s_n\}\rangle$.

To perform this computation we again exploit integrability.  
The ability to express $|n,\{s_n\}\rangle$ as a collection of n
distinct particles allows one to write down a set of algebraic
constraints that the matrix elements must satisfy
\cite{Smirnov92book}.  These constraints encode both the simplified
form the scattering of n particles takes in an integrable model
together with analytic constraints coming from crossing symmetries
present in a relativistic quantum field theory. While these
constraints can be written down for eigenstates involving an arbitrary
number, n, of particles, they become increasingly cumbersome to solve
as $n$ increases.  Fortunately, if we are interested merely in the
behavior of the low energy spectral function, it is only necessary to
compute matrix elements involving a few particles.  The spectral
function is obtained by Fourier transforming \fr{intro1} with respect
to space and imaginary time and then analytically continuing to real
frequencies ($i\omega_n\rightarrow \omega+i0$)
\begin{eqnarray}
\fl
-{1 \over \pi} {\rm Im}\ \chi_{\cal O}(\omega ,q ) &=& 2\pi
\sum_{n=0}^{\infty}\sum_{s_n} \bigg\{ 
\left|\langle 0|{\cal O}(0,0)|n;\{s_n\}\rangle\right|^2
\delta(\omega- E[\{s_n\}])\
\delta(q- P[\{s_n\}])
\cr\cr
&& - \epsilon \left|\langle n;\{s_n\}|{\cal O}^\dagger(0,0)|0\rangle\right|^2
\delta(\omega + E_{s_n})\ \delta(q+ P[\{s_n\}])\bigg\},
\end{eqnarray}
where $P[\{s_n\}]$ denotes the momentum of $|n,\{s_n\}\rangle$ and 
$\epsilon = \pm 1$ depending on whether the field ${\cal O}$ is
bosonic or fermionic.
The presence of the delta function in the above expression for the spectral
function guarantees that at an energy $\omega$ only eigenfunctions with this exact
energy contribute.  As the theory is massive, eigenstates with n-particles
will have a minimum energy, $n\Delta$ (supposing the particles all have mass $\Delta$).
Thus for energies, $\omega < n\Delta$, such eigenstates will not contribute to
the spectral function.  For low energies, only a small finite number of
matrix elements need to be computed in order to obtain exact results for the
spectral function.  Even at higher energies, it has been typically found
that the sum of matrix elements is strongly convergent, and that matrix elements
with higher particle number make only an extremely small contribution to the
spectral function \cite{DelMuss,oldff1,deltwo,CET,review}.

While the above described approach has been successful at computing
zero temperature correlation functions, it is not a settled
question whether the form factor bootstrap approach
can be used universally to gain information about $T>0$ dynamical
correlations. This question has been investigated in a number of
specific instances \cite{ising3,muss,finiteTFF,rmk,AKT}. On the basis
of this past work, there appear, in various forms, two main
difficulties in doing so. These difficulties become particularly
acute in theories which are interacting. To illustrate these problems, 
we write out the corresponding form-factor expansion for a Green's
function at finite temperature:
\begin{eqnarray}\label{FFexpFT}
\fl
-{1 \over \pi} {\rm Im}\ \chi_{\cal O} (\omega,q ) &=& {2\pi\over \cal Z}
\sum_{{n s_n}\atop{m s_m}}\delta (\omega - E[\{s_n\}]+E[\{s_m\}])
\ \delta (q - P[\{s_n\}]+P[\{s_m\}])\cr\cr
&&\times \left[e^{-\beta E[\{s_m\}]}-\epsilon e^{-\beta E[\{s_n\}]}\right]\
\Big|\langle m;\{s_m\}|{\cal O}(0,0)|n;\{s_n\}\rangle\Big|^2.
\end{eqnarray}
The first difficulty can be seen in that the 
form-factor expansion now involves two sums over eigenstates, the first sum
arising from the insertion of the resolution of the identity as before and the
second sum coming from the Boltzmann trace associated with working at finite temperature.
Concomitantly, working at a particular energy, $\omega$, no longer
guarantees that only a finite number of matrix elements will
contribute to the spectral function. This problem was partially
resolved in Ref. \cite{rmk}.  There it was advocated that the
Boltzmann factor provides a natural small parameter, that is, terms
involving eigenstates $|m;\{s_m\}\rangle$ with many particles (and so
$E[\{s_m\}]$ large) play only a specific, limited role, at least at
low temperatures. It was shown there that for a certain class of
correlation functions, it was possible to develop a low
temperature expansion of the response function. However this work
left unresolved how to perform the low temperature expansion in
general. In particular, it did not address how to develop the low
temperature expansion for response functions with singular features at
zero temperature.

Such response functions form a class of great physical interest. As
an important physical example, the spin response, $S(\omega,q)$, 
of a gapped quantum spin chain as represented by either the quantum
Ising model or the O(3) NLSM is given by $S(\omega,q) = Z(q)\delta(\omega
- \epsilon(q))+\dots$ at zero temperature, where the dots indicate
very weak multi-paticle scattering continua.  At finite temperature this
$\delta$-function response broadens out to a lineshape with finite
width. But it has been unclear how to capture this broadening in the
context of the form factor expansion as given in
\fr{FFexpFT}. It is one of the achievements of this paper to
demonstrate how this can be done. 

The second difficulty is that unlike the zero temperature case,
we are now faced with computing matrix elements between states,
$|n;\{s_n\}\rangle$ and $|m;\{s_m\}\rangle$ both with finite 
particle number (as opposed to amplitudes $\langle 0|{\cal
  O}(0,0)|n;\{s_n\}\rangle$ governing the transition between the
vacuum and some eigenstate). In an infinite volume, matrix elements of
the form $\langle m;\{s_m\}|{\cal O}(0,0)|n;\{s_n\}\rangle$ with
$n,m\neq 0$ possess singularities. Moreover these singularities are
`squared' as it is the quantity 
$|\langle m;\{s_m\}|{\cal O}(0,0)|n;\{s_n\}\rangle|^2$ that determines
the correlation function.  Making sense of these squared singularities
amounts to giving a sensible interpretation to terms of the form,
``$\delta(0)$''. 

It is perhaps the main achievement of this paper that we have
demonstrated a set of regularization procedures for such
singularities. These ultimately arise as a consequence
of working in an infinite volume, where the momenta of
particles found in the n-particle and m-particle states,
$|n;\{s_n\}\rangle$ and $|m;\{s_m\}\rangle$, can be identical.  In a
matrix element of the form, $\langle m;\{s_m\}|{\cal
  O}(0,0)|n;\{s_n\}\rangle$, identical momenta lead to divergences.
One approach is thus to work in a large but {\it finite} volume.  In
such cases the momenta (at least in the relevant examples) of the
particles composing the states $|n;\{s_n\}\rangle$ and
$|m;\{s_m\}\rangle$ are never equal leaving 
the matrix element $\langle m;\{s_m\}|{\cal O}(0,0)|n;\{s_n\}\rangle$
finite. Such an approach is feasible as $\langle m;\{s_m\}|{\cal
  O}(0,0)|n;\{s_n\}\rangle$ retains the same momentum dependencies as
in an infinite volume. Provided we work in a large, finite volume, 
the sole difference in evaluating the form-factor $\langle m;\{s_m\}|{\cal
  O}(0,0)|n;\{s_n\}\rangle$ for general momenta between the finite and
infinite volume cases lies in taking into account that in finite
volume the momenta of the states $|n;\{s_n\}\rangle$ and
$|m;\{s_m\}\rangle$ are quantized \cite{takacs}.  Thus we can still
use in a large, finite volume the infinite volume constraints that
govern these matrix elements.

The use of finite volume regularization has precedent.  For the special 
case of the quantum Ising model this particular problem has been
solved by calculating the form factors on the cylinder
\cite{bugrij,doyon} -- that is, the form factors are computed in an
arbitrary, not merely asymptotically large, finite volume. These
results have then been used in Ref. \cite{AKT} to recover the
semiclassical expression obtained previously by Sachdev 
and Young \cite{young} and in Ref. \cite{Reyes06,Reyes06b} to carry out
low and high temperature expansions for the spin-spin correlation
function.  
Form factors as computed on the cylinder have also been exploited 
to derive partial differential equations governing finite temperature
correlation functions in the quantum Ising model \cite{doyongamsa}.
In this work, the large space and time asymptotics of spin correlations
were determined.  Such partial different equations for correlation
functions have been derived in a number of instances \cite{PDE}
and employed very recently \cite{perk} to analyze time dependent
zero temperature correlators in the Ising model.  However like in
Ref. \cite{doyongamsa}, the underlying theory 
always has had trivial scattering, that is the scattering
matrix is momentum independent and diagonal. It is presently not known
how to generalize the results of \cite{bugrij,doyon} to integrable
QFTs with non-trivial S-matrices.

In addition to our use of a finite volume regularization scheme, we 
demonstrate in this paper a new regularization technique that operates
in an infinite volume.  In this regularization scheme, ambiguous terms such
as ``$\delta(0)$'' are absent by construction. We demonstrate
explicitly that our two regularization schemes lead to the same
answers. We do so in two examples, the quantum Ising model and the
O(3) non-linear sigma model (NLSM). That the latter model is
interacting and so highly non-trivial provides a strong indication
that our infinite volume regularization scheme is robust and so
provides a candidate for a regulator that works generally. In
particular it would be useful to check the scheme for cases involving
form factors between two 2-particle states or a 2-particle and a
3-particle state. 

Our work, both in how we develop a low temperature expansion,
and in our particular choice of integrable models, the quantum Ising
model and the O(3) NLSM, is motivated in large degree by 
recent inelastic neutron scattering experiments on several quantum
magnets. A key objective of this recent experimental work
has been to investigate how the spin dynamics crosses over from the
strongly correlated zero temperature quantum regime to the classical
high temperature regime \cite{kenz,xu,zheludev}. In a 
system such as the spin-1 Heisenberg chain that supports a coherent,
gapped, magnetic single-particle excitation at $T=0$, the question
arises of how the dominant feature in the dynamical structure factor,
a delta-function at zero temperature, broadens at finite temperatures. 
As field theories, the quantum Ising model and the O(3) NLSM describe
the scaling limits of the (non-integrable) spin-chain Hamiltonians
used to model these various experiments.

Applying the methodology detailed in this paper, we analyze
this finite temperature lineshape.  As our central finding 
in this regard, we demonstrate that the lineshape is always asymmetric
in energy, a feature that becomes more pronounced as the temperature
increases. For the Ising model we further demonstrate the emergence of
a ``temperature dependent gap''.  A subset of our results on the
lineshape have been previously reported in Ref. \cite{EK08}.

At temperatures $T$ far below the spin gap $\Delta$
our technique  complements previous semiclassical approaches to the
study of the lineshape of quantum spin
chains\cite{young,sachdevbook,damle,semiclassicsSG,zarand}. 
Both in our methodology and in semi-classical approaches,
the lineshape has been shown to be essentially Lorentzian for $T\ll\Delta$
and for energies, $\omega$, in the vicinity of the gap, $\Delta$.
However in our approach we can both study temperatures where the semi-classics
is inaccurate as well as the entire lineshape, not merely $\omega \sim \Delta$.

The outline of this paper is as follows. In Section \ref{sec:general}
we set out our framework for deriving low-temperature expansions of
dynamical correlation functions in massive integrable quantum field
theories.  In particular, we summarize how the space of Hamiltonian
eigenstates in an integrable model is handled as well as how we
develop the low temperature expansion appropriate for computing the
finite temperature lineshape. In Section \ref{sec:ising} we apply the
method to the quantum Ising chain. We introduce how we regularize the squares
of form-factors both by working in finite volume as well as in our new
infinite volume scheme.  We show that in the context of the Ising model,
these schemes are equivalent.
In the following section, Section \ref{sec:Isingresults}, we present a detailed discussion of the results obtained 
for the quantum Ising model as well as a
comparison to the semiclassical results of Sachdev and
Young. In the next two sections
we move on to the O(3) NLSM showing that our methodology also works
for interacting theories with non-trivial (even non-diagonal) scattering
matrices.  In Section \ref{sec:SM} we
consider the case of the retarded Green's function of the vector field
in the O(3) nonlinear sigma model, the quantity that corresponds to the
spin response of a gapped Heisenberg spin chain near wavevector $\pi$. 
We again show that the two regularization schemes yield the same result for
this correlation function.
In Section \ref{sec:O3results} we
detail the results that so arise for the low-temperature dynamics for the O(3) NLSM and compare them to
the semiclassical results of Ref \cite{damle}.
The final section, Section \ref{sec:summ}, presents a summary and discussion of our
results. Computational details on our new method for regularizing form factor squares
directly in the infinite volume are presented in several appendices.

\section{General Formalism}
\label{sec:general}
A defining feature of integrable quantum field theories is a basis of
scattering states of ``elementary'' excitations, which are eigenstates
of the Hamiltonian. It is customary to construct these states from
the so-called Faddeev-Zamolodchikov algebra
\begin{eqnarray}
\label{FZalgebra}
Z_{a} (\th_1 ) Z_{b} (\th_2 ) &=& 
S^{a'b'}_{ab} (\th_1-\th_2) Z_{b'} (\th_2) Z_{a'} (\th_1)\ ,\nn
Z^\dagger_{a} (\th_1 ) Z^\dagger_{b} (\th_2 ) &=& 
S^{a'b'}_{ab} (\th_1-\th_2) 
Z^\dagger_{b'} (\th_2) Z^\dagger_{a'} (\th_1)\ ,\nn
Z_{a} (\th_1 ) Z^\dagger_{b} (\th_2 ) &=& 
2\pi\delta_{ab}\delta(\th_1-\th_2)+
S^{b'a}_{ba'} (\th_1-\th_2) Z^\dagger_{b'} (\th_2) Z_{a'} (\th_1).
\end{eqnarray}
Here $\th_{1,2}$ are rapidity variables, $a,b$ are quantum numbers and
$S$ is the exact two-particle scattering matrix describing the purely
elastic scattering of the elementary excitation.
The S-matrix is a solution to the Yang-Baxter equation, which can be
thought of as a consistency condition for factorizable three-particle
scattering. 
Using the Faddeev-Zamolodchikov operators, a Fock space of states can
be constructed as follows. The vacuum is defined by
\begin{equation}
Z_{a}(\theta) |0\rangle=0 \ .
\label{vac}
\end{equation}
Multiparticle states are then obtained by acting with strings of
creation operators $Z_b^\dagger(\theta)$ on the vacuum
\begin{equation}
|\theta_n\ldots\theta_1\rangle_{a_n\ldots a_1} = 
Z^\dagger_{a_n}(\theta_n)\ldots
Z^\dagger_{a_1}(\theta_1)|0\rangle\ .
\label{states}
\end{equation} 
Energy and momentum of the states \fr{states} are by construction
additive 
\bea
E_s(\th_1,\ldots,\th_s)&=&\sum_{j=1}^s\epsilon(\th_j)\ ,\quad
\epsilon(\th)=\Delta\cosh\th\ ,\nn
P_s(\th_1,\ldots,\th_s)&=&\sum_{j=1}^s\frac{\Delta}{v}\sinh\th_j\ .
\label{ep}
\eea
In terms of this basis the resolution of the identity is given by
\be
1\!\! 1 = |0\rangle\langle 0| \label{id}
+ \sum_{n=1}^\infty\sum_{\{a_i\}}\int_{-\infty}^{\infty}
\frac{d\theta_1\ldots d\theta_n}{(2\pi)^nn!}
|\theta_n\ldots\theta_1\rangle_{a_n\ldots a_1}
{}^{a_1\ldots a_n}\langle\theta_1\ldots\theta_n|\ .
\ee

In the basis of scattering states introduced above, the
following formal spectral representation for the retarded finite
temperature two-point function of the local operator ${\cal O}$ holds
\be
\chi_{\cal O}(\omega,q)=\frac{1}{{\cal Z}}\sum_{r,s=0}^\infty
C^{\cal O}_{r,s}(\omega,q)\ . 
\label{OO}
\ee
Here $C_{r,s}$ denotes the contribution with $r$ particles in the
thermal trace and $s$ in the intermediate state
\bea
C^{\cal O}_{r,s}(\omega,q)&=&\int_0^\beta d\tau\int_{-\infty}^\infty
 dx e^{i\omega_n\tau-iqx} C_{r,s}(\tau,x)\Bigr|_{i\omega_n\rightarrow\omega+i0}\cr\cr
C^{\cal O}_{r,s}(\tau,x) &=&-\sum_{\{a_j\},\{a'_k\}}
\int\frac{d\th_1\ldots d\th_r}{(2\pi)^rr!}
\int\frac{d\th'_1\ldots d\th'_s}{(2\pi)^ss!}
e^{-\beta E_r}e^{-\tau(E_s-E_r)}\nn
&&\quad\times\
e^{-i(P_r-P_s)x}\
|^{a_1\ldots a_r}\langle\theta_1\ldots\th_r|{\cal O}(0,0)|
\th'_s\ldots\th'_1\rangle_{a_s'\ldots a_1'}|^2.
\label{CrsO}
\eea
The partition function can formally be expressed as
\bea
{\cal Z}&=&\langle 0|0\rangle+\sum_b
\int\frac{d\theta}{2\pi}e^{-\beta
\epsilon(\th)}\ {}^b\langle \theta|\theta\rangle_b\nn
&+&\sum_{b_1,b_2}\int\frac{d\th_1d\th_2}{2(2\pi)^2}
e^{-\beta(\epsilon(\th_1)+\epsilon(\th_2))}\
{}^{b_1b_2}\langle \th_1\th_2|\th_2,\th_1\rangle_{b_2b_1}
+\ldots
\equiv\sum_{n=0}^\infty{\cal Z}_n.
\label{Zgen}
\eea
Both the partition function and the Lehmann representations of
correlation functions are ill-defined in the infinite volume limit
as the normalization condition for scattering states is (for
$\th_1>\th_2>\ldots>\th_n$ and $\th'_1>\th'_2>\ldots>\th'_n$)
\be
{}^{a_1\ldots a_n}\langle\theta_1,\ldots,\th_n|
\theta'_n,\ldots,\th'_1\rangle_{a'_n\ldots a'_1}=\prod_{j=1}^n
2\pi\delta(\th_j-\th'_j)\ \delta_{a_j,a'_j}.
\label{gennorm}
\ee
The idea of a low-temperature expansion is to subtract these
divergences in some way \cite{muss,rmk,AKT}. Here we proceed as
follows. We separate the contributions ${\cal C}_{r,s}$ in the Lehmann
representation of the two-point function \fr{OO} according to their 
different formal temperature dependencies into 
\be
C^{\cal O}_{r,s}(\omega,q)=E^{\cal O}_{r,s}(\omega,q)+F^{\cal
  O}_{r,s}(\omega,q), 
\label{CEF}
\ee
where
\bea
E^{\cal O}_{r,s}(\omega,q)&=&\sum_{\{a_j\},\{a'_k\}}
\int\frac{d\th_1\ldots d\th_r}{(2\pi)^rr!}
\int\frac{d\th'_1\ldots d\th'_s}{(2\pi)^ss!}2\pi\delta(q+P_r-P_s)\nn
&\times&
\frac{e^{-\beta E_r}}{\omega+i\delta-E_s+E_r}
\Big|{}^{a_1\ldots a_r}\langle\theta_1\ldots\th_r|{\cal O}(0,0)|
\th'_s\ldots\th'_1\rangle_{a_s'\ldots a_1'}\Big|^2.
\label{Ers}
\eea
The functions $E_{r,s}$, $F_{r,s}$ are related by
\be
E^{\cal O}_{r,s}(\omega,q)=\left[F^{\cal O}_{s,r}(-\omega,-q)\right]^*.
\ee
The matrix elements 
${}^{a_1\ldots a_r}\langle\theta_1\ldots\th_r|{\cal O}(0,0)|
\th'_s\ldots\th'_1\rangle_{a_s'\ldots a_1'}$ can be decomposed into a
{\sl connected} and a {\sl disconnected} contribution. The latter is
characterized by containing factors of $\delta(\th_j-\th'_k)$,
signalling that some of the particles do not encounter the operator
${\cal O}$ in the process described by the matrix
element. A fundamental assumption of our approach is that the
disconnected contributions act to cancel the partition function in the
denominator of \fr{OO}. More precisely we define quantities
\bea
E_{j,k}^{{\cal O},n}&=&E_{j,k}^{\cal O}-\sum_{m=1}^n{\cal Z}_m
E^{{\cal O},n-m}_{j-m,k-m}\ ,\nn
F_{j,k}^{{\cal O},n}&=&F_{j,k}^{\cal O}-\sum_{m=1}^n{\cal Z}_m
F^{{\cal O},n-m}_{j-m,k-m}\ ,\ n=0,1,2\ldots
\eea
\bea
{\cal E}^{\cal O}_{n}&=&\sum_{k=0}^{n-1}E^{{\cal O},k}_{n,k}+
\sum_{m=n}^\infty E_{n,m}^{{\cal O},n}\ ,\ n=0,1,2\dots
\label{El}\\
{\cal F}^{\cal O}_{n}&=&\sum_{k=0}^{n-1}F^{{\cal O},k}_{k,n}+
\sum_{m=n}^\infty F_{m,n}^{{\cal O},n}\ ,\ n=0,1,2\dots
\label{Fl}
\eea

The key assertion of our low-temperature expansion is that the
quantities defined in this way are {\sl finite} in the thermodynamic
limit. Upon re-ordering of the infinite sums the two-point function
\fr{OO} is expressed in terms of the ${\cal E}^{\cal O}_s$ and ${\cal
  F}^{\cal O}_r$ as
\be
\chi_{\cal O}(\omega,q)=\sum_{s=0}^\infty
{\cal E}^{\cal O}_{s}(\omega,q)+{\cal F}^{\cal O}_{s}(\omega,q).
\label{lowTexgen}
\ee
By construction ${\cal E}^{\cal O}_s$ and ${\cal F}^{\cal O}_s$ formally have
a temperature dependence
\be
{\cal E}^{\cal O}_s,{\cal F}^{\cal O}_s\sim {\cal O}\left(e^{-s\beta\Delta}\right),
\ee
and \fr{lowTexgen} hence constitutes a low-temperature expansion of 
$\chi_{\cal O}(\omega,q)$. 
The two-point functions we analyze in detail below have the symmetry
\be
\chi_{\cal O}(\omega,q)=\chi_{\cal O}^*(-\omega,-q),
\ee
which relates the positive and negative frequency regions. It is
useful to combine ${\cal E}_l$ and ${\cal F}_l$
into quantities that exhibit the same symmetry
\be
{\cal C}^{\cal O}_l(\omega,q)={\cal E}^{\cal O}_l(\omega,q)+{\cal F}^{\cal O}_l(\omega,q).
\label{Cl}
\ee
In terms of these quantities the spectral representation takes the form
\be
\chi_{\cal O}(\omega,q)=\sum_{s=0}^\infty
{\cal C}^{\cal O}_{s}(\omega,q).
\label{lowTex}
\ee
A key property of the ${\cal C}^{\cal O}_l$ is that they are {\sl finite} in the
thermodynamic limit. We demonstrate this explicitly for the first
nontrivial terms ${\cal C}^{\cal O}_1$  and ${\cal C}^{\cal O}_2$ below and postulate
that it is true in general.

\subsection{Resummation}
\label{ssec:resum}
Following the procedure set out above, the finite temperature Lehman
representation of the particular two-point functions analyzed below
can be re-expressed in the form \fr{lowTex}, where the quantities
${\cal C}^{\cal O}_{r}(\omega,q)$ are finite in the 
infinite volume limit. However, in the cases we are interested in, the
functions ${\cal C}^{\cal O}_r(\omega,q)$ are not uniformly small. 
In order to make this statement more precise let us denote the
single-particle dispersion relation by
\be
\veps(q)=\sqrt{\Delta^2+v^2q^2}.
\label{veps}
\ee
We observe that as long as both $\omega\pm\veps(q)\sim{\cal
  O}(1)$, the ${\cal C}^{\cal O}_r$ are of order ${\cal O}\Bigl(e^{-\beta
  r\Delta}\Bigr)$ and hence \fr{lowTex} provides a good
low-temperature expansion of the two point functions we are interested
in far away from the mass shell. On the other hand, when we approach
the mass shell we have 
\be
{\cal C}^{\cal O}_r(\omega,q)\propto \left(\omega^2-\varepsilon^2(q)\right)^{r+1}.
\ee
In order to obtain an expression for the susceptibility close to the
mass shell we therefore need to sum up an infinite number of terms in
\fr{lowTex}. For the cases considered below, the zero-temperature
two point function is of the form
\be
{\cal C}^{\cal O}_0(\omega,q)=\frac{Z}
{(\omega+i\delta)^2-\epsilon^2(q)}+\ldots,
\ee
where the corrections are negligible in the regime of temperatures and
frequencies we consider. We then introduce a quantity
$\Sigma^{\cal O}(\omega,q)$ by defining
\begin{equation}
\label{gensigma0}
\chi_{\cal O}(\omega, q) = \frac{{\cal C}^{\cal O}_0(\omega, q)}
{1-{\cal C}^{\cal O}_0(\omega,q)\Sigma^{\cal O}(\omega, q)}.
\end{equation}
The low-temperature expansion \fr{lowTex} for $\chi^{\cal O}(\omega, q)$
then provides a way of determining low temperature approximations to
$\Sigma^{\cal O}(\omega, q)$ in the following way. Assuming that a
low-temperature expansion of the form
\be
\Sigma^{\cal O}(\omega,q)=\sum_{n=1}\Sigma^{\cal O}_n(\omega,q)
\label{gensigma}
\ee
exists, we may determine the leading term at low temperatures by
expanding 
\bea
\chi_{\cal O}(\omega,q)  &=& {\cal C}^{\cal O}_0(\omega,q) 
+ \left[{\cal C}^{\cal O}_0(\omega,q)\right]^2\!
\Sigma^{\cal O}(\omega,q)+ \left[{\cal C}_0^{\cal O}(\omega,q)\right]^3
\!\Big[\Sigma^{\cal O}(\omega,q)\Big]^2\!+\!\ldots\nn
&=&
{\cal C}_0^{\cal O}(\omega,q) + \left[{\cal C}_0^{\cal O}(\omega,q)\right]^2
\Sigma^{\cal O}_1(\omega,q)
+ \ldots
\eea
and then comparing this expansion to \fr{lowTex}. This gives
\bea
\Sigma_1^{\cal O}(\omega,q)&=&{\cal C}^{\cal O}_1(\omega,q)
\left[{\cal C}^{\cal O}_0(\omega,q)\right]^{-2},\nn
\Sigma_2^{\cal O}(\omega,q)&=&-{\cal C}^{\cal O}_0(\omega,q)
\left[\Sigma^{\cal O}_1(\omega,q)\right]^2
+{\cal C}_2(\omega,q)\left[{\cal C}_0^{\cal O}(\omega,q)\right]^{-2},\nn
\Sigma^{\cal O}_3(\omega,q)&=&\ldots
\label{gensigman}
\eea
We now turn to the implementation of the programme set out above to
the case of the quantum Ising model.

\section{Quantum Ising Model}
\label{sec:ising}
\begin{figure}[ht]
\begin{center}
\epsfxsize=0.45\textwidth
\epsfbox{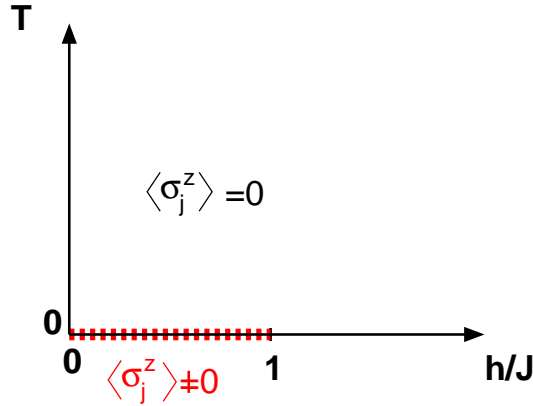}
\end{center}
\caption{Phase diagram of the quantum Ising chain. At zero temperature
ferromagnetic long range order occurs for $h<J$.}
\label{fig:phase}
\end{figure}

The Hamiltonian of the transverse field Ising ferromagnet is given by
\be
H=\sum_n-J\sigma^z_n\sigma^z_{n+1}+h\sigma^x_n\ ,
\label{Hising}
\ee
where we take $J,h>0$. The phase diagram of the model \fr{Hising} is
shown in Fig. \ref{fig:phase}.
At zero temperature the quantum Ising model \fr{Hising} exhibits two
phases: 
\begin{enumerate}
\item{}{\it Ordered Phase: }

This phase occurs for $h<J$ and is characterized by spontaneously
broken $\mathbb{Z}_2$ symmetry associated with long-range magnetic order along
the $z$-direction
\be
\langle \sigma^z_n\rangle\neq 0.
\ee
\item{} {\it Disordered Phase:}

This phase occurs for $h>J$. There is no spontaneous symmetry
breaking and the only ordered moment is along the x-direction.
\end{enumerate}
The two phases are related by the Kramers-Wannier duality 
transformation \cite{kw}, which provides a map of operators and therefore
correlation functions \cite{kogut,fradkin}. By virtue of this map it
is sufficient to consider the regime $h>J$ only. For the remainder of
this paper we will concentrate on this parameter regime.

\subsection{Scaling Limit in the Disordered Phase}
The quantum Ising model \fr{Hising} can be solved exactly
\cite{McCoyWu73,barouch}. However, rather than analyzing the lattice
model \fr{Hising} directly we concentrate on the simpler scaling limit
defined by 
\be
J,h\to\infty\ ,\ a_0\to 0,\qquad |J-h|=\Delta\ {\rm fixed}\ ,\
v=a_0\sqrt{Jh}\ {\rm fixed}.
\ee
Here $a_0$ is the lattice spacing. In the scaling limit the lattice
spin operators turn into continuum fields $\sigma^z_n\rightarrow
C a_0^{1/8}\sigma(x)$, while the Hamiltonian is 
expressed in terms of left and right-moving Majorana fermions
\begin{equation}
H =\int_{-\infty}^\infty \frac{dx}{2\pi}\left[ \frac{iv}{2}({\bar\psi}
\partial_x\bar\psi - \psi\partial_x\psi) -
i\Delta\psi{\bar\psi}\right].
\label{Hscaling}
\end{equation} 
The excitation spectrum of \fr{Hscaling} follows from a mode expansion
of the Majorana field and is given by
\be
\veps(q)=\sqrt{\Delta^2+v^2q^2}.
\ee
In the disordered phase excitations can be thought of in terms of
simple spin flips, as can be seen by considering the limit $h\gg J$.
Let us denote the corresponding annihilation and creation operators by
$Z(\theta)$ and $Z^\dagger(\theta)$ respectively. They fulfil the
simple Faddeev--Zamolodchikov algebra
\begin{eqnarray}
A(\theta_1)A(\theta_2)&=&SA(\theta_2)A(\theta_1),\nonumber\\*
A^\dagger(\theta_1)A^\dagger(\theta_2)&=&
SA^\dagger(\theta_2)A^\dagger(\theta_1),\label{eq:Aalgebra}\\*
A(\theta_1)A^\dagger(\theta_2)&=&2\pi\delta(\theta_1-\theta_2)
+SA^\dagger(\theta_2)A(\theta_1),\nonumber
\end{eqnarray}
where the scattering matrix is $S=-1$. The ground state is then
defined as
\begin{equation}
A(\theta)|0\rangle=0,
\end{equation}
and a basis of scattering states is given by
\begin{equation}
  \label{eq:defA}
  \ket{\theta_1,\ldots,\theta_n}=
  A^\dagger(\theta_1)\ldots A^\dagger(\theta_n)\ket{0}.
\end{equation}
Energy and momentum of the scattering states are by construction
additive 
\bea
E_s(\th_1,\ldots,\th_s)&=&\sum_{j=1}^s\Delta\cosh\th_j\ ,\quad\nn
P_s(\th_1,\ldots,\th_s)&=&\sum_{j=1}^s\frac{\Delta}{v}\sinh\th_j\ .
\eea
In terms of the states (\ref{eq:defA}) the resolution of the identity
reads 
\bea
\mathrm{id}&=&\ket{0}\bra{0}+\sum_{n=1}^\infty
\int_{-\infty}^\infty\frac{d\theta_1\ldots d\theta_{n}}{(2\pi)^n
  n!}\ket{\theta_1,\ldots,\theta_n}\bra{\theta_n,\ldots,\theta_1}.
\label{eq:residentity}
\eea
For the calculation of correlation functions the knowledge of the matrix
elements or form factors of local operators is necessary. In the
{\sl disordered phase } the non-vanishing form factors of
$\sigma^\mathrm{z}$ contain an odd number of particles and are
given by~\cite{barouch,Berg-79,CardyMuss,yurov,Delfino04} 
\begin{equation}
\bra{0}\sigma(0,0)\ket{\theta_{1},\ldots,\theta_{2n+1}}=
i^n\bs\prod_{i<j}^{2n+1}
\tanh\frac{\theta_i-\theta_j}{2}.
\label{eq:muff} 
\end{equation}
It is customary to choose the normalization of the field $\sigma(x)$
such that 
\be
\lim_{x\to 0}\ \langle
0|\sigma(x)\sigma(0)|0\rangle=\frac{1}{|x|^\frac{1}{4}},
\ee
which implies that
\bea
\bar{\sigma}&=&
2^\frac{1}{12}e^{-\frac{1}{8}}{\cal A}^\frac{3}{2}\
\left[\frac{\Delta}{v}\right]^\frac{1}{8}\ ,
\label{sigmabar}\quad
{\cal A}=1.28242712910062...
\eea
We note that in this normalization the continuum field $\sigma(x)$ is
related to the lattice spin operator by 
$\sigma^z_j\rightarrow 2^{1/24}e^{1/8}{\cal A}^{-3/2}a_0^{1/8}\sigma(x)$.
In what follows we need more general matrix elements of the form
\begin{equation}
\bra{\theta'_1,\ldots,\theta'_k}\sigma(0,0)\ket{\theta_{1},\ldots,\theta_{n}}. 
\end{equation}
These can be calculated using crossing relations following
\cite{Smirnov92book}. The necessary identities are summarized in
\ref{app:xing}.

\subsection{Spectral Representation of the Dynamical Susceptibility}
Our main interest is in calculating the retarded dynamical
susceptibility at finite temperature, which is obtained by
analytically continuing the Matsubara two-point function
\bea
\chi_{\sigma}(\omega,q)&=&\int_0^\beta d\tau dx\ e^{i\omega_n \tau-iqx}
\chi_{\sigma}(\tau,x)\Bigr|_{\omega_n\to\delta-i\omega}\ ,\nn
\chi_{\sigma}(\tau,x)&=&-\langle T_\tau \sigma(\tau,x) \sigma(0,0)\rangle.
\eea
In the basis of scattering states introduced above, the
following formal spectral representation for the finite temperature dynamical
susceptibility holds
\be
\chi_{\sigma}(\omega,q)=\frac{1}{{\cal Z}}\sum_{r,s=0}^\infty
C^{\sigma}_{r,s}(\omega,q)\ . 
\label{chiSR}
\ee
Here $C_{r,s}$ denotes the contribution with $r$ particles in the
thermal trace and $s$ in the intermediate state and is of the form
\fr{CrsO} without isotopic quantum numbers $a_j$ and $a'_k$.
As we have already stated, both the partition function and the Lehmann
representations of correlation functions are ill-defined in the
infinite volume limit by virtue of the normalization condition of
states \fr{gennorm}. Setting this issue aside for a moment, we may
cast $C^{\sigma}_{r,s}$ in the form
\bea
C^{\sigma}_{r,s}(\omega,q)&=&
\int\frac{d\th_1\ldots d\th_r}{(2\pi)^rr!}
\int\frac{d\th'_1\ldots d\th'_s}{(2\pi)^ss!}2\pi\delta(q+P_r-P_s)\nn
&&\times\qquad
\frac{e^{-\beta E_r}-e^{-\beta E_s}}{\omega+i\delta-E_s+E_r}
|\langle\theta_1\ldots\th_r|\sigma(0,0)|\th'_s\ldots\th'_1\rangle|^2.
\label{Crs2}
\eea
In order to implement the low temperature expansion we separate
the $C^{\sigma}_{r,s}$ according to their (formal) temperature dependencies
into $E^{\sigma}_{r,s}$ and $F^{\sigma}_{r,s}$ following \fr{CEF}, \fr{Ers}.
At zero temperature the spectral sum simplifies dramatically as only
terms with $r=0$ or $s=0$ remain.

\subsection{Zero Temperature Dynamical Susceptibility}
\begin{figure}[ht]
\begin{center}
\epsfxsize=0.48\textwidth
\epsfbox{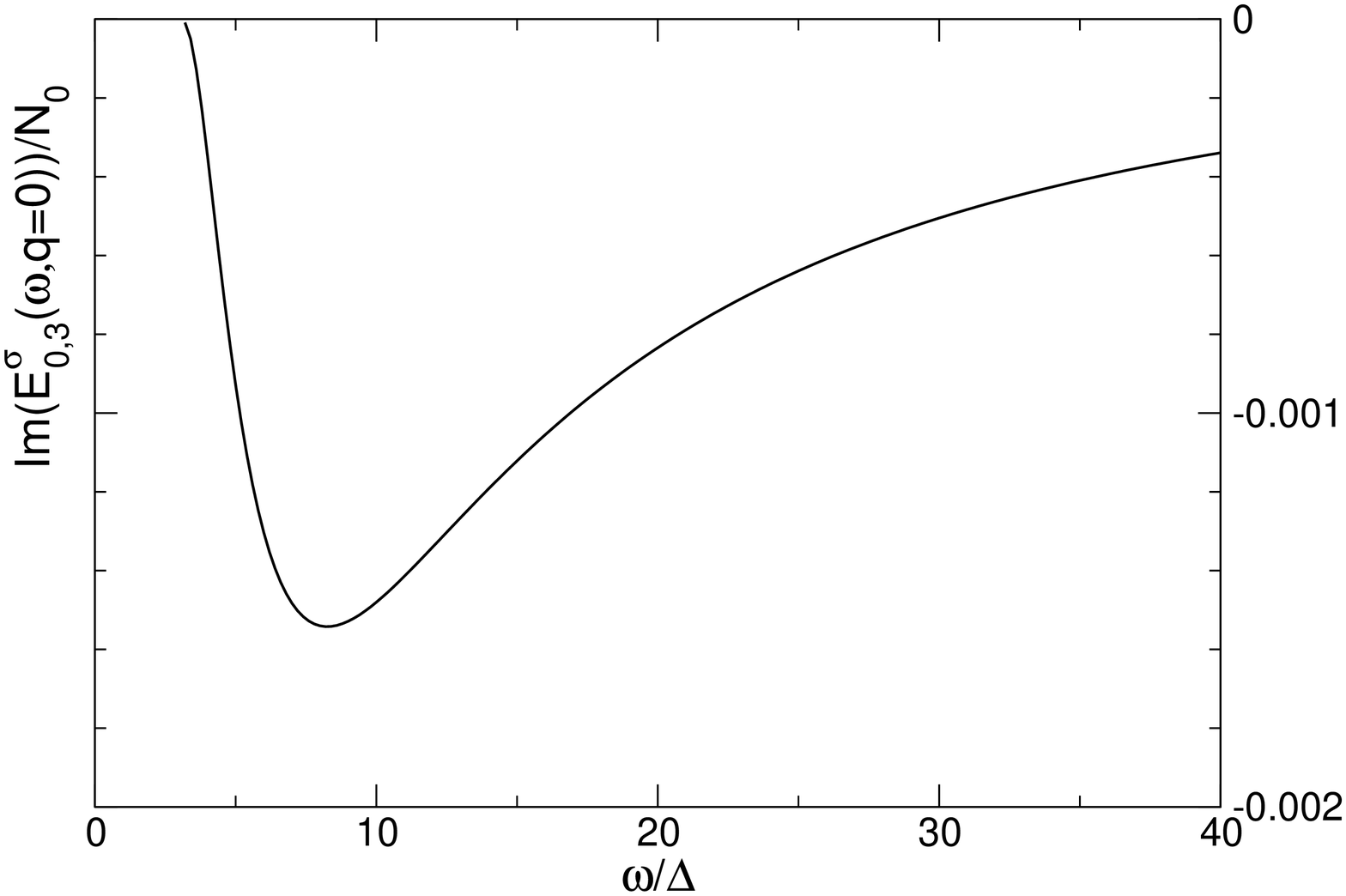}\quad
\epsfxsize=0.48\textwidth
\epsfbox{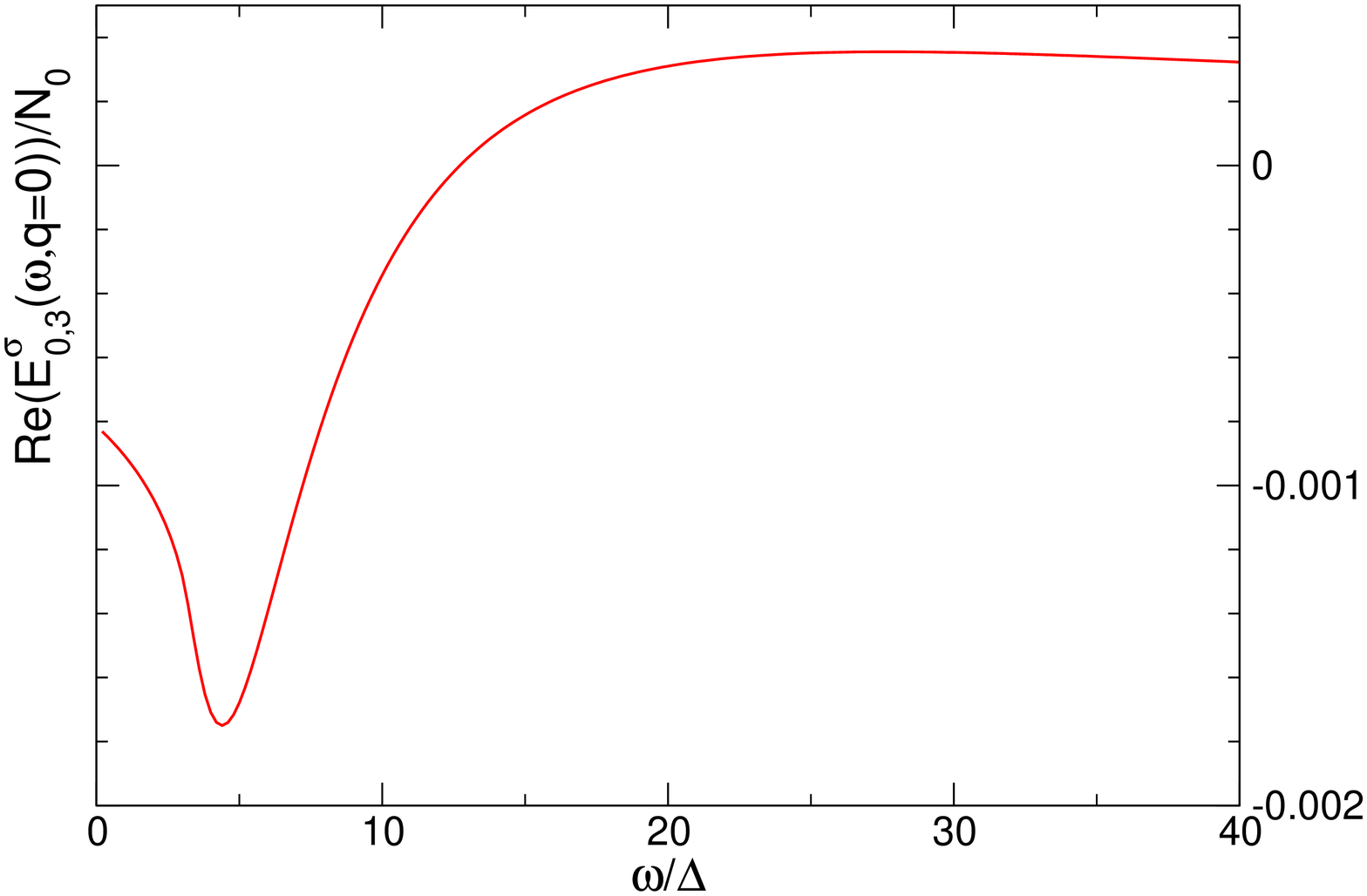}
\end{center}
\caption{Real and imaginary parts of $E^{\sigma}_{03}(\omega,q=0)$.}
\label{fig:isingT=0}
\end{figure}

At $T=0$ the leading contributions to the dynamical susceptibility at
low frequencies are
\bea
\chi_{\sigma}(\omega,q)&\approx&\left[
E^{\sigma}_{0,1}(\omega,q)+F^{\sigma}_{1,0}(\omega,q)
+E^{\sigma}_{0,3}(\omega,q)+F^{\sigma}_{3,0}(\omega,q)\right].
\eea
Here the 1-particle contributions are
\bea\label{op}
E^{\sigma}_{0,1}(\omega,q)&=&\frac{v\bs^2}{\veps(q)}\frac{1}{\omega-\veps(q)+i0},\nn
F^{\sigma}_{1,0}(\omega,q)&=&-\frac{v\bs^2}{\veps(q)}\frac{1}{\omega+\veps(q)+i0}.
\eea
The 3-particle terms can be cast in the form
\bea
&&E^{\sigma}_{0,3}(\omega,q)=
\left[F^{\sigma}_{3,0}(-\omega,-q)\right]^*\nn
&&=\frac{v\bs^2}{\Delta}\int\frac{d\th_1 d\th_2}{6(2\pi)^2}\frac{1}{\cosh\th_3}
\frac{\tanh^2\Bigl(\frac{\th_1-\th_2}{2}\Bigr)
\tanh^2\Bigl(\frac{\th_1-\th_3}{2}\Bigr)
\tanh^2\Bigl(\frac{\th_2-\th_3}{2}\Bigr)}
{\omega-\Delta\sum_{j=1}^3\cosh\th_j+i0},
\eea
where
\be
\th_3={\rm arcsinh}\Bigl(\frac{vq}{\Delta}-\sinh\th_1-\sinh\th_2\Bigr).
\ee
We plot the real and imaginary parts of $E_{0,3}(\omega,q=0)$ in
Fig.\ref{fig:isingT=0}. 
In order to plot
these functions it is useful to separate off a dimensionful
normalization factor 
\be
N_0=\frac{v\bs^2}{\Delta^2}.
\label{normalizationN0}
\ee
We see that by virtue of the smallness of
$E^{\sigma}_{0,3}$ the dynamical response at low energies is dominated by the
coherent single-particle contributions $E^{\sigma}_{0,1}+F^{\sigma}_{1,0}$. 
This yields the following result for the temperature independent part
of the spectral representation for the dynamical susceptibility
\bea
{\cal C}_0^{\sigma}(\omega,q)&\approx&E^{\sigma}_{0,1}(\omega,q)
+F^{\sigma}_{1,0}(\omega,q)=\frac{2v\bs^2}{(\omega+i0)^2-\veps^2(q)},
\eea
where $\veps(q)=\sqrt{\Delta^2+v^2q^2}$.

\subsection{Infinite Volume Regularization}
At low temperatures the next most important contributions arise from
$C^{\sigma}_{1,2}$ and $C^{\sigma}_{2,1}$, which are formally given by
\bea
C^{\sigma}_{1,2}(\omega,q)&=&v
\int\frac{d\theta d\theta_1 d\theta_2}{2(2\pi)^2}
\frac{e^{-\Delta\beta c(\theta)}-e^{-\Delta\beta[c(\th_1)+c(\th_2)]}}
{\omega+i0-\Delta[c(\theta_1)+c(\theta_2)-c(\theta)]}\nn
&&\times\quad
|\langle\theta|\sigma(0)|\theta_2,\theta_1\rangle|^2
\delta\bigl(vq-\Delta[s(\theta_1)+s(\theta_2)-s(\theta)]\bigr),
\label{C12}
\eea
\bea
C^{\sigma}_{2,1}(\omega,q)&=&v
\int\frac{d\theta d\theta_1 d\theta_2}{2(2\pi)^2}
\frac{e^{-\Delta\beta[c(\th_1)+c(\th_2)]}-e^{-\Delta\beta c(\theta)}}
{\omega+i0+\Delta[c(\theta_1)+c(\theta_2)-c(\theta)]}\nn
&&\times\quad
|\langle\theta|\sigma(0)|\theta_2,\theta_1\rangle|^2
\delta\bigl(vq+\Delta[s(\theta_1)+s(\theta_2)-s(\theta)]\bigr),
\label{C21}
\eea
where $c(\theta)=\cosh\theta$ and $s(\theta)=\sinh\theta$.
We note that $C_{2,1}$ can be obtained from $C_{1,2}$ as
\be
C^{\sigma}_{2,1}(\omega,q)=\left[C_{1,2}^{\sigma}(-\omega,-q)\right]^*.
\label{c21c12}
\ee
In order to proceed further we need to evaluate the products of
form factors. Following Smirnov \cite{Smirnov92book}
we have the following crossing
relations for three-particle form factors of the spin field in the
disordered phase
\bea
\langle\theta_1,\theta_2|\sigma(0,0)|\theta_3\rangle&=&
\langle\theta_1-i0,\theta_2-i0|\sigma(0,0)|\theta_3\rangle
+2\pi\bs[\delta(\theta_{32})-\delta(\theta_{31})]\nn
&=&\langle\theta_1+i0,\theta_2+i0|\sigma(0,0)|\theta_3\rangle
-2\pi\bs[\delta(\theta_{32})-\delta(\theta_{31})]\nn
&=&\langle\theta_1+i0,\theta_2-i0|\sigma(0,0)|\theta_3\rangle
+2\pi\bs[\delta(\theta_{32})+\delta(\theta_{31})]\nn
&=&\langle\theta_1-i0,\theta_2+i0|\sigma(0,0)|\theta_3\rangle
-2\pi\bs[\delta(\theta_{32})+\delta(\theta_{31})],
\label{12ising}
\eea
where we have defined $\theta_{jk}=\theta_j-\theta_k$.
The poles occurring in the form factors appearing on the RHS
have been shifted away from the real rapidity axis. The
delta-function contributions correspond to disconnected pieces of the
form factor. It is clear from expression (\ref{12ising}) that the
absolute value squared of a form factor is ill-defined as a
consequence of working in an infinite volume. 
As we will show below, the absolute value squared of a form factor
contains divergent pieces that get canceled by a corresponding
divergence in the partition function. In order to exhibit these
cancellations we need to exhibit the divergences in the form factor
squares explicitly. For the three-particle form factor
of the spin field we do this as follows
\bea
|\langle\theta_3|\sigma(0,0)|\theta_2,\theta_1\rangle|^2
&\equiv&\lim_{\kappa\to 0}
\langle\th_3|\sigma(0,0)|\th_2,\th_1\rangle
\langle\th_1,\th_2|\sigma(0,0)|\th_3+\kappa\rangle.
\eea
The product of form factors on the RHS is now well defined, so that we
can use \fr{12ising} to extract the divergent pieces explicitly. For
form factors involving more than three particles one needs to
introduce several parameters $\kappa_j$, e.g.
\bea
|\langle\theta_4,\theta_5|\sigma(0,0)|\theta_3,\theta_2,\theta_1\rangle|^2
&\equiv&\lim_{\kappa_{1,2}\to 0}
\langle\th_4,\th_5|\sigma(0,0)|\th_3,\th_2,\th_1\rangle\nn
&&\times
\langle\th_1,\th_2,\th_3|\sigma(0,0)|\th_5+\kappa_2,\th_4+\kappa_1\rangle.
\label{kappahigher}
\eea
In
order to calculate the contributions $C^{\sigma}_{1,2}$ and $C^{\sigma}_{2,1}$ by means of
contour integral techniques the following choice of analytically
continuing $\th_{1,2}$ into the complex plane is the most convenient
\bea
&&|\langle\theta_3|\sigma(0,0)|\theta_2\theta_1\rangle|^2=\nn
&&\quad\lim_{\kappa\to 0}\Bigl(
\langle\th_1+i0,\th_2-i0|\sigma(0,0)|\th_3+\kappa\rangle
+2\pi\bs[\delta(\theta_{32}+\kappa)+\delta(\theta_{31}+\kappa)]\Bigr)
\nn
&&\qquad\qquad\times\ \Bigl(\langle\th|\sigma(0,0)|\theta_2-i0,\theta_1+i0\rangle
-2\pi\bs[\delta(\theta_{32})+\delta(\theta_{31})]\Bigr).
\label{ffsq}
\eea
The product of from factors on the RHS of eqn \fr{ffsq} is evaluated
in \ref{app:A}. Using \fr{FF2ising} we can express the
contribution $C^{\sigma}_{1,2}$ in the form
\be
C^{\sigma}_{1,2}(\omega,q)=C_{1,2}^{\rm conn}(\omega,q)+C_{1,2}^{\rm
  dis}(\omega,q),
\label{c12decomp}
\ee
where the ``connected'' $C_{1,2}^{\rm conn}(\omega,q)$ 
and ``disconnected'' $C_{1,2}^{\rm dis}(\omega,q)$ parts correspond to
the contributions of the first and second terms on the rhs of
\fr{FF2ising}. The disconnected part is given by
\bea
C_{1,2}^{\rm dis}(\omega,q)&=&v\bs^2\lim_{\kappa\to 0}
\int d\theta_1 d\theta_2 d\th_3
\frac{e^{-\Delta\beta c(\theta_3)}-e^{-\Delta\beta[c(\th_1)+c(\th_2)]}}
{\omega+i0-\Delta[c(\theta_1)+c(\theta_2)-c(\theta_3)]}\nn
&&\times
\delta\bigl(vq-\Delta[s(\theta_1)+s(\theta_2)-s(\theta_3)]\bigr)
\left[\delta(\kappa)\delta(\th_{31})
+\delta(\th_{32})\delta(\th_{31})\right]\nn
&=&\frac{v\bs^2}{\veps(q)}
\frac{1-e^{-\beta\veps(q)}}{\omega+i0-\veps(q)}\left[
e^{-\beta\veps(q)}+\delta(\kappa)\int d\theta e^{-\beta\Delta
c(\th)}\right]\nn
&=&C^{\sigma}_{0,1}(\omega, q)\left[
e^{-\beta\veps(q)}+\delta(\kappa)\int d\theta e^{-\beta\Delta
c(\th)}\right].
\label{C12dis}
\eea
We note that most of the ``cross-terms'' derived in
\ref{app:A} cancel in the integral as they are antisymmetric in
$\th_1$ and $\th_2$, while the remaining part of the integrand is
symmetric. The connected part of $C_{1,2}$ is given by
\bea
C_{1,2}^{\rm conn}(\omega,q)&=&v\bs^2
\int\frac{d\theta d\theta_+ d\theta_-}{(2\pi)^2}
\left[e^{-\Delta\beta c(\theta)}-e^{-2\Delta c(\th_+)c(\th_-)}\right]\nn
&\times&
K(\theta_-,\theta_+,\theta)\
\frac{\delta\bigl(vq-\Delta[2s(\theta_+)c(\theta_-)-s(\theta)]\bigr)}
{\omega-\Delta[2c(\theta_+)c(\theta_-)-c(\theta)]}
\label{C12conn}
\eea
where we have changed variables to
\be
\theta_\pm=\frac{\theta_2\pm\theta_1}{2},
\ee
and defined a function
\be
K(\theta_-,\theta_+,\theta)=
\frac{\tanh^2(\theta_-)}{
\tanh^2\Bigl(\frac{\theta_--\theta+\theta_+-i0}{2}\Bigr)
\tanh^2\Bigl(\frac{\theta_-+\theta-\theta_+-i0}{2}\Bigr)}.
\ee
We can carry out the integration over $\theta_+$ using the momentum
conservation delta-function. The result is
\bea
C_{1,2}^{\rm conn}(\omega,q)&=&v\bs^2\!\!
\int\frac{d\theta  d\theta_-}{(2\pi)^2}
\frac{e^{-\Delta\beta c(\theta)}-
e^{-\beta u(q,\th,\th_-)}}
{\ot-u(q,\th,\th_-)}\frac{K(\theta_-,\theta_+^0(q,\th,\th_-),\theta)}
{u(q,\th,\th_-)}.
\label{S12conn2}
\eea
Here we have introduced the notations
\bea
\ot&=\omega+i\delta+\Delta\cosh(\theta)\ ,\nn
\qt&=q+\frac{\Delta}{v}\sinh(\theta)\ ,\nn
\theta_+^0(q,\th,\th_-)&={\rm arcsinh}\Bigl(\frac{v\qt}{2\Delta c(\theta_-)}\Bigr),\nn
u(q,\theta,\theta_-)&=\sqrt{(v\qt)^2+(2\Delta c(\theta_-))^2}.
\label{ot}
\eea
In order to proceed it is useful to consider the integrand as a
function of the complex variable $\theta_-$ (the analytic properties
as a function of $\theta$ are not as nice). We want to move the
integration contour away from the real axis in order to avoid the
vicinities of the double poles. The branch point of the square roots
and inverse hyperbolic functions occur only at $|{\rm
  Im}(\theta_-)|=\frac{\pi}{2}$, which allows us to move the contour
to a line parallel to the real axis in the lower half-plane. When
doing this we may encounter a simple pole when 
\be
\ot-u(q,\th,\th_-)=0.
\ee
Two kinds of solutions to this equation exist in the strip 
$-\frac{\pi}{2}<\gamma\leq{\rm Im}(\theta_-)\leq 0$ (for simplicity we
assume $\omega>0$ in the following)
\begin{enumerate}
\item{}
If $\Omega^2-(vQ)^2$ is larger than $4\Delta^2$, we have a
simple pole at 
\bea
\label{tm0}
\alpha(\omega,q,\th)&=&-{\rm arccosh}
\Bigl(\frac{\tilde{s}(\omega,q,\th)}{2\Delta}\Bigr)-i0\ ,\\
\tilde{s}(\omega,q,\theta)&=&\left[(\omega+\Delta\cosh\theta)^2-(vq+\Delta\sinh\theta)^2\right]^\frac{1}{2} .
\label{stilde}
\eea
\item{}
If $4\Delta^2\cos^2\gamma<\Omega^2-(vQ)^2<4\Delta^2$, we have a pole at
\be
\label{tm1}
\alb(\omega,q,\th)=-i\ {\rm arccos}
\Bigl(\frac{\tilde{s}(\omega,q,\th)}{2\Delta}\Bigr)-0\ .
\ee
\end{enumerate}
Defining
\be
\label{t0}
\theta_0(\omega,q,\th)={\rm
arcsinh}\Bigl(\frac{vq+\Delta\sinh\theta}{\tilde{s}(\omega,q,\th)}\Bigr),
\ee
we may cast $C_{12}^{\rm conn}$ in the form
\bea
C_{1,2}^{\rm conn}(\omega,q)&=&
-iv\bs^2\bigl(1-e^{-\beta\omega}\bigr)
\int_{\rm S_+}\frac{d\theta}{2\pi}\frac{e^{-\beta\Delta c(\theta)}\
K(\alpha,\theta_0,\theta)
}
{\tilde{s}(\omega,q,\th)\sqrt{\tilde{s}^2
(\omega,q,\th)-4\Delta^2}}\  
\nn
&-&v\bs^2\bigl(1-e^{-\beta\omega}\bigr)
\int_{T_+^\gamma}\frac{d\theta}{2\pi}\frac{e^{-\Delta\beta c(\theta)}
\ K({\overline\alpha},\theta_0,\theta)}
{\tilde{s}(\omega,q,\th)\sqrt{4\Delta^2-\tilde{s}^2(\omega,q,\th)}}
\nn
&+&
v\bs^2\int\frac{d\theta}{(2\pi)^2}\int_{\rm S}d\theta_-
\frac{e^{-\beta\Delta c(\th)}-e^{-\beta u(q,\th,\th_-)}}
{\ot-u(q,\theta,\theta_-)}
\frac{K(\theta_-,\theta_+^0,\theta)}{u(q,\theta,\theta_-)}.
\label{S12conn4}
\eea
Here $\rm S$ is a straight line in the lower half-plane parallel to
the real axis with imaginary part $-i\gamma$, $\rm S_+$ are the
segments of the real axis such that $\tilde{s}^2>4\Delta^2$ and
$T_+^\gamma$ are the segments of the real axis such that 
$4\Delta^2\cos^2\gamma\leq \tilde{s}^2\leq 4\Delta ^2$. The segments
$\rm S_+$, $\rm T^\gamma_+$ for positive frequencies $\omega>0$ can be
characterized as follows:  
\begin{enumerate}
\item{$\rm S_+$:}

\be
\theta\in\left\{\begin{array}{ll}
(-\infty,\infty) & {\rm if}\ \omega>|vq|\ {\rm and}\ s^2>\Delta ^2\ ,\\
(-\infty,\thh_-]\cup[\thh_+,\infty)& {\rm if}\ \omega>|vq|\ {\rm
      and}\ s^2<\Delta ^2\ ,\\
(-\infty,\thh_-]& {\rm if}\ 0<\omega<vq\ ,\cr
[\thh_+,\infty)& {\rm if}\ 0<\omega<-vq\ ,\\
\end{array}
\right.
\ee
where
\be
\thh_\pm=\ln\left[\frac{3\Delta ^2-s^2\pm\sqrt{s^4-10\Delta ^2s^2+9\Delta ^4}}{2\Delta (\omega-vq)}\right] .
\ee
\item{${T^+_\gamma}$:}

\be
\theta\in\left\{\begin{array}{ll}
[\thh_-,\thh_+] & {\rm if}\ \omega>|vq|\ {\rm and}\ \Delta ^2
\gamma_-^2<s^2<\Delta^2\ ,\\
\lbrack\thh_-,\thh_-'\rbrack\cup\lbrack\thh_+',\thh_+\rbrack& {\rm if}\ \omega>|vq|\ {\rm
      and}\ s^2<\Delta ^2\gamma_-^2\ ,\\
\lbrack\thh_-,\thh_-'\rbrack& {\rm if}\ 0<\omega<vq\ ,\cr
\lbrack\thh_+',\thh_+\rbrack& {\rm if}\ 0<\omega<-vq\ ,\\
\end{array}
\right.
\ee
where $\gamma_\pm=1\pm2\cos\gamma$ (we have assumed that
$0<\gamma<\frac{\pi}{3}$) and 
\be
\thh_\pm'=\ln\left[\frac{-s^2-\Delta ^2\gamma_+\gamma_-\pm
\sqrt{(s^2+\Delta ^2\gamma_+\gamma_-)^2-4\Delta ^2s^2}}{2\Delta
(\omega-vq)}
\right] .
\ee
\end{enumerate}
The real part of $C_{1,2}^{\rm conn}$ is given by the last two terms
in \fr{S12conn4}. For fixed $\gamma$ there always will be a value of
$\theta$ such that the $\theta_-$ integral in the third line of
\fr{S12conn4} is very close to a singularity at $\theta_-=0$. The
problem occurs at $\theta$ such that 
\be
\ot-\sqrt{(v\qt)^2+4\Delta ^2\cos^2(\gamma)}=0.
\ee
For $\omega>|vq|$ there are singularities at $\theta=\thh'_\pm$ if
$s^2<\Delta ^2\gamma_-^2$ and for $\omega<vq$ there is a single singularity
at $\theta=\thh'_-$. From a practical point of view we are
interested only in the regime
\be
s^2\approx \Delta ^2.
\ee
For $\gamma=\frac{\pi}{3}$ we have $\gamma_-=0$, so that we basically
can always stay away from this problem. If we want to know the answer
for $s^2<0$ we can always calculate $f(\theta,\omega,q)$ for different
values of $\gamma$ and in this way always stay away from having to
deal with a singularity. This concludes our evaluation of
$C_{2,1}(\omega,q)$ in the infinite volume regularization.
The contribution $C_{2,1}(\omega,q)$ can be obtained in the same way. 

We are now in a position to calculate the the leading contributions to
the quantity ${\cal C}^{\sigma}_1(\omega,q)$
\be
{\cal C}^{\sigma}_1
\approx E^{\sigma}_{1,0}+F^{\sigma}_{0,1}+
\left(E^{\sigma}_{1,2}-{\cal Z}_1E^{\sigma}_{0,1}\right)
+\left(F^{\sigma}_{2,1}-{\cal Z}_1F^{\sigma}_{1,0}\right).
\ee
The leading corrections to ${\cal C}^{\sigma}_1$ are 
$(E^{\sigma}_{1,4}-{\cal Z}_1E^{\sigma}_{0,3})$ and
$(F^{\sigma}_{4,1}-{\cal Z}_1F^{\sigma}_{3,0})$ and 
these are expected to be small for the same reasons that $E^{\sigma}_{0,3}$ is
negligible for $\omega\approx \Delta$. We further observe that
$F^{\sigma}_{2,1}-{\cal Z}_1F^{\sigma}_{1,0}$ is negligibly small in
the parameter regime we are interested in ($\omega\approx\Delta$ and
low temperatures). We therefore drop it in the following.

As ${\cal Z}_1$ is ill-defined in the infinite-volume limit we
regulate it by ``shifting the trace'' in the same way as \fr{ffsq}
\bea
{\cal Z}_1&=&\lim_{\kappa\to 0}\int\frac{d\th}{2\pi}e^{-\beta\Delta
  c(\th)}\langle \th|\th+\kappa\rangle =\lim_{\kappa\to 0}
\delta(\kappa)\int d\th\  e^{-\beta\Delta c(\th)}. 
\label{z1reg}
\eea
Combining \fr{Cl}, \fr{c12decomp}, \fr{C12dis},
\fr{S12conn4} and \fr{z1reg} we arrive at the following expression for
the first subleading part of the dynamical susceptibility \fr{lowTex}

\bea
{\cal C}_1^{\sigma}(\omega,q)
&\approx&
-iv\bs^2\int_{\rm S_+ }\frac{d\theta}{2\pi}\frac{e^{-\beta\Delta c(\theta)}
\ K(\alpha(\omega,q,\th),
\theta_0(\omega,q,\th),\theta)}
{\tilde{s}(\omega,q,\th)
\sqrt{\tilde{s}^2(\omega,q,\th)-4\Delta^2}}\nn
&&-v\bs^2\int_{T_+^\gamma}\frac{d\theta}{2\pi}\frac{e^{-\Delta\beta
    c(\theta)}
\ K(\alb(\omega, q,\th),
\theta_0(\omega, q,\th),\theta)}
{\tilde{s}(\omega,q,\th)
\sqrt{4\Delta^2-\tilde{s}^2(\omega,q,\th)}}
\nn
&&+v\bs^2
\int\frac{d\theta}{(2\pi)^2}\int_{\rm S}d\theta_-
\frac{e^{-\beta\Delta c(\th)}\ K(\theta_-,\theta_+^0(q,\th,\th_-),\theta)}
{\left[\Omega(\th,\omega)-u(q,\theta,\theta_-)\right]u(q,\theta,\theta_-)}.\nn
\label{C1inf}
\eea
Here $S_+$ and $T_+^\gamma$ are the segments of the real axis
characterized by $\tilde{s}^2(\omega, q,\th)>4\Delta^2$ and
$4\Delta^2\cos^2\gamma\leq\tilde{s}^2(\omega, q,\th)\leq 4\Delta^2$ 
respectively and $S$ is the contour from $-\infty-i\gamma$ to 
$\infty-i\gamma$ parallel to the real axis.

\subsection{Finite Volume Regularization}
Another way to regulate infinities in matrix elements is to work in a
large, finite volume $R$. The Hamiltonian on finite, periodic line of
length $R$ is 
\begin{equation}
\label{eiv}
H = \int^R_0 \frac{dx}{2\pi}\left[
 \frac{iv}{2}({\bar\psi} \partial_x\bar\psi - \psi\partial_x\psi) 
- i\Delta\psi{\bar\psi}\right].
\end{equation}
The Hilbert space of the theory divides itself into two sectors:
Neveu-Schwarz (NS) and Ramond (R).  The NS-sector consists of a Fock
space built with  even numbers of half-integer fermionic modes,
i.e. states of the form 
\be
|p_1\cdots p_{2N}\rangle_{NS} \equiv a^\dagger_{p_1}\cdots
a^\dagger_{p_{2N}}|0\rangle_{NS}\ ,
\ee
where a mode's momentum satisfies
\be
p_i=\frac{2\pi}{R}\Big(n_i+\frac{1}{2}\Big),\quad n_i\in\mathbb{ Z}.
\ee
On the other hand, the R-sector consists of a Fock space composed of
odd numbers of even integer fermionic modes,
\be
|k_1\cdots k_{2M+1}\rangle_R \equiv a^\dagger_{k_1}\cdots
a^\dagger_{k_{2M+1}}|0\rangle_R\ ,\quad
k_i=\frac{2\pi}{R} n_i.
\ee
Energy $E(p_i)$ and momentum $P(p_i)$ of a NS state $|p_1\cdots
p_{2N}\rangle_{NS}$ are given simply by 
\be
E(\{p_i\})= \sum_{i=1}^{2N}\veps(p_i)\ ,\quad
P(\{p_i\})=\sum_{i=1}^{2N}p_i,
\ee
where as before $\veps(p)=\sqrt{\Delta^2+v^2p^2}$.
An identical relation holds for states in the R-sector. 
It is useful to parametrize the momenta in terms of a rapidity
variable as 
\be
p_i = \frac{\Delta}{v}\sinh(\theta_{p_i})\ .
\ee
The finite volume form factors of the spin field have been determined
in Refs \cite{bugrij,zamo}. The non-vanishing form factors in the
disordered phase are
\begin{eqnarray}
\label{ev}
_R\langle k_1\cdots k_{2M+1}|\sigma(0)|p_1\cdots p_{2N}\rangle_{NS}
&=& i^{M+N}C_R \prod_{i,j}g(\theta_{k_i})g(\theta_{p_j})\cr
&& \hskip -2.3in 
\times \prod_{i<j}\tanh\Big(\frac{\theta_{k_i}-\theta_{k_j}}{2}\Big)
\prod_{i<j}\tanh\Big(\frac{\theta_{p_i}-\theta_{p_j}}{2}\Big)
\prod_{i,j} \coth\Big(\frac{\theta_{k_i}-\theta_{p_j}}{2}\Big),
\end{eqnarray}
where for large $\Delta R$ we have
\bea
C_R&=&\bs+{\cal O}\Bigl(e^{-\Delta R}\Bigr)\ ,\nn
g(\theta)&=& \frac{1+{\cal O}\Bigl(e^{-\Delta R}\Bigr)}
{\sqrt{\Delta R v^{-1}\cosh(\theta)}}\ .
\eea
We note that the factors $g(\theta)$ disappear when the states are
normalized in terms of rapidity variables. Importantly, up to
exponentially small corrections, the matrix elements \fr{ev} have the
same functional form as at $R=\infty$. The essential difference
between large but finite $R$ and the infinite volume limit is that at
finite $R$, the momenta are quantized. This is a pattern that repeats
itself for general integrable models, as emphasized in
Ref.\cite{takacs}, and that we will exploit for our analysis of spin-1
chains.  

Crucially, for large but finite $R$ the matrix elements \fr{ev} are
finite. In the infinite volume limit divergences develop in the factor
\be
\prod_{i,j} \coth\Big(\frac{\theta_{k_i}-\theta_{p_j}}{2}\Big),
\ee
and occur when two momenta, $k_i$ and $p_j$, approach one another. 
However, a finite $R$ regulates these divergences by virtue of
$k_i$ lying in the R-sector with integer quantization and $p_j$
in the NS-sector with half-integer quantization respectively.
Hence the two are never exactly equal. The finite temperature Lehmann
representation for the two-point function of the spin field on a ring
of length $R$ takes the form
\be
\chi^R_\sigma(\omega,q)=\frac{1}{{\cal Z}^{R}}\sum_{r,s=0}^\infty
C^{R}_{r,s}(\omega,q)\ ,
\label{chiSRfvol}
\ee
where {\sl e.g.}
\begin{eqnarray}
\fl
\label{eix}
C^{R}_{2M+1,2N}(\omega,q = \frac{2\pi  n_q}{R}) &=& 
\int^R_0dx\int^\beta_0d\tau 
e^{i\omega_n\tau-iqx}C^{R}_{2M+1,2N}(\tau,x)\Bigg|_{\omega_n\rightarrow
\eta-i\omega}\nn
&=& \sum_{\{k_j\},\{p_i\}}
|_R\langle k_1\cdots k_{2M+1}|\sigma(0)|p_1\cdots p_{2N}\rangle_{NS}|^2\nn
&&\times\ \frac{e^{-\beta E(\{k_j\})}-e^{-\beta E(\{p_i\})}}{\omega+i\eta-E(\{p_i\})+E(\{k_j\})}
R\delta_{P(\{p_i\})-P(\{k_j\}),q}\ ,\nonumber
\end{eqnarray}
\be
{\cal Z}^{R}=1+\sum_{p\in R}e^{-\beta\veps(p)}+
\sum_{p_1,p_2\in
  NS}e^{-\beta[\veps(p_1)+\veps(p_2)]}+\ldots\equiv 
\sum_{n=0}^\infty{\cal Z}^{R}_n.
\ee
Here $\eta$ is a positive infinitesimal.
All terms in the expansion \fr{eix} are finite. As before, we re-order
the spectral sums according to \fr{Cl}, which gives
\be
\chi^R_\sigma(\omega,q)=\sum_{r=0}^\infty{\cal C}^{R}_{r}(\omega,q)\ ,
\label{chiSRfvol2}
\ee
where the ${\cal C}_r^{R}$ are defined as the finite volume analogs
of \fr{Cl}. 


\subsection{Comparison Between Infinite and Finite Volume Regularizations}
\label{ssec:compIsing}

In this section we establish the equivalence between the infinite
volume regularization used in Section 3.2 to evaluate the leading
order terms of $\chi_\sigma(\omega,q)$ and the finite volume regularization
(as $R\rightarrow\infty$) introduced in the preceding section.  In
particular we establish that ${\cal C}^R_0$ 
+ ${\cal C}^R_1$, the first two terms in the temperature expansion of $\chi^R(\omega, q)$, and given by
\begin{eqnarray}
{\cal C}^R_0 &\approx& E^R_{0,1}+F^R_{1,0}\ ,\cr\cr
{\cal C}^R_1 &\approx& E^R_{1,0}+F^R_{0,1}+\left(E^R_{1,2}-{\cal Z}^R_1E^R_{0,1}\right)
+\left(F^R_{2,1}-{\cal Z}^R_1F^R_{1,0}\right),
\end{eqnarray}
are equal to their counterparts in infinite volume once we take volume
$R$ to infinity. The equivalence of ${\cal C}_0$ in both schemes is
straightforward.  We have for $C^R_{0,1}$ and $C^R_{1,0}$
the following
\begin{eqnarray}
C^R_{0,1}(\tau,x) &=& E^R_{0,1}(\tau,x) + F^R_{0,1}(\tau ,x) = -\bar\sigma^2v \sum_{p_i\in R}
\frac{e^{-\veps(p_i)\tau+ipx}}{R\veps(p_i)}\ ,\nn
C^R_{1,0}(\tau,x) &=& E^R_{1,0}(\tau,x) + F^R_{1,0}(\tau ,x) =
-\bar\sigma^2v \sum_{p_i\in R}
\frac{e^{-\veps(p_i)(\beta-\tau)-ipx}}{R\veps(p_i)}\ .
\end{eqnarray}
Taking the Fourier transform we find
\begin{eqnarray}
E^R_{0,1}(\omega,q) &=&
\frac{\bar\sigma^2v}{\veps(q)}\frac{1}{\omega+i0-\veps(q)}\ ,\quad
F^R_{0,1}(\omega,q) = \frac{\bar\sigma^2v}{\veps(q)}
\frac{-e^{-\beta\veps(q)}}{\omega+i0-\veps(q)},\ \nn
E^R_{1,0}(\omega,q) &=&
\frac{\bar\sigma^2v}{\veps(q)}\frac{e^{-\beta\veps(q)}}{\omega+i0+\veps(q)}\ ,\quad
F^R_{1,0}(\omega,q) = \frac{\bar\sigma^2v}{\veps(q)}
\frac{-1}{\omega+i0+\veps(q)}.
\end{eqnarray}
In comparing this to (\ref{op}) and using
$F^R_{r,s}(\omega,q)=\left[E^{R}_{s,r}(-\omega,-q)\right]^*$ we see that the finite
and infinite volume regularizations for $C^{\sigma}_0$ lead to the same
result. 

\begin{figure}[ht]
\begin{center}
\epsfxsize=0.4\textwidth
\epsfbox{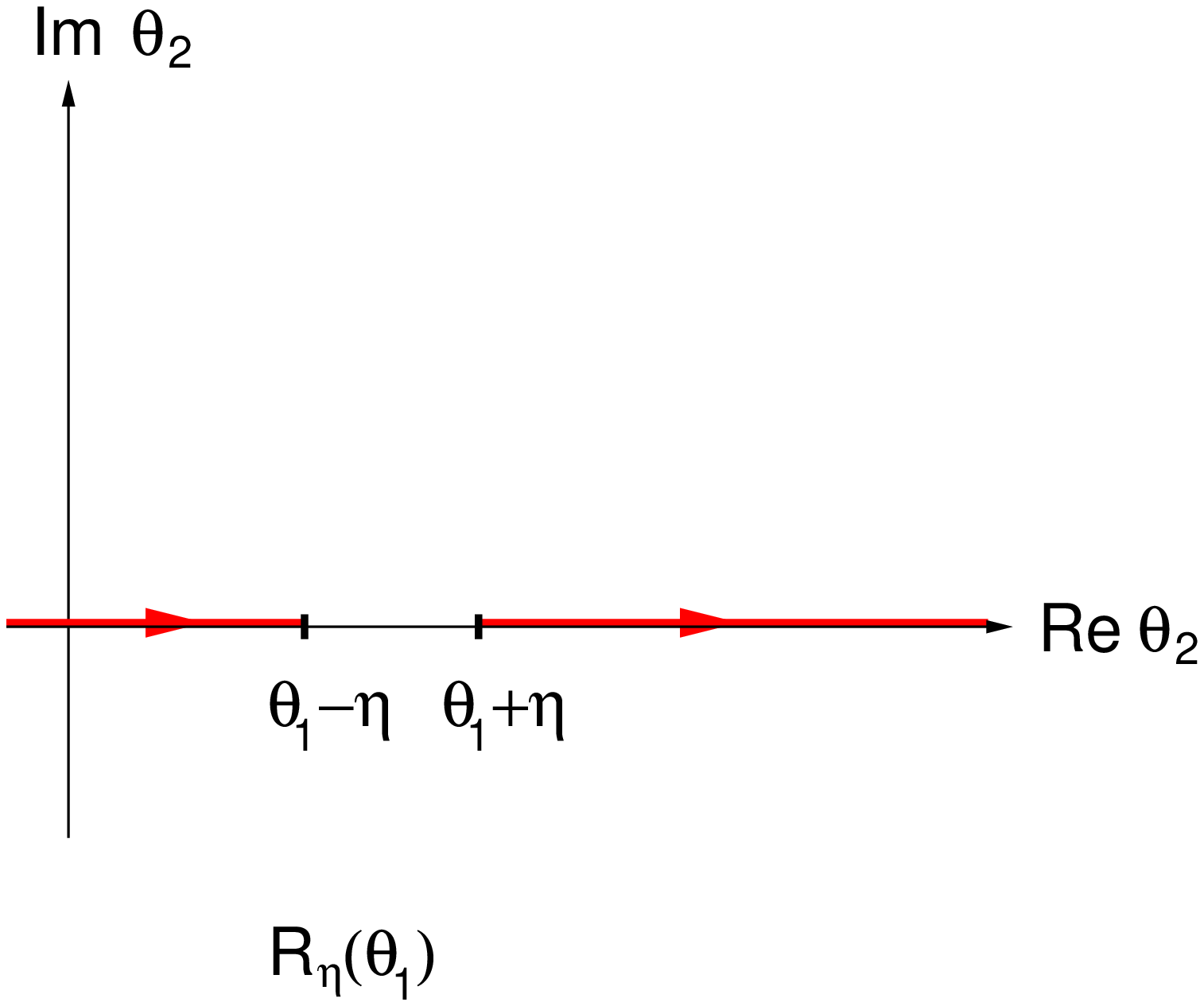}\qquad\qquad
\epsfxsize=0.4\textwidth
\epsfbox{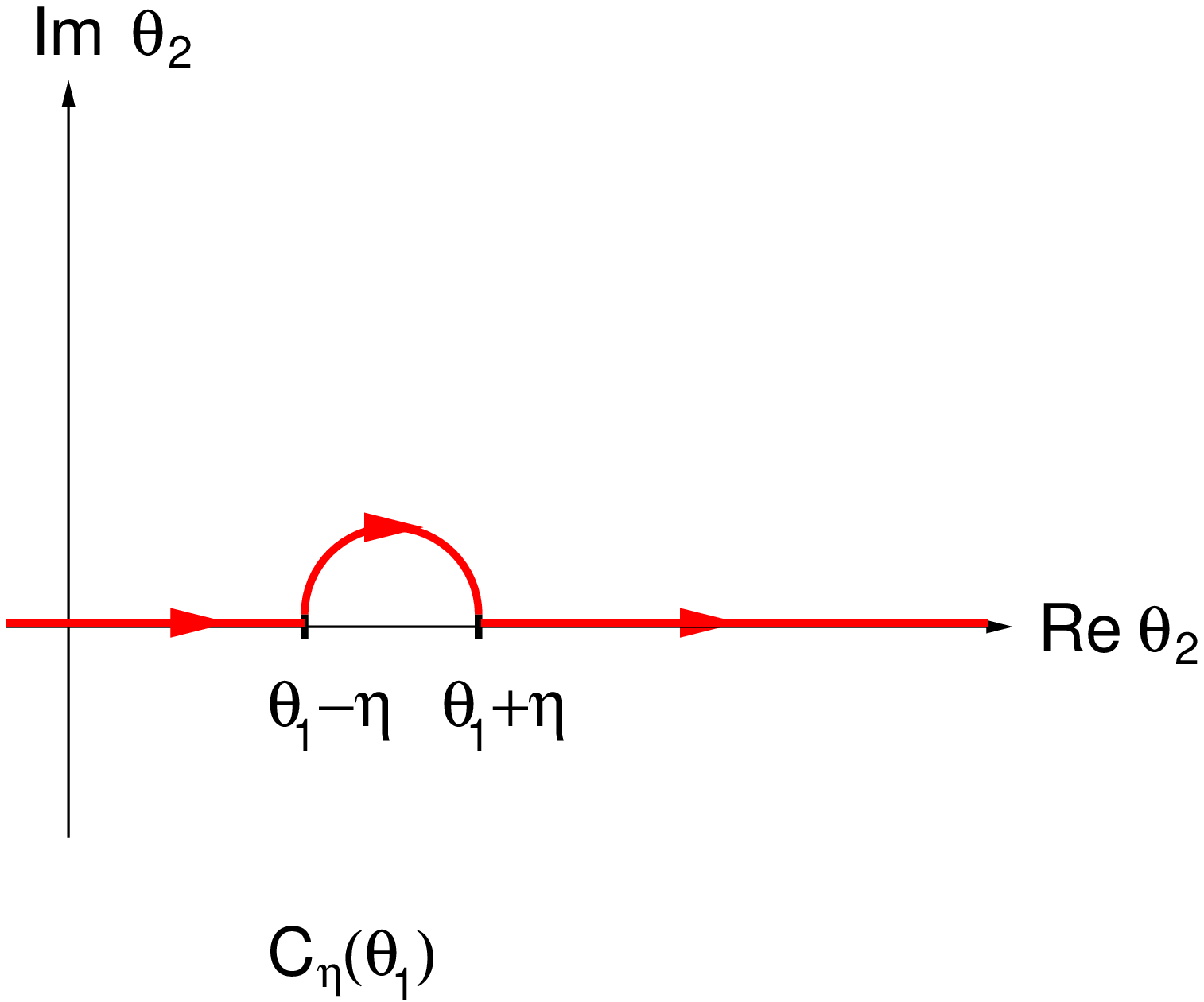}
\end{center}
\caption{Integration contours $R_\eta(\theta)$ and $C_\eta(\theta)$.}
\label{fig:contours}
\end{figure}

The above also establishes that the first two terms of ${\cal C}^{\sigma}_1$,
$E^{\sigma}_{1,0}+F^{\sigma}_{0,1}$, in the two schemes are the same.
It leaves then to verify that   
\bea
\lim_{R\rightarrow\infty}\left(E^R_{1,2}-{\cal Z}^R_1E^R_{0,1}\right)
&=&E^{\sigma}_{1,2}-{\cal Z}_1E^{\sigma}_{0,1},\nn
\lim_{R\rightarrow\infty}\left(F^R_{2,1}-{\cal Z}^R_1F^R_{1,0}\right)
&=&F^{\sigma}_{2,1}-{\cal Z}_1F^{\sigma}_{1,0}.
\eea
We first consider $\left(E^R_{1,2}-{\cal Z}^R_1E^R_{0,1}\right)$.  In
a finite volume this term is equal to
\bea
\label{fvE12}
E^R_{1,2}-{\cal Z}^R_1E^R_{0,1} &=&
\frac{\bar\sigma^2v^3}{2R^2}\Bigg\{
\sum_{p_1\in R}\ \sum_{p_2,p_3 \in NS} 
\frac{\delta_{q,p_2+p_3-p_1}}{\omega+i0-\veps(p_2)-\veps(p_3)+\veps(p_1)}\nn
&&\qquad\times\ 
\frac{e^{-\beta\veps(p_1)}\
\tanh^2\big(\frac{\theta_{23}}{2}\big)\
\coth^2\big(\frac{\theta_{12}}{2}\big)\
\coth^2\big(\frac{\theta_{13}}{2}\big)}
{\veps(p_1)\veps(p_2)\veps(p_3)}
\Bigg\}\nn
&& - \bar\sigma^2v \sum_{p_3\in R}\frac{\delta_{q,p_3}}{\veps(p_3)}\frac{1}{\omega+i0-\veps(p_3)}
\sum_{p_1\in R}e^{-\beta\veps(p_1)}.
\eea
We rewrite ${\cal Z}^R_1$ using the identity
\bea
{\cal Z}^R_1 &=& \sum_{{p_1\in R}} e^{-\beta \veps (p_1)} 
= \sum_{p_1\in R} e^{-\beta \veps (p_1)} 
\frac{1}{\pi^2} \sum_{n\in Z} \frac{1}{(n+1/2)^2}\nn
&=& \sum_{p_1\in R}\sum_{p_2\in NS} \frac{4e^{-\beta \veps
    (p_1)}}{R^2(p_1-p_2)^2}.
\eea
Inserting this into eqn (\ref{fvE12}) we obtain
\bea
\fl
E^R_{1,2}-{\cal Z}^R_1E^R_{0,1}&=& 
\frac{\bar\sigma^2v^3}{2R^2}\sum_{p_1\in R}\ \sum_{p_2,p_3 \in NS}
\Bigg[
\frac{\delta_{q,p_2+p_3-p_1}}{\omega+i0-\veps(p_2)
-\veps(p_3)+\veps(p_1)}\nn
&&\qquad\times\ 
\frac{e^{-\beta\veps(p_1)}\
\tanh^2\big(\frac{\theta_{23}}{2}\big)\
\coth^2\big(\frac{\theta_{12}}{2}\big)\
\coth^2\big(\frac{\theta_{13}}{2}\big)}
{\veps(p_1)\veps(p_2)\veps(p_3)}\nn
&&\qquad - \frac{1}{\veps(p_3+\frac{\pi}{R}))}\frac{\delta_{q,p_3+\frac{\pi}{R}}}{\omega+i0-\veps(p_3+\frac{\pi}{R})}\frac{4e^{-\beta\veps(p_1)}}{v^2(p_1-p_2)^2}\cr\cr
&&\qquad - \frac{1}{\veps(p_2+\frac{\pi}{R}))}\frac{\delta_{q,p_2+\frac{\pi}{R}}}{\omega+i0-\veps(p_2+\frac{\pi}{R})}\frac{4e^{-\beta\veps(p_1)}}{v^2(p_1-p_3)^2}\Bigg].
\eea
We now see that the above expression has no singular pieces of the form 
$(p_1-p_{2})^{-2}$ or $(p_1-p_{3})^{-2}$ as $p_1$ approaches $p_2$ or
$p_3$. There are singular terms of the form $(p_1-p_2)^{-1}$ and
$(p_1-p_3)^{-1}$, but these can be seen to vanish upon summation and
so can be ignored. Given the absence of singular terms we can take the
limit $R\rightarrow \infty$ and convert the summations to principal
value integrals, i.e.  
\begin{eqnarray}
E^R_{1,2}-{\cal Z}^R_1E^R_{0,1}&=&
\frac{v\bar\sigma^2}{8\pi^2}\dashint  
d\th_1d\th_2d\th_3 e^{-\beta\veps(\th_1)}\cr\cr
&& \hskip -2in \times
\Bigg[\frac{\delta(vq-\Delta(s(\th_2)+s(\th_3)-s(\th_1)))}{\omega+i0-\veps(\th_2)-\veps(\th_3)+\veps(\th_1)}
\tanh^2(\frac{\theta_{23}}{2})
\coth^2(\frac{\theta_{12}}{2})\coth^2(\frac{\theta_{13}}{2})\cr\cr
&& \hskip -1.8in -\frac{\delta(vq-\Delta s(\th_3))}{\omega+i0-\veps(\th_3)}\frac{4\veps(\th_1)\veps(\th_2)}{\Delta^2(s(\th_1)-s(\th_2))^2}
- \frac{\delta(vq-\Delta s(\th_2))}{\omega+i0-\veps(\th_2)}\frac{4\veps(\th_1)\veps(\th_3)}{\Delta^2(s(\th_1)-s(\th_3))^2}\Bigg].
\end{eqnarray}
The final step before being able to compare to the results stemming
from our infinite volume regularization scheme is to convert the
principal value integrals into integrals along contours deformed by
$\pm i\eta$'s about the singularities found at $\th_1=\th_2$ and
$\th_1=\th_3$.  To do so we employ the identities
\begin{eqnarray}
\label{equality1}
\fl
\int_{R_\eta(\th_2)}d\theta_1
\frac{f(\th_1,\th_2)}{( s(\th_1)- s(\th_2))^2}
&=&\int_{C_\eta(\theta_2)} d\th_1\ \frac{f(\th_1,\th_2)}{(s(\th_1)-
  s(\th_2))^2}\nn
&&
+\int d\th_1\left[i\pi\frac{\delta'(\th_{21})}{c(\th_2)c(\th_1)}
+
\frac{2}{\eta}\frac{\delta(\th_{12})}{c^2(\th_2)}\right]f(\th_1,\th_2)
+{\cal O}(\eta),
\end{eqnarray}
and
\bea
\fl
\label{equality2}
\int_{R_\eta(\th_1)}
d\th_2\  \frac{f(\th_1,\th_2)}{\tanh^2(\frac{\th_{12}}{2})} &=&
\int_{C_\eta(\th_1)}
d\th_2\  \frac{f(\th_1,\th_2)}{\tanh^2(\frac{\th_{12}}{2})}\nn
&& +
\int d\th_2\ f(\th_1,\th_2) \left[
4\pi i \delta'(\th_{21}) + \frac{8}{\eta}\delta(\th_{12})\right]
+{\cal O}(\eta),
\eea
where $f(\th_1,\th_2)$ is a test function and the integration contours
are defined in Fig.\ref{fig:contours}. Deforming the contours into the
lower half plane instead changes the signs of the $\delta'$-terms. 
Using (\ref{equality1}) and (\ref{equality2}) to deform the $\th_3$
and $\th_2$ integrals into the upper and lower half-plane
respectively, we can rewrite
$E^R_{1,2}-{\cal Z}^R_1E^R_{0,1}$ in the following way
\begin{eqnarray}
&&E^R_{1,2}-{\cal Z}^R_1E^R_{0,1}= \frac{v\bar\sigma^2}{8\pi^2}\int
\!d\th_1d\th_2d\th_3 e^{-\beta\epsilon(\th_1)}
\frac{\delta(vq-\Delta(s(\th_2)+s(\th_3)-s(\th_1)))}
{\omega+i0-\epsilon(\th_2)-\epsilon(\th_3)+\epsilon(\th_1)}
\cr\cr
&& \tanh^2\bigl(\frac{\theta_{23}}{2}\bigr)
\Bigg[ \coth^2(\frac{\theta_{13}-i\eta}{2})
\coth^2(\frac{\theta_{12}+i\eta}{2}) +
16\pi^2\delta'(\th_{12})\delta'(\th_{13})\cr\cr 
&& \hskip .3in + 4\pi i \coth^2(\frac{\th_{12}+i\eta}{2})\delta'(\th_{13}) - 
4\pi i \coth^2(\frac{\th_{13}-i\eta}{2})\delta'(\th_{12})\Bigg]\cr\cr
&& -\frac{v\bar\sigma^2}{2\pi^2}\int 
d\th_1d\th_2d\th_3 e^{-\beta\epsilon(\th_1)}
\frac{\delta(vq-\Delta s(\th_2))}{\omega+i0-\epsilon(\th_2)}
\frac{\epsilon(\th_1)\epsilon(\th_3)}{\Delta^2(s(\th_1-i\eta)-s(\th_3))^2}\nn
&& -\frac{v\bar\sigma^2}{2\pi^2}\int 
d\th_1d\th_2d\th_3 e^{-\beta\epsilon(\th_1)}
\frac{\delta(vq-\Delta s(\th_3))}{\omega+i0-\epsilon(\th_3)}
\frac{\epsilon(\th_1)\epsilon(\th_2)}{\Delta^2(s(\th_1+i\eta)-s(\th_2))^2}
.
\label{e12inter}
\end{eqnarray}
We note that all terms singular in $\eta$ that come about from
deforming the contour of integration of $\th_1$ about the
singularities at $\th_2$ and $\th_3$ vanish (as they must as the
principal value integral is well defined and finite).  
The last two integrals in \fr{e12inter} vanish as may be seen by
deforming $\th_3 \rightarrow \th_3 + i\pi/2$ and $\th_2 \rightarrow
\th_2 - i\pi/2$ respectively. After carrying out the
integrals over the derivatives of delta functions we arrive at
\begin{eqnarray}\label{E12finvol}
&& \hskip -1.in E^R_{1,2}-{\cal Z}^R_1E^R_{0,1} = \frac{v\bar\sigma^2}{8\pi^2}\int
d\th_1d\th_2d\th_3 e^{-\beta\epsilon(\th_1)}
\frac{\delta(vq-\Delta(s(\th_2)+s(\th_3)-s(\th_1)))}
{\omega+i0-\epsilon(\th_2)-\epsilon(\th_3)+\epsilon(\th_1)}
\cr\cr
&& \times\Bigg[\tanh^2(\frac{\theta_{23}}{2})\coth^2(\frac{\theta_{13}-i\eta}{2})
\coth^2(\frac{\theta_{12}+i\eta}{2}) +
8\pi^2\delta(\th_{12})\delta(\th_{13})\Bigg]. 
\end{eqnarray}
We are now in a position to show that $E^R_{1,2}-{\cal  Z}^R_1E^R_{0,1}$ 
equals its value in our infinite volume regularization scheme. Indeed,
combining \fr{C12conn}, \fr{C12dis} and \fr{z1reg} and then keeping
only the part of order ${\cal O}(e^{-\beta\Delta})$ we obtain
\fr{E12finvol}. 

By an analogous consideration we can show that 
$\lim_{R\to\infty} F^R_{2,1}-{\cal Z}^R_1F^R_{1,0}$ recovers the result
of the infinite volume regularization scheme.

\subsection{Resummation}
Above we have argued that the finite temperature Lehmann representation
can be reexpressed as
\be
\chi_{\sigma}(\omega,q)=\sum_{s=0}^\infty {\cal C}^{\sigma}_s(\omega,q),
\label{resum1}
\ee
where the quantities ${\cal C}^{\sigma}_{r}(\omega,q)$ are finite in the
infinite volume limit. In particular 
\be
{\cal C}^{\sigma}_0(\omega,q)=\frac{2v\bs^2}
{(\omega+i\delta)^2-\veps^2(q)}+\ldots
\ee
is the zero-temperature dynamical structure factor.
We emphasize that we only consider frequencies close to $\Delta$ and
low temperatures, so that we can neglect n-particle contributions to
${\cal C}^{\sigma}_0$ with $n\geq 3$ as their contributions are
vanishingly small.  For $\omega-\veps(q)\sim{\cal O}(1)$ ${\cal C}_r^\sigma$ is
of order ${\cal O}\Bigl(e^{-\beta r\Delta}\Bigr)$ and hence
\fr{resum1} provides a good low-temperature expansion of the dynamical
susceptibility far away from the mass shell. On the other hand, when
we approach the mass shell we have
\be
{\cal C}^{\sigma}_s(\omega,q)\propto \left(\omega^2-\veps^2(q)\right)^{s+1}.
\ee
In order to obtain an expression for the susceptibility close to the
mass shell we therefore have to carry out a resummation of terms.
This is achieved by introducing a quantity $\Sigma(\omega,q)$ by
\fr{gensigma0} and using our results for ${\cal C}^{\sigma}_1$ and
${\cal C}^{\sigma}_2$ to carry out a low-temperature expansion of
$\Sigma(\omega,q)$ using 
\fr{gensigman}. 

\section{Results for the Low-Temperature Dynamical 
Susceptibility of the Quantum Ising Model}
\label{sec:Isingresults}
In order to present results obtained from the resummation \fr{gensigma0} it
is useful to define quantities

\begin{equation}
\label{chin}
\chi_\sigma^{(m)}(\omega, q) = \frac{{\cal C}^{\sigma}_0(\omega, q)}
{1-{\cal C}^{\sigma}_0(\omega,q)\sum_{j=1}^m\Sigma^{(j)}(\omega, q)},
\end{equation}
where $\Sigma^{(j)}(\omega,q)$ are given by
\fr{gensigman}. Loosely speaking $\chi^{(n)}_\sigma(\omega,q)$ is the dynamical
susceptibility obtained by calculating the $\Sigma(\omega,q)$ up to
order ${\cal O}\Bigl(e^{-(n+1)\beta\Delta}\Bigr)$. 
From the point of view of applications to experiment the relevant
quantity is the dynamical structure factor, which is related to the
retarded susceptibility by
\begin{equation}
\label{DSF}
S(\omega,q)=-\frac{1}{\pi}\frac{1}{1-e^{-\frac{\omega}{T}}}{\rm
  Im}\ \chi_{\sigma}(\omega, q) . 
\end{equation}
In analogy with the definition of the $m^{\rm th}$ order low
temperature approximation $\chi^{(m)}$ for the dynamical
susceptibility we define
\begin{equation}
\label{Sm}
S^{(m)}(\omega,q)=-\frac{1}{\pi}\frac{1}{1-e^{-\frac{\omega}{T}}}{\rm Im}\ \chi^{(m)}_\sigma(\omega, q) .
\end{equation}

The leading order result $S^{(1)}(\omega,q)$ is most easily calculated
using the infinite-volume regularization scheme and carrying out the
integrals in \fr{C1inf} numerically. In order to determine 
$S^{(2)}(\omega,q)$ we employ the finite volume regularization scheme
instead. We calculate $S^{(2)}(\omega+i\eta,q)$ for system sizes up to
$R=800$. The non-zero imaginary part is introduced in order to
suppress finite-size effects. If $\eta$ is sufficiently large the
numerically accessible values of $R$ will coincide (within numerical
accuracy) with the thermodynamic limit results. In order to obtain
results for real frequencies we compute $S^{(2)}(\omega+i\eta,q)$ for
a sequence of different $\eta$'s and then extrapolate to $\eta\to
0$. We have tested this method for $S^{(1)}(\omega,q)$, which we can
calculate directly in the thermodynamic limit, and found it to work
well.

\begin{figure}[ht]
\begin{center}
\epsfxsize=0.6\textwidth
\epsfbox{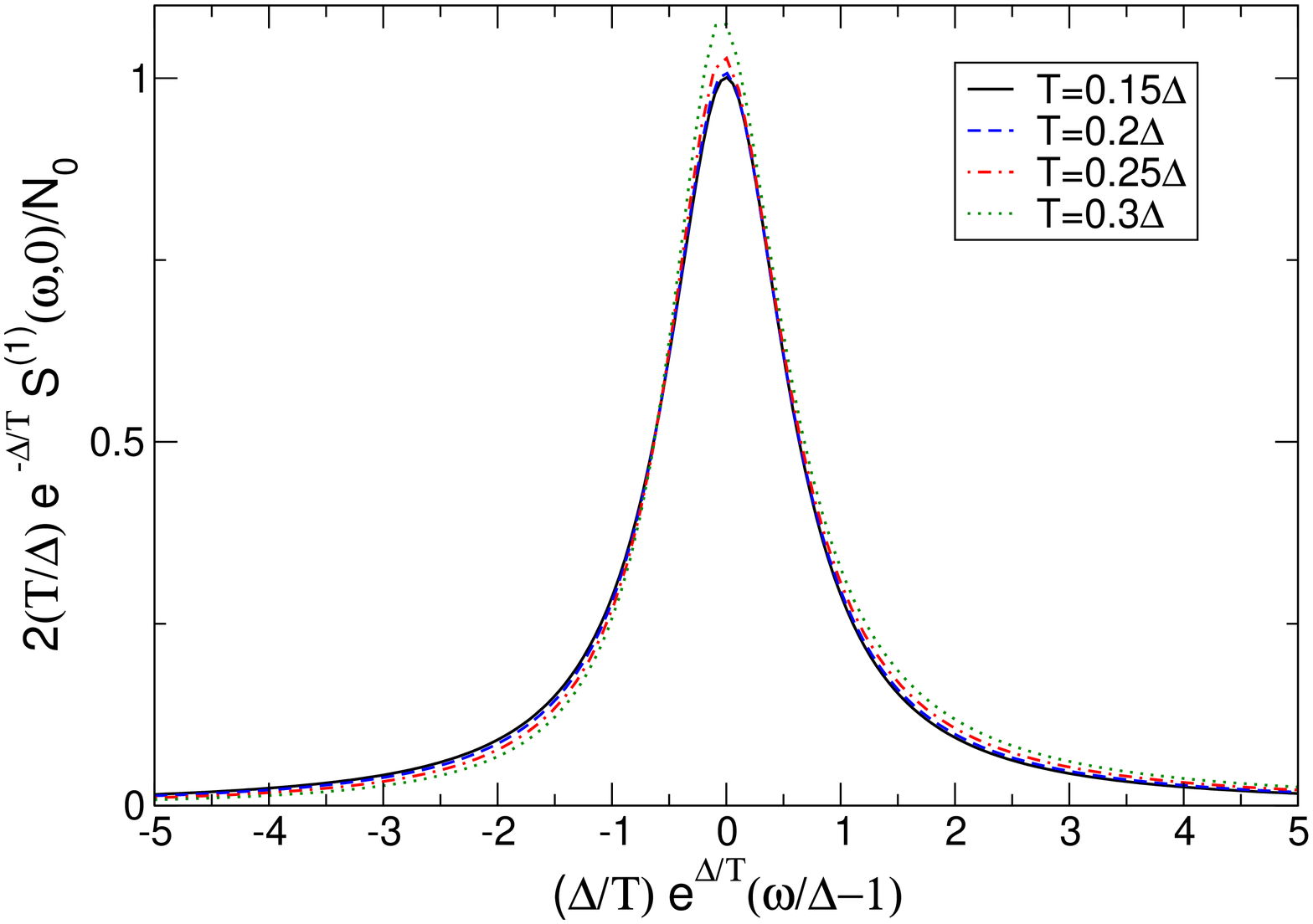}
\end{center}
\caption{Leading low-T approximation $S^{(1)}(\omega,q=0)$ of the
dynamical structure factor for several temperatures. Both axes have
been rescaled as discussed in the text.} 
\label{fig:chi1}
\end{figure}

At very low temperatures it is sufficient to calculate the leading
approximation $S^{(1)}(\omega,q)$ as the corrections are exponentially
small in $e^{-\Delta/T}$. We find that the difference between
$S^{(1)}(\omega,q)$ and $S^{(2)}(\omega,q)$ is negligible for
$T<0.3\Delta$ and we discuss this regime first.
Evaluation of $S^{(1)}(\omega,q)$ for low temperatures shows that, as
expected, the $T=0$ delta function at $\omega=\veps(q)$ broadens
with temperature. We find that the resulting peak scales as
\bea
{\rm peak\ height}&\propto&
\frac{\Delta}{T}\ \exp\Bigl(\frac{\Delta}{T}\Bigr),\nn
{\rm peak\ width}&\propto&
\frac{T}{\Delta}\ \exp\Bigl(\frac{-\Delta}{T}\Bigr).
\label{peakscaling}
\eea
In order to exhibit the evolution of the structure factor as a
function of frequency for fixed momentum $q$ with temperature it is
therefore useful to rescale both the frequency axis and the structure
factor. The result is shown in Fig.\ref{fig:chi1} for a temperature
range $0.15\Delta\leq T<0.3\Delta$, which corresponds to approximately
a factor of $50$ difference in peak height (and width). We see that the
lineshape is {\sl asymmetric} in frequency, with more spectral weight
appearing at higher frequencies. The asymmetry increases with
temperature. This effect is most easily quantified by comparison with
a Lorentzian lineshape, which is done below.

\begin{figure}[ht]
\begin{center}
\epsfxsize=0.48\textwidth
\epsfbox{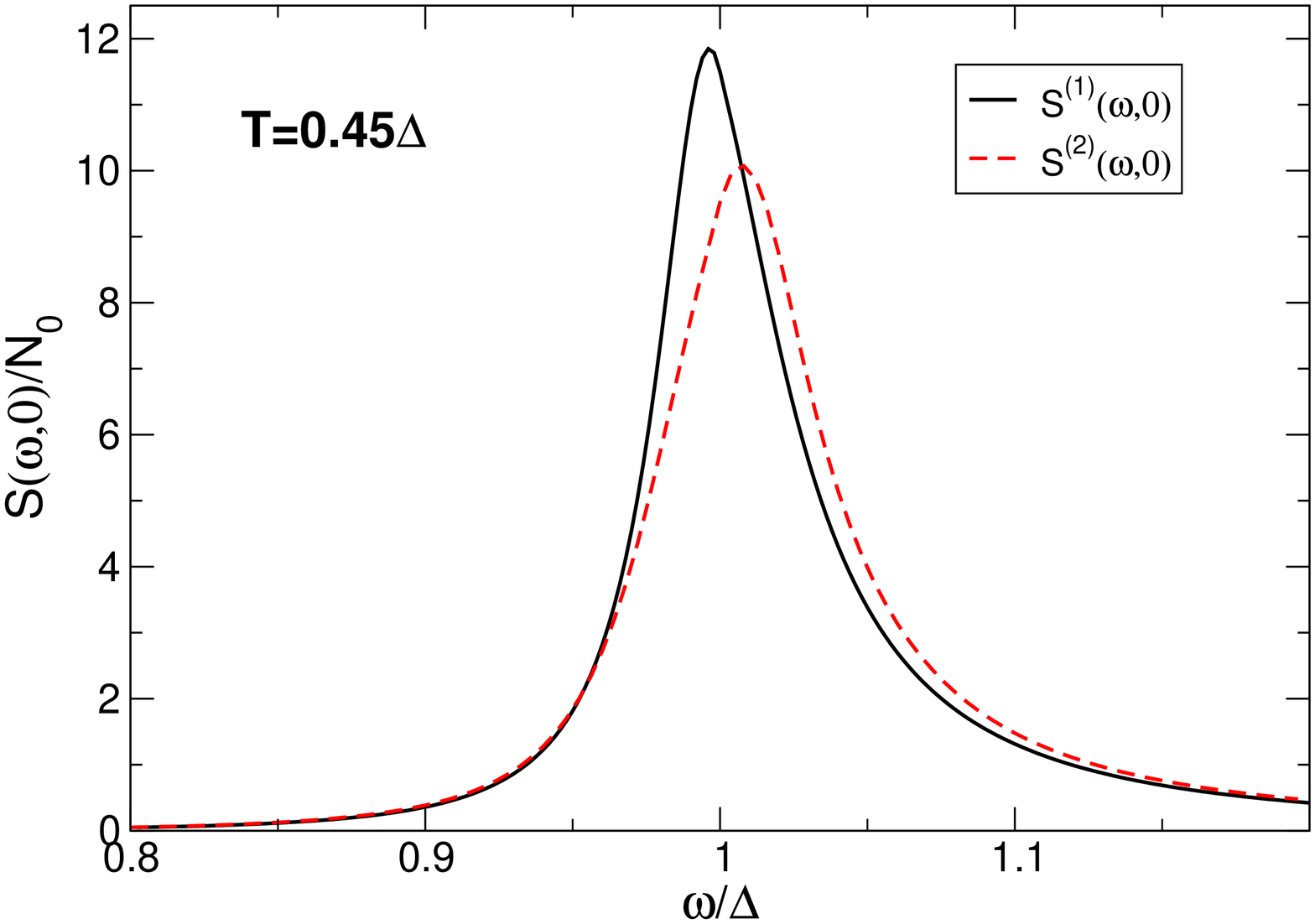}\quad
\epsfxsize=0.48\textwidth
\epsfbox{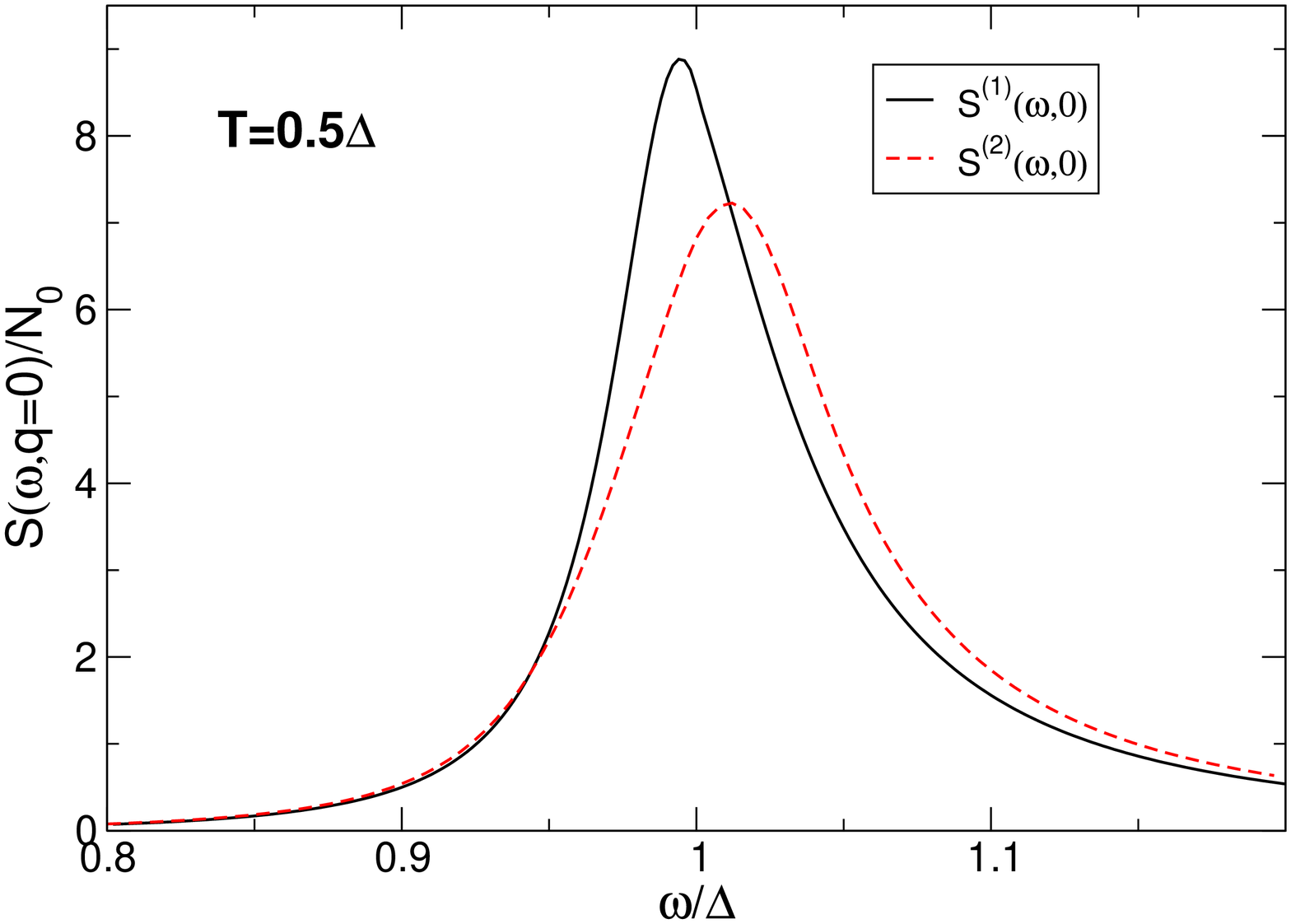}
\end{center}
\caption{Leading $S^{(1)}(\omega,q=0)$ and improved $S^{(2)}(\omega,q=0)$
low-T approximation of the dynamical structure factor for temperatures
$T=0.45\Delta$ and $T=0.5\Delta$ respectively. The most important
effect of the next to leading order approximation is the shift of the
maximum upwards in frequency (``temperature dependent gap'').
}
\label{fig:S045}
\end{figure}

If we increase the temperature beyond $T\approx 0.3\Delta$ the
correction $\Sigma^{(2)}(\omega,q)$ \fr{gensigman} in the expression for
the dynamical susceptibility \fr{gensigma0}, \fr{gensigman} is no longer
negligible compared to the leading term $\Sigma^{(1)}(\omega,q)$.
In Fig.\ref{fig:S045} we show the leading $S^{(1)}(\omega,q)$ and 
improved $S^{(2)}(\omega,q)$ low-temperature approximation to the
dynamical structure factor for $T=0.45\Delta$ and $T=0.5\Delta$. 
As expected the difference between the two is small sufficiently far
away from the mass shell. The most important effect is the shift of
the maximum to higher frequencies. This effect is known as
``temperature dependent gap''.

\subsection{Comparison to Semiclassical Results of Sachdev and Young}
In \cite{young,sachdevbook} Sachdev and Young have developed a
semiclassical approach to determine the dynamical structure factor.
Their result \footnote{Our definition of the structure factor differs
from \cite{sachdevbook} by a factor of $2\pi$.}, valid for $T\ll \Delta$, is
\bea
S_{\rm sc}(\omega,q)
&=&
\int_{-\infty}^\infty \frac{dt}{2\pi}
dx\ e^{i\omega t-iqx}K(x,t)R(x,t)\ ,
\label{SSY}
\eea
where
\bea
K(x,t)&=&\frac{\bs^2}{\pi}K_0(|\Delta|\sqrt{(x/c)^2-t^2})\ ,\nn
R(x,t)&=&\exp\left(-\int_{-\infty}^\infty
\frac{dk}{\pi}e^{-\veps(k)/T}|x-v(k)t|\right),\nn
v(k)&\equiv&\frac{d\veps(k)}{dk}=\frac{v^2k}{\veps(k)}.
\eea
Here $\bs$ is defined in \fr{sigmabar}. At sufficiently low
temperatures and when $|vq|\ll\sqrt{T\Delta}$
the semiclassical result is well approximated by a Lorentzian
\cite{sachdevbook} 
\be
S_{\rm Lor}(\omega,q)=
\frac{\bs^2 v}{\pi\veps(q)}
\frac{1/\tau_\phi}{(\omega-\veps(q))^2+1/\tau_\phi^2},
\label{Lorentzian}
\ee
where
\be
\tau_\phi=\frac{\pi}{2T}e^{\Delta/T}\ .
\ee

\begin{figure}[ht]
\begin{center}
\epsfxsize=0.48\textwidth
\epsfbox{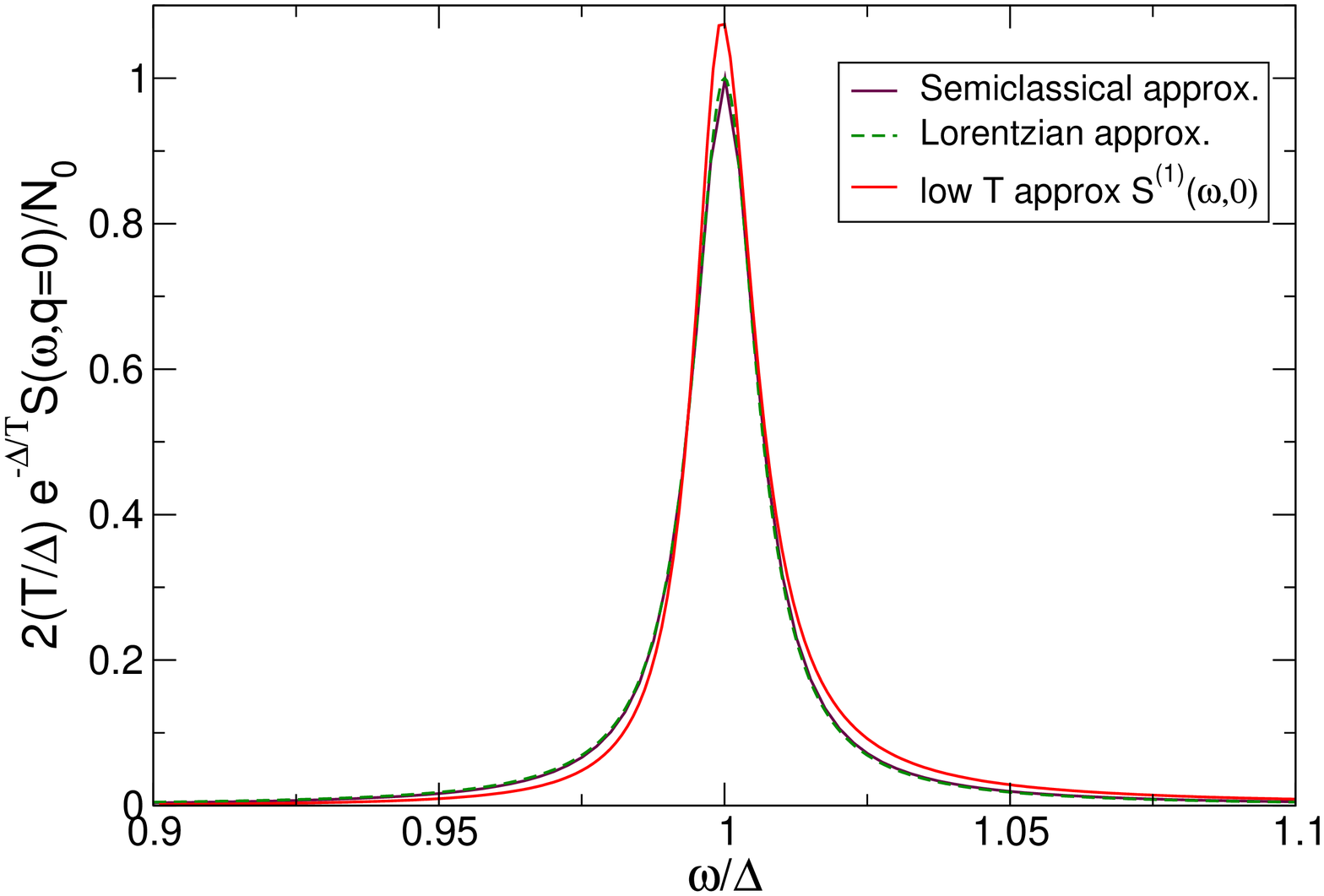}\
\epsfxsize=0.48\textwidth
\epsfbox{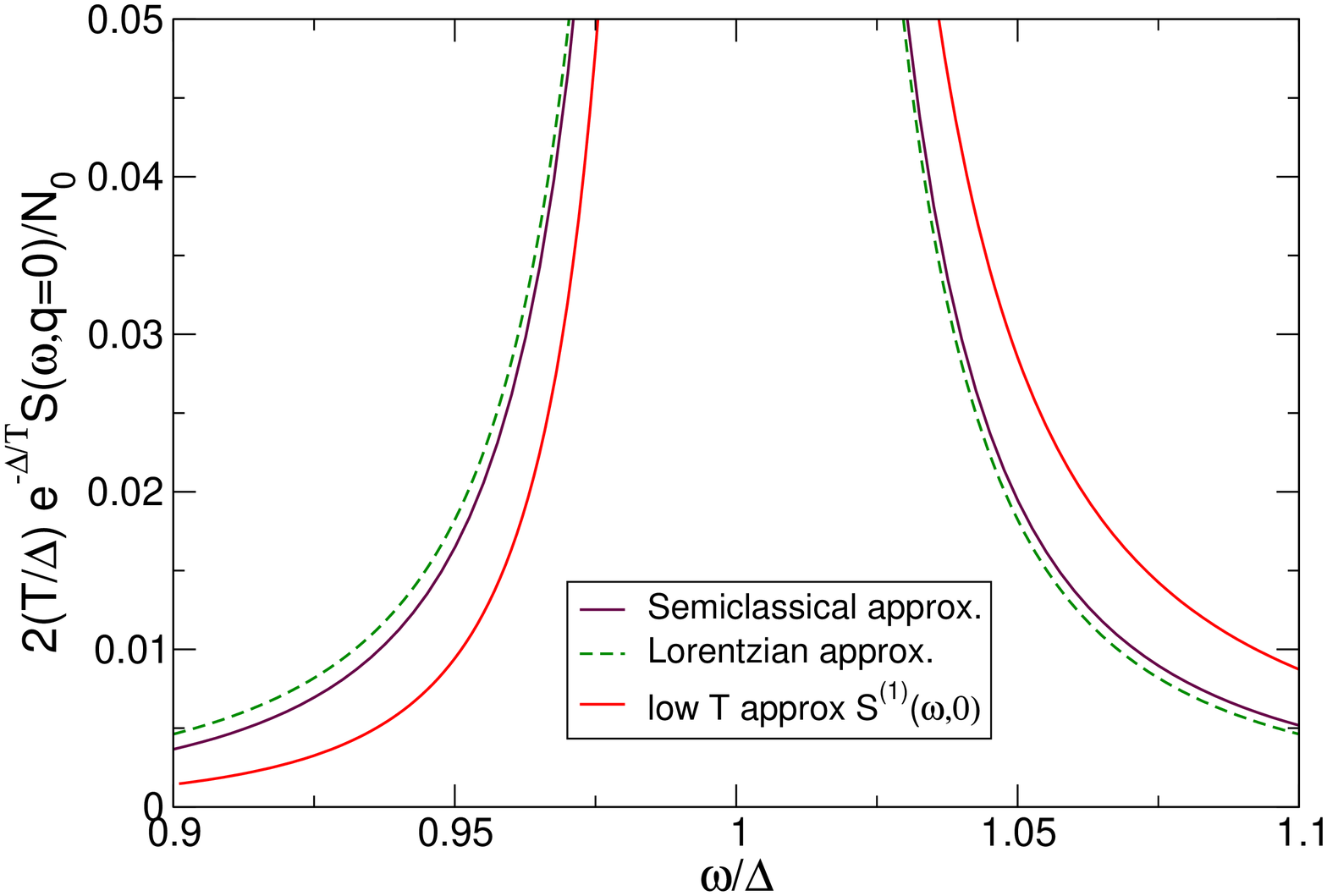}
\end{center}
\caption{Dynamical structure factor for $T=0.3\Delta$, $q=0$. The
  result of the present work (red line) is still in good agreement
  with the semiclassical result of Sachdev and Young as well as the
  Lorentzian approximation to the latter. The asymmetry of the
  lineshape is more pronounced than the one predicted by the
  semiclassical result. }
\label{fig:isingT03a}
\end{figure}

We expect our low-temperature expansion to reproduce the semiclassical
results at sufficiently low temperatures $T\ll\Delta$. The comparison of
the improved result $S^{(2)}(\omega,q)$ to the leading order 
$S^{(1)}(\omega,q)$ suggests that the latter is a good approximation
up to temperatures of $T\approx 0.3\Delta$. In Fig.\ref{fig:isingT03a}
we show a comparison of our low temperature approximation to the
semiclassical result \fr{SSY} and the Lorentzian approximation
\fr{Lorentzian} for $T=0.3\Delta$. We see that while the gross
structure of the lineshapes is quite similar, there are considerable
differences away from the mass shell. This implies that the
semiclassical approximation is no longer quantitatively accurate for
$T=0.3\Delta$. 

\begin{figure}[ht]
\begin{center}
(a)\epsfxsize=0.45\textwidth
\epsfbox{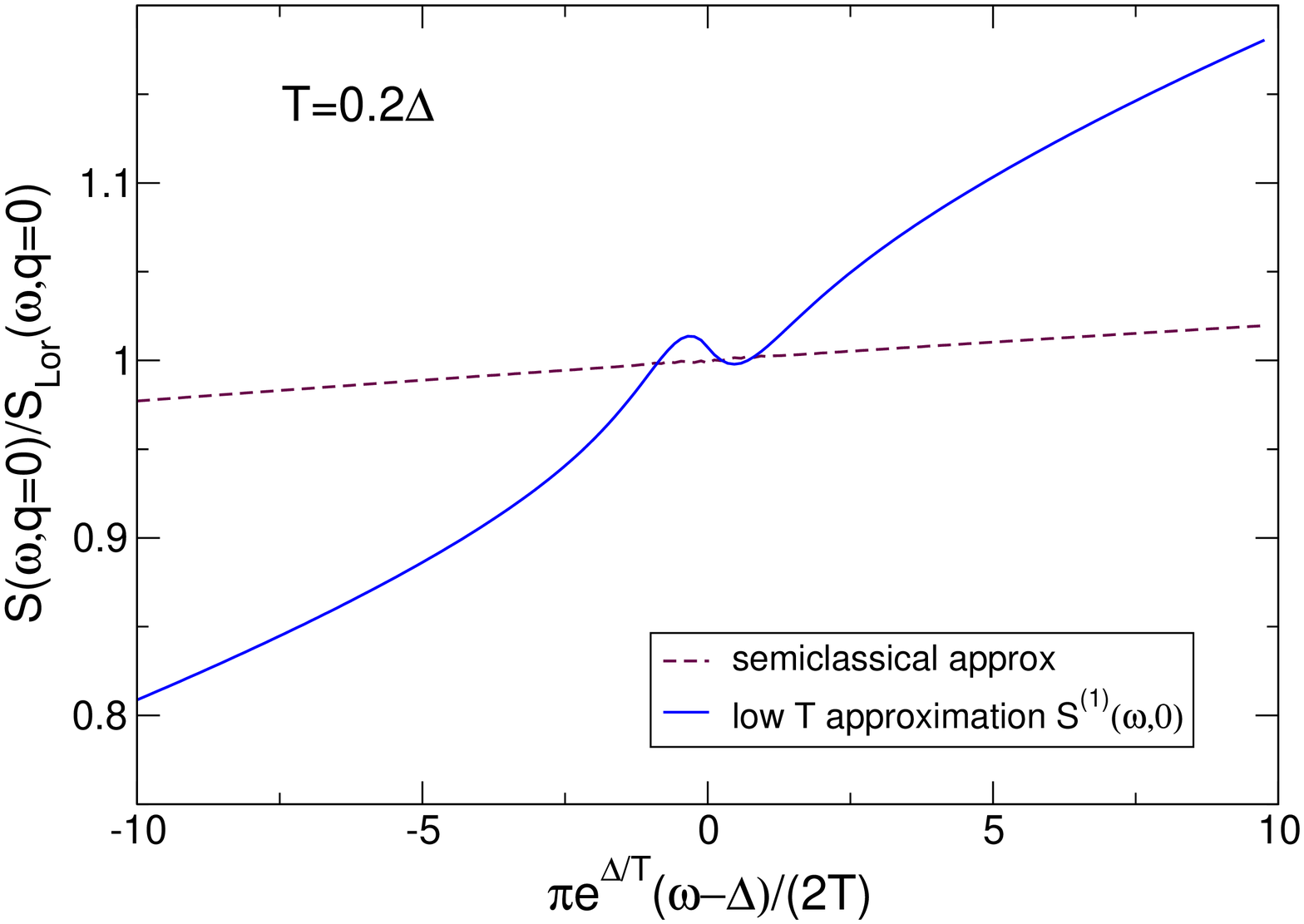}\quad
(b)\epsfxsize=0.45\textwidth
\epsfbox{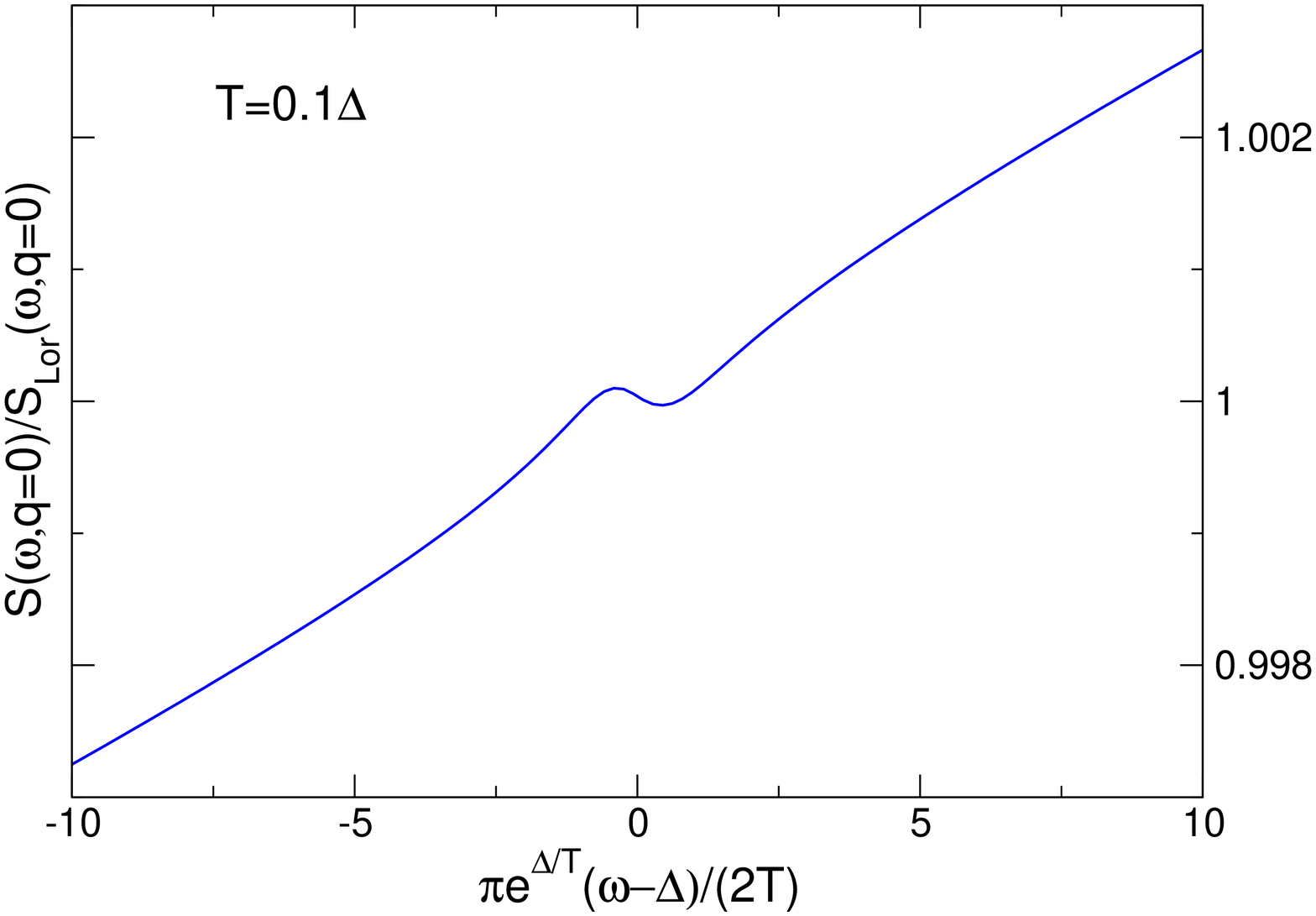}\
\end{center}
\caption{(a) Ratios of the dynamical structure factors calculated from
  eq. \fr{gensigma0} (this work) and \fr{SSY} (Sachdev and Young) to the
  Lorentzian approximation \fr{Lorentzian} for $T=0.2\Delta$ and
  $q=0$. The non-monotonic behaviour very close to the
  mass shell indicates a small shift of the maximum of
  $S^{(1)}(\omega,q=0)$ away from the $T=0$ gap.
(b) Ratio of $S^{(1)}(\omega,q=0)$ to the  Lorentzian approximation
  \fr{Lorentzian} for $T=0.1\Delta$ and $q=0$. We see that the low
  temperature approximation recovers the semiclassical result at
  sufficiently low temperatures.
}
\label{fig:asymm}
\end{figure}

In order to exhibit the differences between the various approximations
more clearly, we plot the ratios $S^{(1)}(\omega,q)/
S_{\rm Lor}(\omega,q)$ and $S_{\rm sc}(\omega,q)/S_{\rm
  Lor}(\omega,q)$ in Fig.\ref{fig:asymm}(a) for $T=0.2\Delta$. We see
that the semiclassical approximation underestimates the asymmetry of
the lineshape. On the other hand, for sufficiently small temperatures
the low-temperature expansion indeed recovers the semiclassical result
as can be seen from Fig.\ref{fig:asymm}(b).

The ratio $S^{(1)}(\omega,q)/S_{\rm Lor}(\omega,q)$ displays
non-monotonic behaviour close to the mass shell. This suggests that
the maximum of $S^{(1)}(\omega,q)$  is in fact very slightly shifted
away from the $T=0$ gap. This is indeed the case.

\subsection{Comparison to Exact Finite Temperature Form Factors}
In Refs \cite{AKT,doyon} the following form factor expansion for the
two point function of the spin field in the quantum Ising chain was
derived 
\bea
\label{chiFT}
\chi_\sigma(\tau,x)&=&\sum_{r,s}D^\sigma_{r,s}(\tau,x)\ ,\\
D^\sigma_{r,s}(\tau,x)&=&-
\frac{C^2(\beta)}{r!s!}e^{-n(\beta)|x|}
\int\prod_{j=1}^r\left[
\frac{d\th_j}{2\pi} f(\th_j)e^{\tau\Delta
  c(\th_j)-i\frac{\Delta}{v}s(\th_j)x-\eta_+(\th_j)}\right]\nn
&\times&\
\int\prod_{l=1}^s
\frac{d\th'_l}{2\pi} f(\th'_l)e^{(\beta-\tau)\Delta
  c(\th'_l)+i\frac{\Delta}{v}s(\th'_l)x-\eta_-(\th'_l)}
\prod_{n>m=1}^r\tanh^2\Bigl(\frac{\th_{nm}}{2}\Bigr)\nn
&&\times
\prod_{p>q=1}^s\tanh^2\Bigl(\frac{\th'_{pq}}{2}\Bigr)
\prod_{p,n}\coth^2\Bigl(\frac{\th_n-\th'_p+i0}{2}\Bigr),
\eea
where $c(\th)=\cosh\th$, $s(\th)=\sinh\th$, $\th_{jk}=\th_j-\th_k$ and
\bea
C(\beta)&=&\bs\exp\left[\frac{(\Delta\beta)^2}{2}
\int\frac{d\th_1 d\th_2}{(2\pi)^2}\frac{s(\th_1)s(\th_2)}
{s\Bigl(\Delta\beta c(\th_1)\Bigr)s\Bigl(\Delta\beta c(\th_2)\Bigr)}
\ln\left|\coth\frac{\th_{12}}{2}\right|
\right]
,\nn
n(\beta)&=&\frac{2\Delta}{\pi v}\sum_{k=1}^\infty\frac{1}{2k-1}
K_1\Big((2k-1)\beta\Delta\Big)\approx
e^{-\beta\Delta}\sqrt{\frac{2\Delta}{\pi\beta v^2}},\nn
\eta_\pm(\th)&=& \pm 2\int_{-\infty\mp i0}^{\infty\mp
  i0}\frac{d\th'}{2\pi i\sinh(\th-\th')}
\ln\left[\frac{1+e^{-\beta\Delta c(\th')}}{1-e^{-\beta\Delta c(\th')}}\right].
\nonumber
\eea
At low temperatures we have
\bea
C(\beta)&\approx&\bs,\nn
e^{\eta_\pm(\th)}&\approx& \left[1-2e^{-\beta\Delta
  c(\th)}\right]
\exp\left[\mp \frac{2i}{\pi}\int_0^\infty\frac{dx}{\sinh
    x}\big(e^{-\beta\Delta c(\th-x)}-e^{-\beta\Delta c(\th+x)}\big)\right],\nn
n(\beta)&\approx& e^{-\beta\Delta}\sqrt{\frac{2\Delta}{\pi\beta v^2}}.
\nonumber
\eea
An important question is in which way this expansion is related to
the low-temperature expansion based on the zero temperature form
factors. Fourier transforming and analytically continuing to real
frequencies we find
\bea
D^\sigma_{0,1}(\omega,q)&=&C^2(\beta)\int\frac{d\th}{2\pi}
\frac{e^{-\eta_-(\th)}}{\omega+i0-\Delta c(\th)}\
\frac{2n(\beta)}
{n^2(\beta)+(q-\frac{\Delta}{v} s(\th))^2},\nn
D^\sigma_{1,0}(\omega,q)&=&-C^2(\beta)\int\frac{d\th}{2\pi}
\frac{e^{-\eta_+(\th)}}{\omega+i0+\Delta c(\th)}\
\frac{2n(\beta)}
{n^2(\beta)+(q+\frac{\Delta}{v} s(\th))^2}.
\label{D0110}
\eea
The main difference between $D_{r,s}$ and the corresponding zero
temperature quantities is the replacement of the momentum conservation
delta function by a Lorentzian of width $n(\beta)$. We note that the
imaginary part of $D_{0,1}$ has a square root divergence for
$\omega\to\Delta$ and it is necessary to sum an infinite number of
terms in \fr{chiFT} to get a meaningful answer \cite{AKT}.
In order to recover our low-temperature expansion from the Lehmann
representation in terms of exact finite temperature form factors we
should expand the Lorentzian expressing approximate momentum conservation
under the integral, e.g. 
\bea
\frac{2n(\beta)}{n^2(\beta)+q^2(\th)}&=&2\pi\delta(q(\th))
+4n(\beta)\frac{q^2(\th)-\eps^2}{(q^2(\th)+\eps^2)^2}\nn
&&-4\pi n^2(\beta)\delta^{\prime\prime}(q(\th))+
{\cal O}\big(n^3(\beta)\big),
\nonumber
\eea
where $q(\th)=q-\frac{\Delta}{v}s(\th)$.
\section{O(3) Nonlinear $\sigma$-Model}
\label{sec:SM}
We now apply the methods outlined above to the O(3) nonlinear sigma
model. Unlike the quantum Ising model the sigma model describes a
strongly interacting theory featuring dynamical mass generation.
The Lagrangian of the sigma model is given by
\begin{equation}
\label{exii}
{\cal L} = \frac{1}{2g}\int dx \left[
\frac{1}{v}\partial_t{\bf n}\cdot\partial_t{\bf n}-
v \partial_x{\bf n}\cdot\partial_x{\bf  n}
\right]. 
\end{equation}

The O(3) nonlinear sigma model describes the scaling limit of integer
spin-S Heisenberg models \cite{haldane}
\be
H=J\sum_{j}{\bf S}_j\cdot{\bf S}_{j+1},\quad {\bf S}_j^2=S(S+1).
\ee
The velocity and coupling constant of the sigma model are related to
the lattice model parameters by
\be
g=\frac{2}{S}\ ,\quad v=2JSa_0,
\ee
where $a_0$ is the lattice spacing \cite{o3}. The lattice spin
operators, ${\bf S}_j$, are related to the continuum fields by 
\be
{\bf S}_j \simeq S (-1)^{j} {\bf n}(x) + {\bf l}(x)
\ ,\quad x=ja_0,
\label{latticespin}
\ee
where ${\bf l}(x)\propto\frac{a_0}{vg}\,{\bf
  n}(x)\times\frac{\partial{\bf n}(x)}{\partial t}$. The dynamical
susceptibilities of the lattice model are given by 
\bea
\chi^{ab}_{\rm lat}(\omega,Q)&=&-\int_0^\beta d\tau \sum_l\ e^{i\omega_n \tau-iQla_0}
\langle T_\tau S^a_{l+1}(\tau) S^b_1\rangle
\Bigr|_{\omega_n\to\delta-i\omega}\ .
\label{latticechi}
\eea
Substituting \fr{latticespin} into \fr{latticechi} we see that in the
vicinity of the antiferromagnetic wave number ($Q=\frac{\pi}{a_0}+q$ 
with $|q|\ll\frac{\pi}{a_0}$) the
lattice susceptibility at low energies $\omega\ll v/a_0$ can be
expressed in terms of the two-point function of the ${\bf n}$-field
\bea
\chi^{zz}_{\rm lat}(\omega,Q)
&\propto&-\int_0^\beta d\tau\int dx \ e^{i\omega_n \tau-iqx}
\langle T_\tau n^z(\tau,x) n^z(0)\rangle
\Bigr|_{\omega_n\to\delta-i\omega}\ .
\label{latticechi2}
\eea
The ${\bf n}$-field needs to be renormalized and in an appropriate
scheme is related to the spin field ${\bf\Phi}$ by
\be
n^a=\zeta \Phi^a\ .
\ee
In what follows we analyze the two-point function of the sigma model
spin field
\bea
\chi_{_\Phi}(\omega,q)&=&-\int_0^\beta d\tau dx\ e^{i\omega_n \tau-iqx}
\langle T_\tau \Phi^a(\tau,x) \Phi^a(0,0)\rangle
\Bigr|_{\omega_n\to\delta-i\omega}\ ,
\eea
keeping in mind that at low energies we have
\be
\chi^{zz}_{\rm lat}\Bigl(\omega,\frac{\pi}{a_0}+q\Bigr)\propto
\chi_{_\Phi}(\omega,q).
\ee

The O(3) nonlinear sigma model is integrable
\cite{luescher,o3smat,O3integrability} and the exact spectrum and
scattering matrix have been known for a long time. The elementary
excitations of the sigma model are a triplet of massive particles with
spins $S^z=\pm 1,0$. It is useful to parametrize their energy and
momentum in terms of a rapidity variable $\th$ 
\be
\epsilon(\theta)=\Delta\cosh\th\ ,\quad
p(\th)=\frac{\Delta}{v}\sinh\th.
\ee
It is convenient to choose a basis such that the spin operators $S^a$
act on single-particle states as
\be
S^a|\th\rangle_b=i \epsilon_{abc}|\th\rangle_c,\ a=1,2,3.
\label{spin}
\ee

A convenient basis for the Hilbert space is formed by scattering
states of these elementary excitations. In order to describe it one
introduces creation and annihilation operators
$Z_a(\th)$ and $Z^\dagger_a(\th)$ respectively, fulfilling the 
Faddeev-Zamolodchikov algebra \fr{FZalgebra}. Here $a,b$ are O(3)
quantum numbers and $S$ is the exact two-particle scattering matrix
\cite{o3smat}  
\bea
S_{ab}^{cd}(\theta)&=&\sigma_1(\theta)\delta_{ab}\delta_{cd}
+\sigma_2(\theta)\delta_{ac}\delta_{bd}
+\sigma_3(\theta)\delta_{ad}\delta_{bc}\ ,\nn
\sigma_1(\t)&=&\frac{2\pi i\t}{(\t+i\pi)(\t-2\pi i)}\ ,\nn
\sigma_2(\t)&=&\frac{\t(\t-i\pi)}{(\t+i\pi)(\t-2\pi i)}\ ,\nn
\sigma_3(\t)&=&-\frac{2\pi i(\t-i\pi)}{(\t+i\pi)(\t-2\pi i)}.
\label{smat}
\eea
We note that the S-matrix is a solution to the Yang-Baxter equation
and fulfils
\be
\left[S_{ab}^{cd}(\t)\right]^*=S_{ab}^{cd}(-\t).
\ee
Using the Faddeev-Zamolodchikov operators, a Fock space of states can
be constructed as in \fr{vac}, \fr{states}. Energy and momentum of are
by construction additive and given by \fr{ep}.
In the basis of scattering states introduced above, formally the
following spectral representation for the finite temperature dynamical
susceptibility holds
\be
\chi_{_\Phi}(\omega,q)=\frac{1}{{\cal Z}}\sum_{r,s=0}^\infty
C^\Phi_{r,s}(\omega,q)\ . 
\label{SMchiSR}
\ee
Here $C_{r,s}$ are given by \fr{CrsO}
\bea
C^\Phi_{r,s}(\omega,q)&=&\int_0^\beta d\tau\int dx e^{i\omega_n\tau-iqx}
C^\Phi_{r,s}(\tau,x)\Bigg|_{\omega_n\to\delta-i\omega}\nn
C^\Phi_{r,s}(\tau,x) &=&
-\int\frac{d\th_1\ldots d\th_r}{(2\pi)^rr!}
\int\frac{d\th'_1\ldots d\th'_s}{(2\pi)^ss!}
e^{-\beta E_r}e^{-\tau(E_s-E_r)-i(P_r-P_s)x}\nn
&\times&\sum_{b_1,\ldots,b_r}\sum_{b_1',\ldots,b_s'}
|_{b_1\ldots b_r}\langle\theta_1\ldots\th_r|\Phi^a(0,0)|
\th'_s\ldots\th'_1\rangle_{b_s'\ldots b_1'}|^2.
\label{SMCrs}
\eea
The partition function can formally be expressed as in \fr{Zgen}.
The low-temperature expansion is constructed by following the steps
set out in section \ref{sec:general}, e.g. the functions $E^\Phi_{r,s}$
used in the decomposition \fr{CEF} of $C^\Phi_{r,s}$ are
\bea
E^\Phi_{r,s}(\omega,q)&=&
\sum_{b_1,\ldots,b_r}\sum_{b_1',\ldots,b_s'}
\int\frac{d\th_1\ldots d\th_r}{(2\pi)^rr!}
\int\frac{d\th'_1\ldots d\th'_s}{(2\pi)^ss!}2\pi\delta(q+P_r-P_s)\nn
&\times&
\frac{e^{-\beta E_r}}{\omega+i\delta-E_s+E_r}
|{}_{b_1\ldots b_r}\langle\theta_1\ldots\th_r|
\Phi^a(0,0)|\th'_s\ldots\th'_1\rangle_{b_s'\ldots b_1'}|^2,
\label{SMErs}
\eea
where $E_r$ and $P_r$ are defined in \fr{ep}. Our aim is to determine
the functions ${\cal E}^\Phi_l$, ${\cal F}^\Phi_l$ and ${\cal C}^\Phi_l$ defined by
\fr{El},  \fr{Fl} and \fr{Cl} respectively and to use them to obtain a
low-temperature expansion \fr{lowTex} of the susceptibility.

\subsection{Zero Temperature Dynamical Response}
The low energy dynamical response in the O(3) nonlinear sigma model
has been worked out in \cite{affwes,3particle}.
The zero-temperature form factors in the infinite volume have been
determined by several groups \cite{BH,BN,ks}. 
The one and three-particle form factors of the spin field are \cite{BN}
\bea
&&\langle 0|\Phi^a(0)|\theta\rangle_b=\delta_{ab}\ ,\\
&&\langle 0|\Phi^a(0)|\theta_3,\theta_2,\theta_1\rangle_{a_3a_2a_1}
=\frac{\pi^{3}}{2}\psi(\theta_{32})\psi(\theta_{31})\psi(\theta_{21})\nn
&&\qquad\qquad\times
\Bigl[
\delta_{aa_1}\delta_{a_2a_3}\theta_{23}
+\delta_{aa_2}\delta_{a_1a_3}(\theta_{31}-2\pi i)
+\delta_{aa_3}\delta_{a_1a_2}\theta_{12}
\Bigr],
\label{SM3part}
\eea
where
\be
\psi(\theta)=\frac{\theta-i\pi}{\theta(2\pi i-\theta)}
\tanh^2\bigl(\frac{\theta}{2}\bigr).
\ee
Note that these differ slightly from \cite{BN} because we use the a
different normalization condition for the scattering states
\be
{}_a\langle\theta|\theta'\rangle_b=2\pi\delta_{ab}\ \delta(\theta-\theta').
\ee
Some useful identities involving the function $\psi(\th)$ are 
\bea
\psi(\theta+i\pi)&=&-\psi(i\pi-\theta)=-\frac{\theta\coth^2(\theta/2)}{\theta^2+\pi^2},\nn
\psi(\theta)\ \psi(\theta+i\pi)&=&\frac{1}{(\theta+i\pi)(\theta-2\pi
  i)}\ ,\nn
\psi(\theta+i\pi^+)&=&\psi(\theta+i\pi^-)+\frac{8i}{\pi}\delta(\theta).
\eea
Here $\pi^\pm=\pi\pm 0$. 

\begin{figure}[ht]
\begin{center}
\epsfxsize=0.48\textwidth
\epsfbox{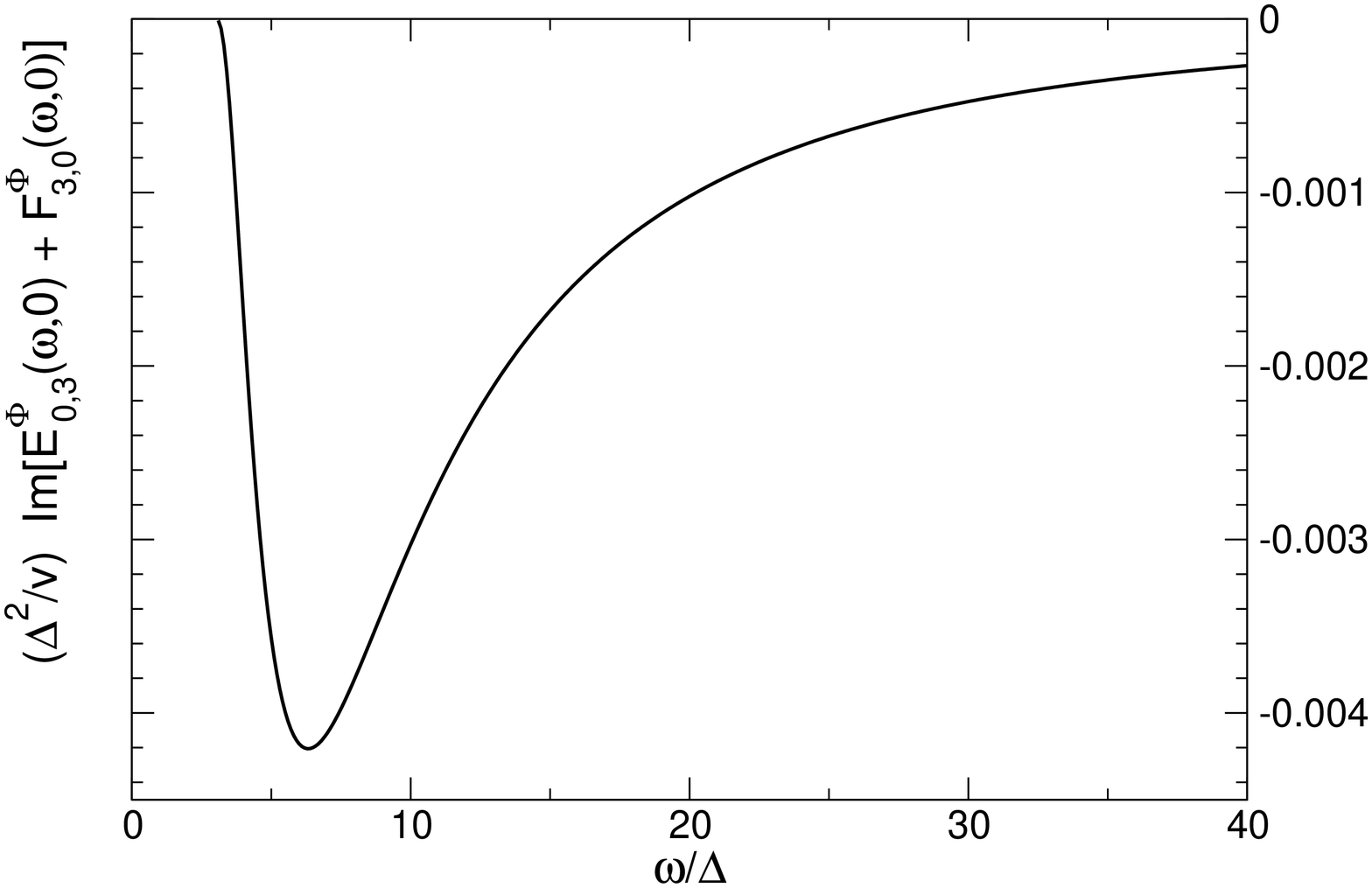}\quad
\epsfxsize=0.48\textwidth
\epsfbox{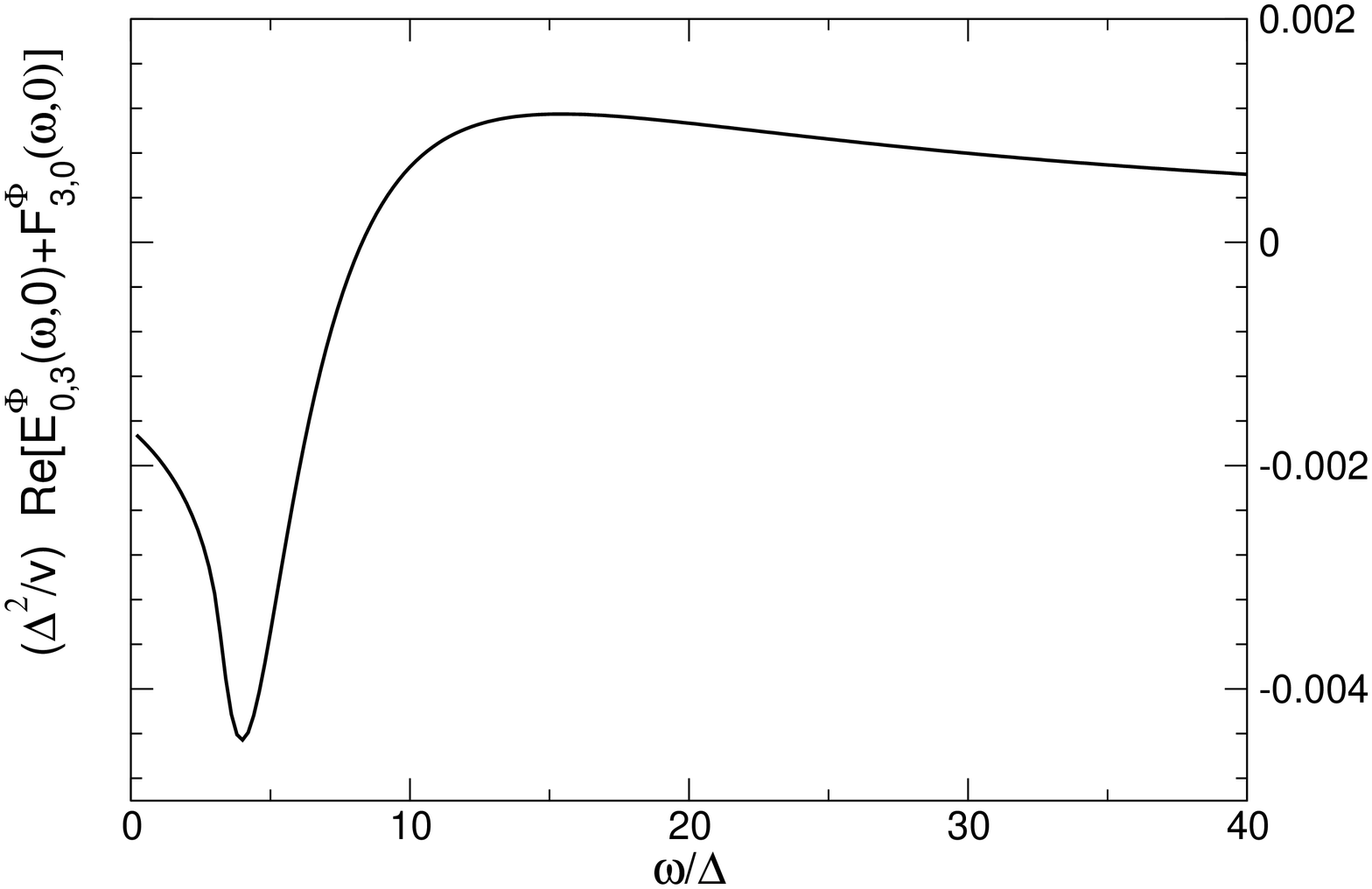}
\end{center}
\caption{Real and imaginary parts of $E^\Phi_{03}(\omega,q=0)$. At low
frequencies both are small compared to $E^\Phi_{01}(\omega,q=0)$.}
\label{fig:SMT=0}
\end{figure}

At $T=0$ the leading contributions to the dynamical susceptibility at
low frequencies are
\bea
\chi_{_\Phi}(\omega,q)\Bigr|_{T=0}&\approx&\left[
E^\Phi_{0,1}(\omega,q)+F^\Phi_{1,0}(\omega,q)
+E^\Phi_{0,3}(\omega,q)+F^\Phi_{3,0}(\omega,q)\right].
\eea
Here the 1-particle contributions are
\bea
E^\Phi_{0,1}(\omega,q)&=&\left[F^\Phi_{1,0}(-\omega,-q)\right]^*=
\frac{v}{\veps(q)}\frac{1}{\omega-\veps(q)+i0}.
\eea
The 3-particle terms can be cast in the form \cite{3particle}
\bea
&&E^\Phi_{0,3}(\omega,q)+
F^\Phi_{3,0}(\omega,q)\nn
&&\qquad=\frac{2v\pi^6}{3}\int_{-\infty}^\infty\frac{dy\ dz}{(2\pi)^2}
\frac{f(y+z)f(y-z)f(2z)\ [3\pi^2+3z^2+y^2]}
{s^2-\Delta^2-4\Delta^2\cosh(z)[\cosh(z)+\cosh(y)]},
\eea
where
\bea
f(z)&=&\frac{z^2+\pi^2}{z^2(z^2+4\pi^2)}\left[\tanh(z/2)\right]^4,\nn
s^2&=&(\omega+i0)^2-v^2q^2.
\eea
We plot the real and imaginary parts of
$E^\Phi_{0,3}(\omega,q=0)+F_{0,3}(\omega,q=0)$ in 
Fig.\ref{fig:SMT=0}. We see that by virtue of the smallness of
$E^\Phi_{0,3}+F^\Phi_{3,0}$ the dynamical response at low energies is
dominated by the coherent single-particle contributions
$E^\Phi_{0,1}+F^\Phi_{1,0}$.  
This yields the following result for the temperature independent part
of the spectral representation for the dynamical susceptibility
\bea
{\cal C}^\Phi_0(\omega,q)&\approx&E^\Phi_{0,1}(\omega,q)
+F^\Phi_{1,0}(\omega,q)=\frac{2v}{(\omega+i0)^2-\veps^2(q)},
\label{SMC0}
\eea
where $\veps(q)=\sqrt{\Delta^2+v^2q^2}$.
\subsection{Infinite volume regularization}
At low temperatures the next most important contributions arise from
$E^\Phi_{1,2}$ and $F^\Phi_{2,1}$, which are formally given by
\bea
E^\Phi_{1,2}(\omega,q)&=&v\sum_{b,b_1,b_2}
\int\frac{d\theta d\theta_1 d\theta_2}{2(2\pi)^2}
\frac{e^{-\beta\Delta c(\th)}}
{\omega+i0-\Delta[c(\theta_1)+c(\theta_2)-c(\theta)]}\nn
&&\times
|_b\langle\theta|\Phi^a(0)|\theta_2,\theta_1\rangle_{b_2b_1}|^2
\delta\bigl(vq-\Delta[s(\theta_1)+s(\theta_2)-s(\theta)]\bigr),
\label{SMC12}
\eea
\bea
F^\Phi_{2,1}(\omega,q)&=&-v\sum_{b,b_1,b_2}
\int\frac{d\theta d\theta_1 d\theta_2}{2(2\pi)^2}
\frac{e^{-\beta\Delta c(\th)}}
{\omega+i0+\Delta[c(\theta_1)+c(\theta_2)-c(\theta)]}\nn
&&\times
|_{b_1b_2}\langle\theta_1,\th_2|\Phi^a(0)|\theta\rangle_{b}|^2
\delta\bigl(vq+\Delta[s(\theta_1)+s(\theta_2)-s(\theta)]\bigr),
\label{SMC21}
\eea
where $c(\theta)=\cosh\theta$ and $s(\theta)=\sinh\theta$.
In order to proceed further we need to evaluate the absolute value
squares of form factors. The individual form factors can be
analytically continued following Smirnov \cite{Smirnov92book}, see
\ref{app:xing} for a summary. The various ways of analytically
continuing the 3-particle form factors are  
\bea
{}_{b_1b_2}\langle\theta_1,\theta_2|\Phi^a(0)|\theta_3\rangle_b
&=&{}_{b_1b_2}\langle\theta_1+i0,\theta_2-i0|\Phi^a(0)|\theta_3\rangle_b\nn
&&\quad +2\pi\delta(\theta_{32})\delta_{bb_2}\delta_{ab_1}
+2\pi\delta(\theta_{31})\delta_{ab_2}\delta_{bb_1}\nn
&=&{}_{b_1b_2}\langle\theta_1-i0,\theta_2+i0|\Phi^a(0)|\theta_3\rangle_b\nn
&&\quad +2\pi\delta(\theta_{32})S_{b_1b_2}^{ab}(\theta_{12})
+2\pi\delta(\theta_{31})S^{ab}_{b_2b_1}(\theta_{12})\nn
&=&{}_{b_1b_2}\langle\theta_1+i0,\theta_2+i0|\Phi^a(0)|\theta_3\rangle_b\nn
&&\quad+2\pi\delta(\theta_{32})S_{b_1b_2}^{ab}(\theta_{12})
+2\pi\delta(\theta_{31})\delta_{ab_2}\delta_{bb_1}\nn
&=&{}_{b_1b_2}\langle\theta_1-i0,\theta_2-i0|\Phi^a(0)|\theta_3\rangle_b\nn
&&\quad+2\pi\delta(\theta_{32})\delta_{ab_1}\delta_{bb_2}+2\pi\delta(\theta_{31})
S^{ab}_{b_2b_1}(\theta_{12}).
\label{ACFF}
\eea
Clearly the absolute value squared of \fr{ACFF} is ill-defined as it
contains squares of delta functions. We can circumvent this problem by
introducing infinitesimal shifts of rapidities in one of the form
factors. In the case at hand this leads to
\begin{eqnarray}
&&\fl
|{}_{b_1b_2}\langle\theta_1,\theta_2|\Phi^a(0)|\theta_3\rangle_b|^2\equiv
\lim_{\kappa\to 0}\
{}_b\langle\th_3|\Phi^a(0)|\theta_2,\theta_1\rangle_{b_2b_1}\
{}_{b_1b_2}\langle\theta_1,\theta_2|\Phi^a(0)|\theta_3+\kappa\rangle_b
\nn
&&\fl=\lim_{\kappa\to 0}\Biggl\{\Bigl[
{}_{b_1b_2}\langle\theta_1-i0,\theta_2+i0|\Phi^a(0)|\theta_3\rangle_b^*
+2\pi\delta(\th_{32}) S_{b_1b_2}^{ab}(\theta_{21})
+2\pi\delta(\th_{31}) S_{b_2b_1}^{ab}(\theta_{21})
\Bigr]\nn
&&\fl\ \times\Bigl[
{}_{b_1b_2}\langle\theta_1+i0,\theta_2-i0|\Phi^a(0)|\theta_3+\kappa\rangle_b
+2\pi\delta(\theta_{32}+\kappa)\delta_{bb_2}\delta_{ab_1}
+2\pi\delta(\theta_{31}+\kappa)\delta_{ab_2}\delta_{bb_1}\Bigr]
\Biggr\},\nn
\label{SMFFsq}
\end{eqnarray}
where in the second line we have used \fr{ACFF}.
Using the results of \ref{app:B} we can separate \fr{SMFFsq} into a
connected contribution $\Gamma^{\rm conn}$ given by \fr{Gconn},
\fr{SMconnFF} and disconnected contributions $\Gamma^{\rm dis,1}$ and
$\Gamma^{\rm dis,2}$ given by \fr{Gdis1} and \fr{Gdis2}
respectively. Concomitantly $E_{1,2}$ can be split into a connected
and a disconnected part 
\be
E^\Phi_{1,2}(\omega,q)=E_{1,2}^{\rm conn}(\omega,q)+E_{1,2}^{\rm
  dis}(\omega,q).
\label{SMc12decomp}
\ee
The connected part is
\bea
E_{1,2}^{\rm conn}(\omega,q)&=&v\int
\frac{d\th_1d\th_2d\th_3}{2(2\pi)^2} 
\frac{e^{-\beta\Delta c(\th_3)}}
{\omega+i0-\Delta[c(\theta_1)+c(\theta_2)-c(\theta_3)]}\nn
&&\times\delta\bigl(vq-\Delta[s(\theta_1)+s(\theta_2)-s(\theta_3)]\bigr)\
L(\th_1,\th_2,\th_3),
\label{SMc12conn}
\eea
where
\bea\label{defL}
\fl
L(\th_1,\th_2,\th_3)&=&\frac{\pi^6}{2}\left[\th_{12}^2+\th_{13}^2+\th_{23}^2+4\pi^2\right]
\frac{\th_{12}^2+\pi^2}{\th_{12}^2(\th_{12}^2+4\pi^2)}
\tanh^4\Bigl(\frac{\th_{12}}{2}\Bigr) \nn
&\times& \frac{(\th_{13}+i0)^2}{(\th_{13}^2+\pi^2)^2}
\coth^4\Bigl(\frac{\th_{13}+i0}{2}\Bigr) 
\frac{(\th_{23}-i0)^2}{(\th_{23}^2+\pi^2)^2}
\coth^4\Bigl(\frac{\th_{23}-i0}{2}\Bigr).
\eea
The disconnected term, $E^{\rm dis}_{1,2}$ is equal to
\bea
E_{1,2}^{\rm dis}(\omega,q)&=&v\int
\frac{d\th_1d\th_2d\th_3}{2(2\pi)^2} 
\frac{e^{-\beta\Delta c(\th_3)}}
{\omega+i0-\Delta[c(\theta_1)+c(\theta_2)-c(\theta_3)]}\nn
&&\times\delta\bigl(vq-\Delta[s(\theta_1)+s(\theta_2)-s(\theta)]\bigr)\
\sum_{j=1}^2\Gamma^{\rm dis,j}_+(\th_1,\th_2,\th_3),
\eea
where $\Gamma^{\rm dis,j}_+$ are given by \fr{gdis1+} and \fr{gdis2+}
respectively.
Upon evaluation, this term simplifies to
\begin{eqnarray}\label{SMc12dis}
E^{\rm dis}_{1,2}(\omega,q) &=& v\int d\th_1d\th_2d\th_3
\frac{\delta\bigl(vq-\Delta[s(\theta_1)+s(\theta_2)-s(\theta_3)]\bigr)
e^{-\beta\Delta    c(\th_3)}}
{\omega+i0-\Delta[c(\theta_1)+c(\theta_2)-c(\theta_3)]}\nn
&&\times \frac{4}{(\th^2_{12}+\pi^2)(\th^2_{12}+4\pi^2)}\Bigg[
\pi^4\delta(\th_{31})\delta(\th_{32})+
\pi^2\th_{21}\delta'(\th_{32})\nn
&&+\delta(\th_{32})\Big(
\frac{4\pi^2\th_{21}}{\sinh\th_{21}}+\frac{\th_{21}^2(\th_{21}^2+5\pi^2}
{\th_{21}^2+\pi^2}\Big)\Bigg]
+ E_{0,1}(\omega,q){\cal Z}_1,
\end{eqnarray}
where ${\cal Z}_1$ is defined below in \fr{SMZ1}.
Note that unlike the Ising model, the finite
disconnected terms here involve derivatives of $\delta$-functions.

To evaluate the connected term, $E^{\rm conn}_{1,2}$, we now proceed in complete analogy with the Ising case. We change
variables to $\th_\pm=(\th_2\pm\th_1)/2$, carry out the
$\th_+$-integral using the momentum conservation delta-function and
then shift the $\th_-$ integration contour down in the complex plane.
This results in
\bea
E_{1,2}^{\rm conn}(\omega,q)&=&
-iv
\int_{\rm S_+}\frac{d\theta}{2\pi}\frac{e^{-\beta\Delta c(\theta)}\
\ks(\alpha(\omega,q,\th),\theta_0(\omega,q,\th),\theta)
}
{\tilde{s}(\omega,q,\th)\sqrt{\tilde{s}^2
(\omega,q,\th)-4\Delta^2}}\  
\nn
&-&v
\int_{T_+^\gamma}\frac{d\theta}{2\pi}\frac{e^{-\Delta\beta c(\theta)}
\ \ks(\alb(\omega,q,\th),\theta_0(\omega,q,\th),\theta)}
{\tilde{s}(\omega,q,\th)\sqrt{4\Delta^2-\tilde{s}^2(\omega,q,\th)}}
\nn
&+&
v\int\frac{d\theta}{(2\pi)^2}\int_{\rm S}d\theta_-
\frac{e^{-\beta\Delta c(\th)}\ \ks(\theta_-,\theta_+^0(q,\th,\th_-),\theta)}
{\left[\ot-u(q,\theta,\theta_-)\right]u(q,\theta,\theta_-)},
\label{SMc12conn2}
\eea
where 
$\ot$, $\th_+^0(q,\th,\th_-)$ and $u(q,\th,\th_-)$ are given by \fr{ot},
$\alpha$ by \fr{tm0}, $\tilde{s}(\omega,q,\th)$ by \fr{stilde},
$\alb$ by \fr{tm1} and
\be
\ks(\th_+,\th_-,\th_3)=L(\th_+-\th_-,\th_++\th_-,\th_3).
\ee
The remaining integrals in \fr{SMc12conn2}
are easily evaluated numerically. 

We can further simplify the disconnected terms
by performing some of the integrals.  Doing so we
obtain
\bea
E_{1,2}^{\rm  dis}(\omega,q)&=&\frac{v}{\veps(q)}
\frac{3\delta(\kappa)}{\omega+i0-\veps(q)}
\int_{-\infty}^\infty d\th\ e^{-\beta\Delta c(\th)}\nn
&+&\frac{v}{\veps(q)}\frac{1}{\omega+i0-\veps(q)}
\left[e^{-\beta\veps(q)}+\int_{-\infty}^\infty d\th e^{-\beta\Delta\ c(\th+\th_q)}\
g_1(\th)\right]\nn
&+&\frac{v}{\veps(q)}\frac{\Delta}{(\omega+i0-\veps(q))^2}
\int_{-\infty}^\infty d\th\ e^{-\beta\Delta\ c(\th+\th_q)}\
g_2(\th),
\eea
\bea
g_1(\th)&=&
\frac{4}{(\th^2+\pi^2)(\th^2+4\pi^2)}\left[
\th^2+\frac{4\pi^2\th}{\sinh\th}+\frac{4\pi^2\th^2}{\th^2+\pi^2}
\right]\nn
&+&4\pi^2\left[\Big(\Delta\beta s(\th+\th_q)
+\frac{\Delta vq}{\veps^2(q)}c(\th+\th_q)\Big)f(\th)+
\frac{\Delta c(\th+\th_q)}{\veps(q)} 
\ \frac{df(\th)}{d\th}\right]\nn
g_2(\th)&=&4\pi^2f(\th)
\left[s(\th+\th_q)-\frac{vq}{\veps(q)}c(\th+\th_q)\right].
\eea
Here
\bea
f(\th)&=&\frac{\th}{(\th^2+\pi^2)(\th^2+4\pi^2)},\nn
\th_q&=&{\rm arcsinh}\Bigl(\frac{vq}{\Delta}\Bigr).
\label{thq}
\eea
The next step in the low-temperature expansion is to subtract the
contributions due to the partition function following \fr{El},
\fr{Zgen}. The relevant contribution to the partition function is
\bea
{\cal Z}_1&\equiv&\lim_{\kappa\to 0}\sum_b
\int_{-\infty}^\infty \frac{d\th}{2\pi}\ e^{-\beta\Delta c(\th)}\
{}_b\langle\th|\th+\kappa\rangle_b\nn
&=&\lim_{\kappa\to 0}3\delta(\kappa)
\int_{-\infty}^\infty d\th\ e^{-\beta\Delta c(\th)}.
\label{SMZ1}
\eea
Using that ${\cal E}^\Phi_1\simeq E^\Phi_{1,0}+E^\Phi_{1,2}-{\cal Z}_1E^\Phi_{0,1}$ we obtain
\bea
{\cal E}^\Phi_1&=&\frac{v}{\veps(q)}\frac{1}{\omega+i0-\veps(q)}
\left[e^{-\beta\veps(q)}+\int_{-\infty}^\infty d\th e^{-\beta\Delta\
    c(\th+\th_q)}\ g_1(\th)\right]\nn
&+&\frac{v}{\veps(q)}\frac{\Delta}{(\omega+i0-\veps(q))^2}
\int_{-\infty}^\infty d\th\ e^{-\beta\Delta\ c(\th+\th_q)}\
g_2(\th)\nn
&+&\frac{v}{\veps(q)}\frac{e^{-\beta\Delta\veps(q)}}{\omega+i0+\veps(q)}+
E_{1,2}^{\rm conn}(\omega,q).
\eea
Starting from $C^\Phi_{2,1}(\omega,q)$ we can determine the contribution
${\cal F}^\Phi_1$ in an analogous way. 
This then leads to the following result for the leading
finite temperature contribution to the expansion \fr{lowTex} for
frequencies $\omega\approx\Delta$
\bea
\label{C1SMinf}
{\cal C}^\Phi_1(\omega,q)&\approx&\frac{2v}{(\omega+i0)^2-\veps^2(q)}
\int_{-\infty}^\infty d\th e^{-\beta\Delta\ c(\th+\th_q)}\
g_1(\th)\nn
&+&\frac{2v\Delta\Bigl((\omega+i0)^2+\veps^2(q)\Bigr)}
{\veps(q)\Bigl((\omega+i0)^2-\veps^2(q)\Bigr)^2}
\int_{-\infty}^\infty d\th\ e^{-\beta\Delta\ c(\th+\th_q)}\
g_2(\th)\nn
&-&iv\int_{\rm S_+}\frac{d\theta}{2\pi}\frac{e^{-\beta\Delta c(\theta)}\
\ks(\alpha(\omega, q,\th),
\theta_0(\omega, q,\th),\theta)}
{\tilde{s}(\omega, q,\th)\sqrt{\tilde{s}^2
(\omega, q,\th)-4\Delta^2}}
\nn
&-&v\int_{T_+^\gamma}\frac{d\theta}{2\pi}\frac{e^{-\Delta\beta c(\theta)}
\ \ks(\alb( \omega, q,\th),
\theta_0(\omega, q,\th),\theta)}
{\tilde{s}(\omega, q,\th)\sqrt{4\Delta^2-
\tilde{s}^2(\omega, q,\th)}}
\nn
&+&
v\int\frac{d\theta}{(2\pi)^2}\int_{\rm S}d\theta_-
\frac{e^{-\beta\Delta c(\th)}\ \ks(\theta_-,\theta_+^0(q,\th,\th_-),\theta)}
{\left[\Omega(\th,\omega)-u(q,\theta,\theta_-)\right]u(q,\theta,\theta_-)}.
\eea
Here $S_+$ and $T_+^\gamma$ are the segments of the real axis
characterized by $\tilde{s}^2(\omega,q,\th)>4\Delta^2$ and
$4\Delta^2\cos^2\gamma\leq\tilde{s}^2(\omega, q,\th)\leq4\Delta^2$ 
respectively, and $S$ is the contour from $-\infty-i\gamma$ to 
$\infty-i\gamma$ parallel to the real axis.

\subsection{Finite volume regularization}
A second way of regularizing infinities in matrix elements is to work
in a large, finite volume R. As was pointed out in Ref.\cite{takacs},
up to corrections that are exponentially small in system size the
functional form of matrix elements remains the same as in the
thermodynamic limit. The main effect of the finite volume is to
quantize to momenta or equivalently the rapidities, that parametrize
the basis states. The quantization conditions in the finite volume are
\be
e^{iR\frac{\Delta}{v}\sinh\th_1}|\th_1\ldots\th_n\rangle_{b_1\ldots
b_n}=S^{a_nc_{n-1}}_{b_nb_1}(\th_{n1})\ldots
S^{a_2a_1}_{b_2c_1}(\th_{21})
|\th_1\ldots\th_n\rangle_{a_1\ldots a_n},
\ee
where $\th_{kl}=\th_k-\th_l$. In the one-particle sector we simply have
\be
e^{iR\frac{\Delta}{v}\sinh\th}=1.
\ee
This is readily solved
\be
\th_j={\rm arcsinh}\left(\frac{2\pi vj}{R\Delta}\right)\ ,\
j\in {\mathbb Z}.
\label{BAE1}
\ee
The density of such one particle states is
\be
\rho_1(\th)=\frac{\Delta R}{v}\cosh\th.
\ee

In the two-particle sector the finite volume quantization conditions are
obtained from the S-matrix eigenvalues $S_a(\th)$ corresponding to
spin singlet, triplet and quintet representations 
\be
e^{iR\frac{\Delta}{v}\sinh\th_1}=
e^{-iR\frac{\Delta}{v}\sinh\th_2}=
S^{(a)}(\th_2-\th_1)\ ,\ a=0,1,2.
\label{BAE2a}
\ee
We have
\bea\label{2Smat}
S^{(0)}(\th)&=&\frac{\th+2\pi i}{\th-2\pi i}\ ,\\
S^{(1)}(\th)&=&\frac{\th-\pi i}{\th+\pi i}\ \frac{\th+2\pi i}{\th-2\pi i}\ ,\\
S^{(2)}(\th)&=&\frac{\th-\pi i}{\th+\pi i}\ .
\eea
In practice it is useful to consider the logarithmic form of the
quantization conditions
\bea
Y^{(S)}_j(\th_1,\th_2)=\frac{R\Delta}{v}\sinh\th_j-
\sum_{k\neq j}\delta^{(S)}\left(\th_j-\th_k\right)=2\pi I^{(S)}_j\
,
\label{BAE2}
\eea
where $S=0,1,2$, $I_j^{(S)}\in {\mathbb Z}+\frac{S+1}{2}$ and
\bea
\delta^{(0)}(\th)&=&2\arctan\left(\frac{\th}{2\pi}\right),\nn
\delta^{(1)}(\th)&=&2\arctan\left(\frac{\th}{2\pi}\right)
-2\arctan\left(\frac{\th}{\pi}\right),\nn
\delta^{(2)}(\th)&=&-2\arctan\left(\frac{\th}{\pi}\right).
\eea
For later use we define the density of two-particle Bethe ansatz states
with total spin $S$
\be
\rho_2^{(S)}(\th_1,\th_2)=\det\frac{\partial
  Y_j^{(S)}(\th_1,\th_2)}{\partial \th_k}.
\ee

In Bethe ansatz solvable models each solution of the quantization
conditions (Bethe ansatz equations) gives rise to the highest weight
state of an entire multiplet of the global symmetry algebra
\cite{HWS}. For the sigma model this means that each solution of the
of \fr{BAE2} gives the highest-weight state of a O(3) multiplet. The
entire multiplet is constructed from the highest weight state by
acting with the spin lowering operator. This leaves the spatial part
of the wave function unchanged, and all states in the multiplet are
characterized by the same set of quantized rapidities.

Recalling the action of spin operators on single-particle states
\fr{spin}, we may construct the following basis of two-particle
eigenstates with definite values of total spin
\bea
|\th_1,\th_2;2,\pm 2\rangle&=&\frac{1}{2}\Bigl\{|\th_1\th_2\rangle_{11}
-|\th_1\th_2\rangle_{22}\pm i |\th_1\th_2\rangle_{12}
\pm i|\th_1\th_2\rangle_{21}\Bigr\},\nn
|\th_1,\th_2;2,\pm 1\rangle&=&\frac{1}{2}\Bigl\{|\th_1\th_2\rangle_{13}
+|\th_1\th_2\rangle_{31}\pm i |\th_1\th_2\rangle_{32}
\pm i|\th_1\th_2\rangle_{23}\Bigr\},\nn
|\th_1,\th_2;2,0\rangle&=&\frac{1}{\sqrt{6}}\Bigl\{2|\th_1\th_2\rangle_{33}
-|\th_1\th_2\rangle_{11}- |\th_1\th_2\rangle_{22}\Bigr\},
\label{quintet}
\eea

\bea
|\th_1,\th_2;1,\pm 1\rangle&=&\frac{1}{2}\Bigl\{
|\th_1\th_2\rangle_{13}- |\th_1\th_2\rangle_{31}\pm i
|\th_1\th_2\rangle_{23} \mp i |\th_1\th_2\rangle_{32}\Bigr\}
\ ,\nn
|\th_1,\th_2;1,0\rangle&=&\frac{1}{\sqrt{2}}\Bigl\{|\th_1\th_2\rangle_{12}
-|\th_1\th_2\rangle_{21}\Bigr\},
\label{triplet}
\eea

\bea
|\th_1,\th_2;0,0\rangle&=&\frac{1}{\sqrt{3}}\Bigl\{
|\th_1\th_2\rangle_{33} + |\th_1\th_2\rangle_{11}
+|\th_1\th_2\rangle_{22}\Bigr\}.
\label{singlet}
\eea
The form factors in the basis \fr{quintet}, \fr{triplet}, \fr{singlet}
are readily obtained. Using the crossing relations for different
rapidities $\th_{1,2}\neq\th_3$
\be
{}_{b_1b_2}\langle\th_1,\th_2|\Phi^a(0,0)|\th_3\rangle_b=
\langle 0|\Phi^a(0,0)|\th_1+i\pi,\th_2+i\pi,\th_3\rangle_{b_1b_2b},
\ee
we obtain the following result for the form factor squares involving
the two-particle singlet, triplet and quintet states
\bea
\sum_b|\langle\th_1,\th_2;0,0|\Phi^a(0,0)|\th_3\rangle_b|^2&=&
-\frac{\pi^6}{3}\psi(\th_{12})\psi(-\th_{12})(\th_{12}^2+\pi^2)\nn
&&\times\psi(\th_{13}+i\pi)^2\psi(\th_{23}+i\pi)^2,
\eea

\bea
\sum_b\sum_{\sigma=-1}^1|\langle\th_1,\th_2;1,\sigma|\Phi^a(0,0)|\th_3\rangle_b|^2&=&
-\frac{\pi^6}{4}
\psi(\th_{12})\psi(\th_{21})
(\th_{13}+\th_{23})^2\nn
&&\times\psi(\th_{13}+i\pi)^2\psi(\th_{23}+i\pi)^2,
\eea

\bea
\sum_b\sum_{\sigma=-2}^2|\langle\th_1,\th_2;2,\sigma|\Phi^a(0,0)|\th_3\rangle_b|^2&=& -
\frac{5\pi^6}{12}\psi(\th_{12})\psi(\th_{21})(\th_{12}^2+4\pi^2)\nn
&&\times\psi(\th_{13}+i\pi)^2\psi(\th_{23}+i\pi)^2.
\eea
In order to proceed we now assume that the form factors in a large
finite volume have the same functional form as in the infinite volume
up to exponentially small corrections. This is true for the quantum
Ising model, for which the exact finite volume form factors are known,
and support in favour of this hypothesis for general massive
integrable QFTs has been provided in Ref. \cite{takacs}. 
We are now in a position to evaluate the leading contributions ${\cal
  C}_0$, ${\cal C}_1$ \fr{Cl} to the low-temperature expansion
\fr{lowTex} of the two-point function \fr{SMchiSR} of the nonlinear
sigma model in the finite volume regularization scheme.
The leading contribution is
\bea
\!{\cal C}_0^R(\omega,q)&\approx&E^R_{0,1}(\omega,q)
+F^R_{1,0}(\omega,q)=
\frac{v}{\veps(q)}\left[\frac{1}{\omega+i0-\veps(q)}
-\frac{1}{\omega+i0+\veps(q)}\right]\nn
&=&\frac{2v}{(\omega+i0)^2-\veps^2(q)},
\eea
where the momentum is quantized $q=2\pi j/R$ with $j$ an integer. The
first subleading contribution is 
\be
{\cal C}^R_1(\omega,q)={\cal E}^R_1(\omega,q)+{\cal F}^R_1(\omega,q).
\ee
Here ${\cal E}_1^R(\omega,q)\approx E_{1,0}^R+E_{1,2}^R(\omega,q)-{\cal
  Z}_1^RE_{0,1}(\omega,q)$, where
\bea
E^R_{1,2}(\omega,q) &\equiv& \sum_{S=0}^2 E^{RS}_{1,2}(\omega,q)
=\sum_{S=0}^2 \int_0^R dx\ e^{-iqx} E^{RS}_{1,2}(\omega,x)\  ,\nn
E^{RS}_{1,2}(\omega,x) &=& \frac{1}{2} \sum_{b}\sum_{\sigma=-S}^S
\sum_{\th_3}\sum_{\th_1\neq\th_2}
\frac{W(\omega,x,\th_1,\th_2,\th_3)}{\rho_1(\th_3)\ \rho_2^{(S)}(\th_1,\th_2)}
\nn
&& \hskip 1.2in \times\
|\langle\theta_1,\th_2;S,\sigma| \Phi^a(0,0)|\th_3\rangle_{b}|^2\ ,
\label{SMCrsfvol}
\eea
\be
W(\omega, x,\th_1,\th_2,\th_3) = 
\frac{e^{-\beta\Delta c(\th_3)}e^{-ix\Delta [s(\th_3)-s(\th_1)-s(\th_2)]}}
{\omega+i\delta-\Delta[ c(\th_1)+ c(\th_2)-c(\th_3)]}\ ,
\label{W}
\ee
\be
{\cal Z}^{R}_1=3\sum_{j} \exp\Big(-\beta\veps\Bigl(\frac{2\pi
  j}{R}\Bigr)\Big).
\label{SMZR}
\ee
Here the sums are over solutions to the Bethe ansatz equations
\fr{BAE1}, \fr{BAE2}. The contribution ${\cal F}_1^R(\omega,q)$ is
obtained from ${\cal E}_1^R(\omega,q)$ using the symmetry
\be
{\cal F}^R_1(\omega,q)=\left[{\cal E}^R_1(-\omega,-q)\right]^*.
\ee
\subsection{Evaluating $E^R_{1,2}$ as $R\rightarrow\infty$: Comparing
the Finite and Infinite Volume 
Regularization Scheme}
In this section we compute $E^{R}_{1,2}(\omega,q)$ as
$R\rightarrow\infty$.  We will show that
$\lim_{R\rightarrow\infty}E^{R}_{1,2}(\omega,q) -
E^R_{0,1}(\omega,q){\cal Z}^R_1$ is the same as $E_{1,2}(\omega,q) -
E_{0,1}(\omega,q){\cal Z}_1$ as computed in the infinite volume
scheme.  This then establishes again (as with the quantum Ising model)
that the two regularization schemes are equivalent. However in this
case we will have shown this to be true in a non-trivial setting: that
of an interacting theory with non-diagonal scattering.

To evaluate $E^{R}_{1,2}(\omega,q)$ we first examine the contributions
of each spin sector, i.e. the terms in the sum $E^{R}_{1,2}(\omega,q)
= \sum_S E^{RS}_{1,2}(\omega,q)$ (see
Eqn. (\ref{SMCrsfvol})). We will take the Fourier transform in $q$ in
the end and consider $E^{RS}_{1,2}(\omega,x)$ for the time being. The
latter takes the form 
\be
\fl
E^{RS}_{1,2}(\omega,x) = \frac{1}{2}\sum_{\th_3}\sum_{\th_1\neq\th_2}
\frac{G^{(S)}(\omega,x,\sth)}{\rho_1(\th_3) \rho_2^{(S)}(\th_1,\th_2)}
\tanh^2(\frac{\th_{12}}{2})\coth^2(\frac{\th_{13}}{2})
\coth^2(\frac{\th_{23}}{2}) \ ,
\ee
\begin{eqnarray}
G^{(S)}(\omega,x,\sth) &\equiv& -\psi(\th_{12})\psi(\th_{21})
\psi(\th_{13}+i\pi)^2\psi(\th_{23}+i\pi)^2\nn
&\times&\coth^2(\frac{\th_{12}}{2})\tanh^2(\frac{\th_{13}}{2})
\tanh^2(\frac{\th_{23}}{2}) U^{(S)}(\omega,q,\sth),
\end{eqnarray}
where
\begin{eqnarray}
U^{(0)}(\omega,x,\sth) &=& \frac{\pi^6}{3} (\th^2_{12}+\pi^2)
W(\omega,x,\sth)\ ,\nn
U^{(1)}(\omega,x,\sth) &=& \frac{\pi^6}{4} (\th_{13}+\th_{23})^2
W(\omega,x,\sth)\ ,\nn
U^{(2)}(\omega,x,\sth) &=& \frac{5\pi^6}{12} (\th^2_{12}+4\pi^2) 
W(\omega,x,\sth).
\end{eqnarray}
Like with the Ising model in Section \ref{ssec:compIsing}, our aim is
in the $R\rightarrow\infty$ limit to convert these sums into integrals.  To
do so we must first isolate and subtract out the singular terms in
$E^{RS}_{1,2}(\omega,x)$ in order to define finite integrals. 
Expanding $E^{RS}_{1,2}(\omega,x)$ in $\th_3$ about $\th_1$ and
$\th_2$, we obtain its singular pieces  
\begin{eqnarray}
E_{{\rm sing}1,2}^{RS}(\omega,x) &=& \frac{1}{2}\sum_{\th_3}\sum_{\th_1\neq\th_2}
\frac{1}{\rho_1(\th_3)\ \rho_2^{(S)}(\th_1,\th_2)}
\sum_{a=1}^2
\Gamma^{(S)}_a(\omega,x,\sth) \nn
&\equiv& \sum_{a=1}^2E_{\rm sing}^{(a)}(\omega,x),
\eea
\bea
\label{doublepole}
\Gamma^{(S)}_{2}(\omega,x,\sth)&=& 
\frac{16}{\pi^4}\frac{U^{(S)}(\omega,x,\th_1,\th_2,\th_2)}{(\th^2_{12}+\pi^2)(\th^2_{12}+4\pi^2)}
\frac{\cosh(\th_2)\cosh(\th_3)}{(\sinh(\th_2)-\sinh(\th_3))^2}\cr\cr
&& + \big(\th_1\leftrightarrow\th_2)\ ,\\ \cr
\Gamma^{(S)}_{1}(\omega,x,\sth)&=&
\frac{16}{\pi^4}\frac{1}{(\th^2_{12}+\pi^2)^2(\th^2_{12}+4\pi^2)\th_{21}}
\frac{\cosh(\th_3)}{\sinh(\th_3)-\sinh(\th_2)}\cr\cr
&& \hskip -1.in \times\Bigg[-2U^{(S)}(\omega,x,\th_1,\th_2,\th_2)
\bigg[2\frac{\th_{21}}{\sinh(\th_{21})}(\th_{12}^2+\pi^2)+(\th^2_{12}-\pi^2)\bigg]\cr\cr
&& \hskip -0.5in
+(\th^2_{12}+\pi^2)\th_{21}\frac{\partial}{\partial\th_3}\Bigg|_{\th_3=\th_2}
U^{(S)}(\omega,x,\sth)\Bigg]
+ \big(\th_1\leftrightarrow\th_2).
\label{singlepole}
\end{eqnarray}
The singular pieces, as indicated by the introduction of the quantities
$E_{\rm sing}^{(a)}(\omega,x)$ come in the form of both single ($a=1$) and
double poles ($a=2$) in $\th_3$ about $\th_1$ and $\th_2$. 
To regularize the sum we add and subtract the singular pieces from
$E^{RS}_{1,2}(\omega,x)$: 
\begin{eqnarray}
E^{RS}_{1,2}(\omega,x) &=&
\frac{1}{2}\sum_{\th_3}\sum_{\th_1\neq\th_2}
\Bigg[\frac{G^{(S)}(\omega,x,\sth)
\tanh^2(\frac{\th_{12}}{2})\coth^2(\frac{\th_{13}}{2})\coth^2(\frac{\th_{23}}{2})}
{\rho_1(\th_3) \rho_2^{(S)}(\th_1,\th_2)}\cr\cr
&& \hskip 1in -\sum_{a=1}^2
\frac{\Gamma^{(S)}_a(\omega,x,\th_1,\th_2,\th_3)}
{\rho_1(\th_3)\ \rho_2^{(S)}(\th_1,\th_2)}\Bigg]
+ E_{{\rm sing}1,2}^{RS}(\omega,x)\cr\cr
&\equiv& E^{RS}_{{\rm finite 1,2}}(\omega,x) + E_{{\rm sing}1,2}^{RS}(\omega,x).
\end{eqnarray}
The first term in the above equation is singularity free (i.e. the
summand is finite as $\th_3$ approaches either $\th_1$ or $\th_2$). 
We can then take $R\rightarrow\infty$, turning the sum into a
principal value integral after which the integration contours can be
modified so that they deform about the singularities in the same
fashion as was done  for the Ising model (see Section
\ref{ssec:compIsing}). The result of doing so is
\begin{eqnarray}
\fl
E^{RS}_{{\rm finite 1,2}}(\omega,x)\! &=& \!
\int\!\frac{d\th_1d\th_2d\th_3}{2(2\pi)^3}
G^{(S)}(\omega,x,\sth)\tanh^2(\frac{\th_{12}}{2})
\coth^2(\frac{\th_{31}-i\eta}{2})
\coth^2(\frac{\th_{32}+i\eta}{2})\cr\cr
&+& \int 
\frac{d\th_1d\th_2d\th_3}{2\pi}\ \delta(\th_{32})\delta(\th_{31})\
G^{(S)}(\omega,x,\sth)\ .
\end{eqnarray}
We note that the principal part integrals over the single-pole terms
vanish. If we now sum over the spin sectors, $S$, then take the limit
$R\to\infty$ and finally Fourier transform with respect to $x$
we find
\begin{eqnarray}
\sum_S E^{RS}_{{\rm finite 1,2}}(\omega,q) &=& 
\frac{1}{2}\int \frac{d\th_1d\th_2d\th_3}{(2\pi)^3}
\widetilde{W}(\omega,q,\th_1,\th_2,\th_3)L(\th_1,\th_2,\th_3)\cr\cr
&+& 8\pi^2\int\frac{d\th_1d\th_2d\th_3}{(2\pi)^3}
\widetilde{W}(\omega,q,\th_1,\th_2,\th_3)\delta(\th_{12})\delta(\th_{13}),
\end{eqnarray}
where $L(\th_1,\th_2,\th_3)$ is defined in \fr{defL} and
$\widetilde{W}(\omega,q,\th_1,\th_2,\th_3)$ is
\begin{equation}
\widetilde{W}(\omega, q,\th_1,\th_2,\th_3) = \frac{e^{-\beta\Delta c(\th_3)}2\pi\delta(vq+\Delta (s(\th_3)-s(\th_1)-s(\th_2)))}
{\omega+i\delta-\Delta[ c(\th_1)+ c(\th_2)-c(\th_3)]}.
\end{equation}
We see that we have reproduced $E^{\rm conn}_{1,2}(\omega,q)$ as given in Eqn. (\ref{SMc12conn}) plus a disconnected
term.  The remaining disconnected terms together with a term proportional to the partition function
are found $E_{{\rm sing}1,2}^{RS}(\omega,q)$ which we now turn to evaluate.
We are able to carry out the sum over $\th_3$ courtesy of the identities
\begin{eqnarray}
\label{identities}
\sum_{\th_3}
\frac{\cosh(\th_3)}{\rho_1(\th_3)(\sinh(\th_3)-\sinh(\th_2))}
\Bigg|_{\th_1\neq\th_2\in S} &=& 
\frac{i}{2}\frac{1+S^{(S)}(\th_{12})}{1-S^{(S)}(\th_{12})}\ ,\nn
\sum_{\th_3}
\frac{\cosh(\th_2)\cosh(\th_3)}{\rho_1(\th_3)
(\sinh(\th_3)-\sinh(\th_2))^2}\Bigg|_{\th_1\neq\th_2\in
  S} &=&  
\frac{\Delta R}{2v}
\frac{\cosh(\th_2)}{1-{\rm Re}\ S^{(S)}(\th_{21})}\ .
\end{eqnarray}
Here $\theta_1\neq\theta_2\in S$ indicates that that $\th_{1,2}$ are
solutions of the Bethe ansatz equations \fr{BAE2} in the spin-S sector
and $S^{(S)}(\th )$ are the S-matrices \fr{2Smat}. The first identity
is established by replacing $\th_3$ by its corresponding integer
through the quantization conditions \fr{BAE1} and carrying out the
resulting sum using
\be
\cot(\pi z)=\frac{1}{\pi z}+\frac{2z}{\pi}\sum_{k=1}^\infty\frac{1}{z^2-k^2}.
\ee
Finally, the Bethe ansatz equations \fr{BAE2} in the two-particle
sector are used to rewrite the result. The second identity in
\fr{identities} can be established by using the derivative of the
first. 

The second identity in \fr{identities} allows us to evaluate the
double pole term in $E^{RS}_{{\rm sing}1,2}(\omega,x)$ with the result
\bea
E^{(2)}_{\rm sing}(\omega,x) &=& \frac{2S+1}{3v}
\sum_{\th_1\neq\th_2\in S}
\frac{\Delta
  R\cosh(\th_2)W(\omega,x,\th_1,\th_2,\th_2)}
{\rho^{(S)}_2(\th_1,\th_2)}\nn
&=& \frac{2S+1}{3v}\sum_{\th_1,\th_2\in S}
\frac{\Delta R\cosh(\th_2)
W(\omega,x,\th_1,\th_2,\th_2)}{\rho^{(S)}_2(\th_1,\th_2)} \nn
&& \hskip 0in - \delta_{S,1}
\frac{2S+1}{3}\sum_{\th_1}
\frac{W(\omega,x,\th_1,\th_1,\th_1)}{\rho_1(\th_1)}\nn
&=& \frac{2S+1}{3v}\Delta R\int \frac{d\th_1 d\th_2}{(2\pi)^2}
\cosh(\th_2)W(\omega,x,\th_1,\th_2,\th_2) \nn
&& \hskip 0in - \delta_{S,1}
\frac{2S+1}{3}\sum_{\th_1}
\frac{W(\omega,x,\th_1,\th_1,\th_1)}{\rho_1(\th_1)}.
\eea
Here we need to subtract the term $\th_1=\th_2$ only in the triplet
sector because in the $S=0,2$ sectors solutions of the Bethe ansatz
equations \fr{BAE2a}, \fr{2Smat} with coinciding
rapidities do not occur. In the triplet sector the spatial part of the
wave function has to be antisymmetric (as the spin part is and we are
dealing with bosons), which forbids solutions with coinciding
rapidities. Turning the above integral back into a sum over
rapidities subject to free quantization conditions \fr{BAE1} we
arrive at
\bea
E^{{(2)}}_{\rm sing}(\omega,x) &=&\frac{2S+1}{3}
\sum_{\th_1}\frac{\exp\big(ix\frac{\Delta}{v}
  s(\th_1)\big)}{\rho_1(\th_1)[\omega+i\delta-\Delta c(\th_1)]}
\sum_{\th_2}e^{-\beta\Delta c(\th_2)}\nn
&& - \delta_{S,1} 
\int \frac{d\th_1}{2\pi} W(\omega,x,\th_1,\th_1,\th_1).
\eea
Carrying out the Fourier transform in x and summing over spin sectors
we obtain 
\bea
E^{(2)}_{\rm sing}(\omega,q) &=&
E^R_{0,1}(\omega,q){\cal Z}_1^R 
- \int \frac{d\th_1}{2\pi}\ \widetilde{W}(\omega,q,\th_1,\th_1,\th_1),
\end{eqnarray}
where $\widetilde{W}$ is the spatial Fourier transform of $W$.
We thus obtain a term proportional to $R$ as $R\rightarrow\infty$ but
which will be cancelled off by a corresponding term,
$E^R_{0,1}(\omega, q) {\cal Z}^R_1$, arising from the expansion 
of the partition function.

The first identity in \fr{identities} allows us to evaluate the single
pole term in  $E^{RS}_{{\rm sing}1,2}(\omega,q)$.  Unlike the double
pole term, this leads to an expression entirely finite in the
$R\rightarrow\infty$ limit: 
\begin{eqnarray}
\sum_SE^{(1)}_{\rm sing}(\omega,x) &=& 16\pi^2 \int \frac{d\th_1d\th_1d\th_3}{(2\pi)^3}
\frac{W(\omega,x,\th_1,\th_2,\th_3)}{(\th^2_{12}+\pi^2)^2(\th^2_{12}+4\pi^2)}\cr\cr
&& \hskip -1.5in \times\Bigg[\delta(\th_{32})
\bigg\{2\pi^2\bigg(2\frac{\th_{21}(\th^2_{12}+\pi^2)}{\sinh(\th_{21})} + (\th^2_{12}-\pi^2)\bigg)
+(\th^2_{12}+\pi^2)(2\pi^2+\th^2_{12})\bigg\}\cr\cr
&& \hskip .5in + \pi^2\th_{21}(\th^2_{12}+\pi^2)\delta'(\th_{32})\Bigg].
\end{eqnarray}
This then allows us to write down a complete expression for
$E^{R}_{1,2}(\omega,q)$ as $R\rightarrow\infty$: 
\begin{eqnarray}
E^{R}_{1,2}(\omega,q) &=& \frac{1}{2}\int 
\frac{d\th_1d\th_2 d\th_3}{(2\pi)^3}
\widetilde{W}(\omega,q,\th_1,\th_2,\th_3)L(\th_1,\th_2,\th_3)\nn
&& + 4\pi^2 \int \frac{d\th_1d\th_2d\th_3}{(2\pi)^3} 
\widetilde{W}(\omega,q,\th_1,\th_2,\th_3)\delta(\th_{12})\delta(\th_{13})\nn
&&
+ \sum_SE^{(1)}_{\rm sing}(\omega,q) + E^R_{0,1}(\omega,q){\cal Z}^R_1.
\end{eqnarray}
We see that this expression agrees with that of $E_{1,2}(\omega,q)$
using the infinite volume scheme, see \fr{SMc12conn}, \fr{defL} and
\fr{SMc12dis}.
\subsection{Resummation}
In the above we have calculated the first two terms in the
expansion of the dynamical susceptibility
\be
\chi_{_\Phi}(\omega,q)= {\cal C}^\Phi_0(\omega,q)+{\cal C}^\Phi_1(\omega,q)
+\ldots .
\label{chiSM}
\ee
The calculation of ${\cal C}^\Phi_2$ is not possible in practice because
it involves the 5-particle form factor, which is tremendously
complicated for the nonlinear sigma model \cite{BN}.
We have shown that ${\cal C}^\Phi_1(\omega,q)$ exhibits a quadratic
divergence when the frequency approaches the mass shell
$\omega\to\veps(q)$, whereas ${\cal C}^\Phi_0(\omega,q)$ is only linearly
divergent. We have argued that the higher order (in $\exp(-\Delta/T)$)
terms in \fr{chiSM} exhibit stronger divergences and hence a
resummation is required to get meaningful results for
$\omega\approx\veps(q)$. Following the resummation procedure set out
in section \ref{ssec:resum} we determine the quantity
\be
\Sigma^{(1)}(\omega,q)=\frac{{\cal C}^\Phi_1(\omega,q)}
{\left({\cal C}_0^\Phi(\omega,q)\right)^2},
\ee
and then use it to obtain a resummed low-temperature approximation
(see \fr{gensigma0})
\bea
\chi^{(1)}_{_\Phi}(\omega, q) &=& \frac{{\cal C}^\Phi_0(\omega, q)}
{1-{\cal C}^\Phi_0(\omega,q)\Sigma^{(1)}(\omega, q)},
\label{chi1SM}\\
S^{(1)}(\omega,q)&=&-\frac{1}{\pi}\frac{1}{1-e^{-\frac{\omega}{T}}}{\rm Im}\ \chi^{(1)}_{_\Phi}(\omega, q) .
\label{S1SM}
\eea
\section{Results for the Low-Temperature Dynamical Susceptibility of
  the O(3) Nonlinear Sigma Model}
\label{sec:O3results}

\begin{figure}[ht]
\begin{center}
\epsfxsize=0.7\textwidth
\epsfbox{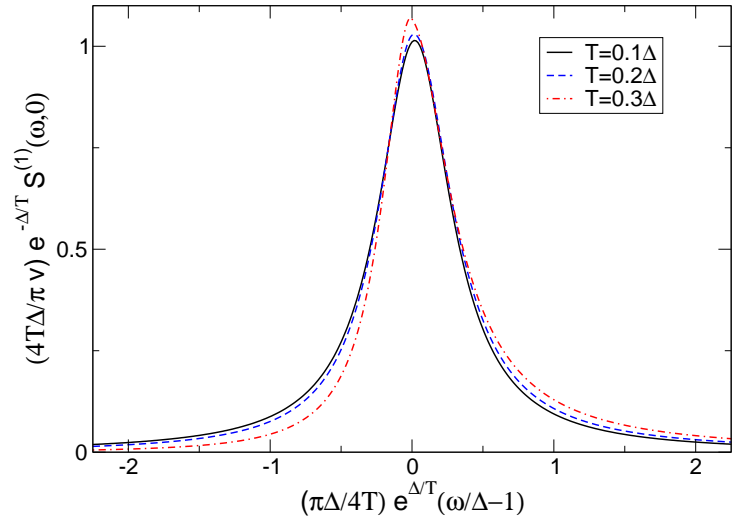}\qquad
\end{center}
\caption{Rescaled structure factor for the O(3) nonlinear sigma model
  for three different temperatures.}
\label{fig:SMscaled}
\end{figure}

The leading order result $S^{(1)}(\omega,q)$ is most easily calculated
using the infinite-volume regularization scheme and carrying out the
integrals in \fr{C1SMinf} numerically. Evaluation of
$S^{(1)}(\omega,q)$ for low temperatures shows that, as expected, the
$T=0$ delta function at $\omega=\veps(q)$ broadens with
temperature. We find that the resulting peak scales as 
\bea
{\rm peak\ height}&\propto&
\frac{\Delta}{T}\ \exp\Bigl(\frac{\Delta}{T}\Bigr),\nn
{\rm peak\ width}&\propto&
\frac{T}{\Delta}\ \exp\Bigl(\frac{-\Delta}{T}\Bigr).
\label{peakscalingSM}
\eea
In order to exhibit the evolution of the structure factor as a
function of frequency for fixed momentum $q$ with temperature it is
therefore useful to rescale both the frequency axis and the structure
factor. The result is shown in Fig.\ref{fig:SMscaled} for a temperature
range $0.1\Delta\leq T<0.3\Delta$, which corresponds to approximately
a factor of $2000$ difference in peak height (and width). We see that the
lineshape is {\sl asymmetric} in frequency, with more spectral weight
appearing at higher frequencies. The asymmetry increases with
temperature. This effect is most easily quantified by comparison with
a Lorentzian lineshape, which is done below.

\begin{figure}[ht]
\begin{center}
\epsfxsize=0.7\textwidth
\epsfbox{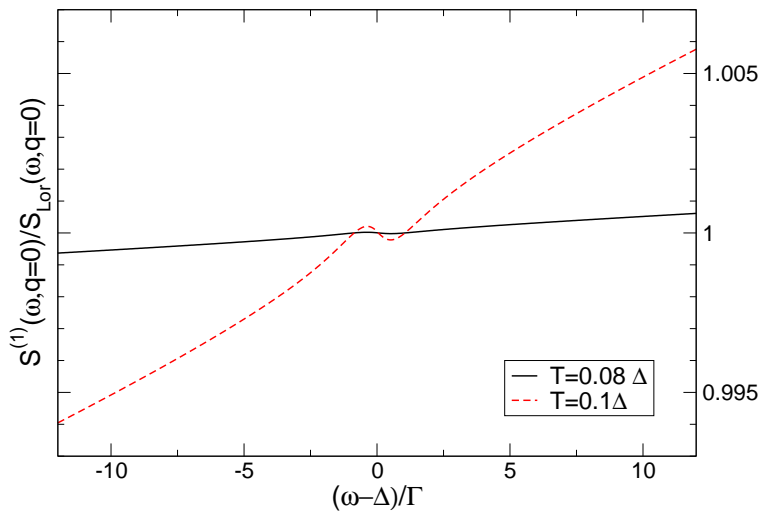}\qquad
\end{center}
\caption{Comparison of the structure factor to the Lorentzian
  approximation \fr{SMLor}.}
\label{fig:SMasymm}
\end{figure}

At very low temperatures we find that the line shape at $q=0$ is
well-approximated by a Lorentzian
\be
S^{(1)}(\omega,0)\approx
\frac{v}{\pi\Delta}
\frac{\Gamma(T)}{(\omega-\Delta(T))^2+\Gamma^2(T)},
\label{SMLor}
\ee
A comparison of $S^{(1)}(\omega,0)$ to \fr{SMLor} is shown in
Fig.\ref{fig:SMasymm} for two temperatures. We see that the agreement
improves with decreasing temperature.

\subsection{Comparison to Semiclassical Results}
In Ref. \cite{damle} Damle and Sachdev carried out a semiclassical
analysis of the dynamical structure factor of the O(3) nonlinear sigma
model. In contrast to the case of the quantum Ising model, the
scattering matrix is not taken into account fully but approximated by
its zero rapidity limit
\be
S_{ab}^{cd}(\theta)\longrightarrow-\delta_{ad}\delta_{bc}.
\ee
We note that compared to the Ising case this additional approximation
may impose tighter restrictions on the window of applicability of the
semiclassical result. 
Damle and Sachdev find the following form for the dynamical structure
factor 
\bea
S_{\rm sc}(\omega,q)
&=&
\int_{-\infty}^\infty \frac{dt}{2\pi}
dx\ e^{i\omega t-iqx}K(x,t)R(x,t)\ ,
\label{SDS}
\eea
where the relaxation function $R(x,t)$ is determined numerically and 
\bea
K(x,t)&=&\frac{Z}{\pi}K_0(|\Delta|\sqrt{(x/v)^2-t^2})\ .
\eea
Here $Z$ is a normalization factor.
At sufficiently low temperatures and when $|vq|<\sqrt{T\Delta}$
the semiclassical result is approximately Lorentzian in form
\cite{damle,sachdevbook} 
\be
S_{\rm sc}(\omega,q)\approx S_{\rm Lor}(\omega,q)=
\frac{vZ}{\pi\veps(q)}
\frac{\alpha/\tau_0}{(\omega-\veps(q))^2+(\alpha/\tau_0)^2},
\label{SMLorentzian}
\ee
where $\alpha\approx 0.72$ \cite{sachdevbook} and 
\be
\tau_0=\frac{\sqrt{\pi}}{3T}e^{\Delta/T}\ .
\ee
We have shown above that our result is well approximated by a
Lorentzian at low temperatures. However, unlike \fr{SMLorentzian}, the
best fit to our result involves a temperature dependent gap, see
\fr{SMLor}. The width of the Lorentzian is quite close to Damle and
Sachdev's result, e.g. we find that
$\Gamma(0.1T)=0.736/\tau_\phi$. However, our value of $\alpha$ is
found to increase as $T\to 0$. In Ref. \cite{zarand} an analytic
expression for the relaxation function $R(x,t)$ was given
\bea
R(x,t)&=&C\int_{-\pi}^\pi \frac{d\phi}{2\pi}
\frac{(1-\gamma^2)\cos(\bar{x}\sin\phi)}{\gamma^2+2\gamma\cos\phi+1}
\nn
&&\times\
\exp\left[-(1-\cos\phi)|\bar{t}|\Big(\frac{e^{-u^2}}{\sqrt{\pi}}+u\ {\rm
    erf}(u)\Big)\right],
\eea
where $C$ is a normalization constant, $\gamma=\frac{1}{3}$,
$\bar{x}=x/\xi_0$, $\bar{t}=t/\tau_0$, $u=\bar{x}/\bar{t}$ and
\be
\xi_0=\frac{1}{3}\sqrt{\frac{2\pi v^2}{\Delta T}}e^{\Delta/T}\ .
\ee
At low temperatures and for $\omega$ sufficiently close to $\veps(q)$
the Fourier transform \fr{SDS} is well approximated by \cite{damle}
\bea
S_{\rm sc}(\omega,q)
&\approx&
\int_{-\infty}^\infty \frac{dt}{2\pi}
dx\ e^{i\omega t-iqx}K(x,t)R(0,t)\ .
\label{SDS2}
\eea
Carrying out the integrals we obtain
\bea
S_{\rm sc}(\omega,q)
&\approx&\frac{ZCv}{\pi\veps(q)}{\rm Re}\left[
\frac{\overline{\Gamma}(T)+i(\omega-\veps(q))}{(\omega-\veps(q))^2+
\overline{\Gamma}^2(T)}\left(1+\frac{1-\gamma^2}{2\gamma\sqrt{\alpha^2-1}}
\right)
\right],
\eea
where 
\be
\overline{\Gamma}(T)=\frac{\gamma+\frac{1}{\gamma}+2}{2\sqrt{\pi}\tau_0}
\ ,\quad
\alpha=1-i\sqrt{\pi}\tau_0(\omega-\veps(q))\ .
\ee
While this expression is very close to being a Lorentzian sufficiently
far away from the mass shell, it exhibits a square root divergence as
$\omega\rightarrow\veps(q)$. This precludes a comparison with the
results of our low-temperature expansion.

\section{Summary and Discussion}
\label{sec:summ}
In this work we have proposed a general method for determining
frequency and momentum dependent two-point functions of local
operators in massive integrable quantum field theories at low
temperatures. We have applied this method to the
calculation of response functions in the disordered phase of the
quantum Ising model and the O(3) nonlinear sigma model.
The methodology in this particular application possesses two crucial
ingredients. First we have shown that there exists a systematic
expansion of the spin-spin response function in terms of the small
parameter $\exp(-\Delta/T)$, where $\Delta$ is the spectral gap and
$T$ is the temperature.  While such expansions have been known to
exist previously \cite{rmk}, they were limited to certain 
correlation functions.  Here we have shown that the spin-spin response
function can be described sensibly by performing a low temperature
expansion of the spin-spin response's ``self-energy'' (or a quantity
analogous to such). The second ingredient was a procedure to make
sense of infinities that appear in the squares of matrix elements that
arise for a Lehmann expansion of the correlators. 
In this paper we have developed a new regulator for the infinities
that appear when working in an infinite volume and shown that this
regulator reproduces results found by working in a large, finite
volume and taking the thermodynamic limit only at the end of the
calculation. We have accomplished this both in the quantum Ising
model, a free fermionic theory, and the O(3) NLSM where the elementary
excitations are strongly interacting with a non-diagonal, momentum
dependent scattering matrix.  

A number of open problems remain on the technical level. In the
present manuscript we have not analyzed the situation where we need
more than one $\kappa$ parameter in the infinite volume regularization
scheme (see e.g. \fr{kappahigher}). It is important to establish the
equivalence of infinite and finite volume regularization schemes in
this more general case as well. We also have not presented a general
proof that {\sl all} terms \fr{El},\fr{Fl} in the low-temperature
expansion \fr{lowTexgen} are finite. We hope that these questions will
be addressed in future work.

At zero temperature the response functions in both the Ising model and
the O(3) NLSM are dominated by a delta function peak at the position of
the single particle dispersion. We have determined how this peak
broadens at finite temperatures smaller than the gap. Our main result
is that the lineshape at $T>0$ exhibits a pronounced asymmetry in
energy that increases with temperature. At very low temperatures
$T<0.1 \Delta$ our results essentially reduce to those of previous
semiclassical analyses \cite{young,damle,sachdevbook}, which concluded
that the lineshape is to a good approximation Lorentzian. This shows that
while the semiclassical approximation gives a good account of the
width and height of the lineshape at $T>0$ \footnote{Indeed, as was
  shown in Ref. \cite{zheludev}, the width of the lineshape   obtained in the
  semiclassical approximation is in agreement with experiments on a
  number of different materials (with gaps that are small compared to
  the magnetic exchange constant) even at elevated temperatures
  $T\approx\Delta$.}, it is quantitatively accurate with regards to
the overall lineshape only at very low temperatures. 

A second important feature seen in our low-temperature expansion is
what is known as the temperature dependent gap. For sufficiently high
temperatures the maximum of the lineshape is seen to shift upwards in
energy when compared to the $T=0$ gap. We found that this phenomenon
emerges in the quantum Ising model once subleading terms in our
low-temperature expansion are taken into account. The calculation of
these terms involves five-particle form factors. While these are
known for the O(3) NLSM as well, their complexity
puts the calculation of subleading terms in the low temperature
expansion beyond the scope of this work. 

The O(3) NLSM describes the scaling limit of integer
spin Heisenberg chains. While the agreement between the field theory and
the Heisenberg lattice model is best for large spins (and low
energies), the sigma model has been found to provide a reasonable
approximation of the structure factor even in the extreme $S=1$ case
\cite{igor}. A number of inelastic neutron scattering experiments have
measured the temperature dependence of the dynamical structure factor
of quasi one dimensional spin-1 Heisenberg magnets\cite{kenz,xu}. Our
finding of an asymmetric lime shape are relevant to these
experiments. In particular, the excess of spectral weight at high
energies reported in \cite{kenz} for ${\rm   CsNiCl_3}$ should at least
be partly accounted for by a lineshape asymmetric in energy. However,
a quantitative comparison of our theory to inelastic neutron
scattering data on ${\rm CsNiCl_3}$ is precluded by the presence of a
N\'eel transition, driven by non-negligible interchain coupling, at a
temperature comparable to the gap. It is likely that the interchain
coupling will affect the precise lineshape in a significant way
\cite{huang}. 
Similarly, in ${\rm YBaNiO_5}$ a quantitative comparison to the finite
temperature structure factor is not straightforward because of the presence
of an exchange anisotropy.

The presence of an asymmetric lineshape even at low temperatures is
expected to be a general feature in quantum magnets that support
coherent single particle excitations at zero temperature. 
Theoretical studies of the alternating Heisenberg chain \cite{mikeska}
suggest that asymmetric lineshapes occur in dimer systems. This has
been confirmed by recent experiments on the quasi one dimensional 
alternating Heisenberg chain copper nitrate \cite{Alan}. It would be
interesting to investigate to what extent the same holds for the two
and three dimensional cases \cite{ruegg}.

The methods we have developed have a wider scope for applications. With
regards to low temperature dynamics in gapped integrable models, it
would be interesting to investigate the case of the spin-1/2
Heisenberg-Ising chain. Here the dynamical structure factor has been
measured for a number of materials \cite{XXZ}. In the limit of large
gaps the transverse component of the structure factor has been
determined by diagrammatic methods \cite{Andrew2}. The excitation
spectrum is quite different compared to the O(3) NLSM
and the disordered phase of the quantum Ising model in that the lowest
excitations are two-parametric. Concomitantly the dynamical structure
factor is dominated by an incoherent two-particle continuum at zero
temperature. At finite temperature a low-frequency resonance, known as
``Villain mode'' develops \cite{Villain}. It should be possible to
adapt our methods to analyze this case.

Our method of regularizing general matrix elements could be useful for 
studying non-integrable perturbations of integrable models, which is an
area of considerable importance \cite{decay}. In fact shortly after
our work a preprint by Takacs appeared (arXiv:0907.2109),
which addresses this problem for the double sine-Gordon model and obtains
an independent derivation of the finite volume regularization scheme.

\ack
The work was supported by the EPSRC under grant EP/D050952/1 (FHLE),
by the DOE under contract number DE-AC02 -98 CH 10886 and by the ESF
network INSTANS. We thank B. Doyon, W. G\"otze,
G. Mussardo, H. Saleur and A.M. Tsvelik for helpful discussions.

\appendix
\section{Crossing Relations}
\label{app:xing}
In this Appendix we summarize identities first given by
Smirnov in \cite{Smirnov92book} that that allow us to analytically
continue form factors. Let 
$A=\{\theta_1,\ldots,\th_n\}$ with $\th_1<\th_2<\ldots<\th_n$ and 
$B=\{\beta_1,\ldots,\beta_m\}$ with $\beta_1<\beta_2<\ldots<\beta_m$
and introduce notations
\bea
Z[\overrightarrow{A}]_{a_1\ldots a_n}&\equiv& 
Z_{a_1}(\th_1)Z_{a_2}(\th_2)\ldots Z_{a_n}(\th_n)\
,\nn
Z^\dagger[\overleftarrow{A}]_{a_n\ldots a_1}
&\equiv&Z^\dagger_{a_n}(\th_n)
Z^\dagger_{a_{n-1}}(\th_{n-1})\ldots Z^\dagger_{a_1}(\th_1)\ .
\eea
Now let $A_1$ and $A_2$ be a partition of $A$. As a consequence of the
periodicity axiom we have
\be
Z[\overrightarrow{A}]_{a_1\ldots
  a_n}=Z[\overrightarrow{A_2}]_{c_1\ldots c_r}
Z[\overrightarrow{A_1}]_{c_{r+1}\ldots c_n}
\ S(\overrightarrow{A}|\overrightarrow{A_1})^{c_1\ldots
  c_n}_{a_1\ldots a_n},
\ee
where $S(\overrightarrow{A}|\overrightarrow{A_1})$ is the product of
two-particle scattering matrices needed to rearrange the order of
Faddeev-Zamolodchikov operators in $Z[\overrightarrow{A}]$ to arrive at
$Z[\overrightarrow{A_2}]Z[\overrightarrow{A_1}]$.
Similarly we have
\be
Z^\dagger[\overleftarrow{B}]_{b_m\ldots b_1}=
Z^\dagger[\overleftarrow{B_1}]_{d_m\ldots d_{s+1}}
Z^\dagger[\overleftarrow{B}_2]_{d_s\ldots d_1}
\ S(\overleftarrow{B_1}|\overleftarrow{B})_{b_m\ldots b_1}^{d_m\ldots d_1}.
\ee
Finally we define
\be
\delta[\overrightarrow{A},\overrightarrow{B}]_{a_1\ldots a_n
\atop b_1\ldots b_m}
=\delta_{n,m}\prod_{j=1}^n
2\pi\delta(\th_j-\beta_j)\delta_{a_j,b_j}.
\ee
For a local operator ${\cal O}$ we then can analytically continue
form factors as
\bea
&&\langle 0|Z[\overrightarrow{A}]_{a_1\ldots a_n}\ {\cal O}(0,0)\
Z^\dagger[\overleftarrow{B}]_{b_m\ldots b_1}|0\rangle\nn
&&=\sum_{{A=A_1\cup A_2}\atop{B=B_1\cup B_2}}
S(\overrightarrow{A}|\overrightarrow{A_1})^{c_1\ldots
  c_n}_{a_1\ldots a_n}
\ S(\overleftarrow{B_1}|\overleftarrow{B})_{b_m\ldots b_1}^{d_m\ldots
  d_1}\ \delta[\overrightarrow{A_2},\overrightarrow{B_2}]_{c_1\ldots c_r
\atop d_s\ldots d_q}
\nn
&&\qquad\qquad\times\ 
\langle 0|Z[\overrightarrow{A}_1+i0]_{c_{r+1}\ldots c_n}
{\cal O}(0,0)
Z^\dagger[\overleftarrow{B}_1]_{d_m\ldots d_{s+1}}|0\rangle.
\eea
Similarly, we could choose to analytically continue to the lower
half-plane 
\bea
&&\langle 0|Z[\overrightarrow{A}]_{a_1\ldots a_n}\ {\cal O}(0,0)\
Z^\dagger[\overleftarrow{B}]_{b_m\ldots b_1}|0\rangle\nn
&&=\sum_{{A=A_1\cup A_2}\atop{B=B_1\cup B_2}}
S(\overrightarrow{A}|\overrightarrow{A_2})^{c_1\ldots
  c_n}_{a_1\ldots a_n}
\ S(\overleftarrow{B_2}|\overleftarrow{B})_{b_m\ldots b_1}^{d_m\ldots
  d_1}\ \delta[\overrightarrow{A_2},\overrightarrow{B_2}]_{c_1\ldots c_r
\atop d_s\ldots d_q}
\nn
&&\qquad\qquad\times\ 
\langle 0|Z[\overrightarrow{A}_1-i0]_{c_{r+1}\ldots c_n}
{\cal O}(0,0)
Z^\dagger[\overleftarrow{B}_1]_{d_m\ldots d_{s+1}}|0\rangle.
\eea
For semi-local operators the above identities needs to be modified as
discussed in \cite{Smirnov92book}.

\section{Products of Form Factors for the Ising Model}
\label{app:A}
Let us consider the product of form factors
\bea
|\langle\theta_3|\sigma(0,0)|\theta_2\theta_1\rangle|^2
&=&
\lim_{\kappa\to 0}
\langle\theta_3|\sigma(0,0)|\theta_2\theta_1\rangle
\langle\th_1,\th_2|\sigma(0,0)|\th_3+\kappa\rangle.
\eea
Using the crossing relations \fr{12ising} this can be rewritten as
\bea
\fl
|\langle\theta_3|\sigma(0,0)|\theta_2\theta_1\rangle|^2
&=&\lim_{\kappa\to 0}\Bigl\{
\Bigl(\langle\th_3|\sigma(0,0)|\theta_2-i0,\theta_1+i0\rangle
-2\pi\bs[\delta(\theta_{32})+\delta(\theta_{31})]\Bigr)\nn
&\times&
\Bigl(
\langle\th_1+i0,\th_2-i0|\sigma(0,0)|\th_3+\kappa\rangle
+2\pi\bs[\delta(\theta_{32}+\kappa)+\delta(\theta_{31}+\kappa)]\Bigr).
\Bigr\}\nn
\eea
Multiplying out the various terms then gives
\bea
\fl
|\langle\theta_3|\sigma(0,0)|\theta_2\theta_1\rangle|^2
&=&\lim_{\kappa\to 0}\Bigl\{
\langle\th_1+i0,\th_2-i0|\sigma(0,0)|\th_3+\kappa\rangle
\langle\th_1-i0,\th_2+i0|\sigma(0,0)|\th_3\rangle^*\nn
&&+2\pi\bs\left[\delta(\th_{32}+\kappa)+\delta(\th_{31}+\kappa)\right]
\ \langle\th_1-i0,\th_2+i0|\sigma(0,0)|\th_3\rangle^*\nn
&&-2\pi\bs\left[\delta(\th_{32})+\delta(\th_{31})\right]
\ \langle\th_1+i0,\th_2-i0|\sigma(0,0)|\th_3+\kappa\rangle\nn
&&-(2\pi\bs)^2\left[\delta(\th_{32}+\kappa)+\delta(\th_{31}+\kappa)\right]
\left[\delta(\th_{32})+\delta(\th_{31})\right]\Bigr\}\nn
&=&\lim_{\kappa\to 0}\left[\Gamma^{\rm conn}+\Gamma^{\rm
    dis,1}+\Gamma^{\rm dis,2}\right].
\label{appA:ffsq}
\eea
The connected part
\bea
\Gamma^{\rm  conn}&=&\langle\th_1+i0,\th_2-i0|\sigma(0,0)|\th_3+\kappa\rangle\nn
&&\times\
\langle\th_1-i0,\th_2+i0|\sigma(0,0)|\th_3\rangle^*
\eea
does not contain any divergent pieces and the limit $\kappa\to 0$ can
be taken straightforwardly. The product of delta-functions gives
\bea
\Gamma^{\rm dis,1}&=&-(2\pi\bs)^2\Bigl\{\delta(\kappa)\left[\delta(\th_{32})
+\delta(\th_{31})\right]\nn
&&+\delta(\th_{31})\delta(\th_{32}+\kappa)
+\delta(\th_{32})\delta(\th_{31}+\kappa)\Bigr\},
\label{Gdis}
\eea
where $\theta_{jk}=\th_j-\th_k$. The cross-terms are
\bea
\Gamma^{\rm dis,2}&=&2\pi\bs\Gamma_{\rm cross,1}+
2\pi\bs\Gamma_{\rm cross,2}
\eea
where
\bea
\Gamma_{\rm cross,1}&=&\delta(\th_{32}+\kappa)
\langle\t_1-i\eta_1,\t_2+i\eta_2|\sigma|\t_3\rangle^*
\nn
&&
-\delta(\th_{32})\langle\t_1+i\eta_1',\t_2-i\eta_2'|\sigma|\t_3
+\kappa\rangle,\nn 
\Gamma_{\rm cross,2}&=&\delta(\th_{31}+\kappa)
\langle\t_1-i\eta_1,\t_2+i\eta_2|\sigma|\t_3\rangle^*\nn
&&-\delta(\th_{31})\langle\t_1+i\eta_1',\t_2-i\eta_2'|\sigma|\t_3+\kappa\rangle.
\eea
Here $\eta_{1,2}$ are positive infinitesimals and we are interested in
the limit $\eta_{1,2}\to 0$ at fixed $\kappa$. Using the explicit form
of the form factors we obtain
\bea
\fl
&&\Gamma_{\rm cross,1}=-i\bs\delta(\th_{32}+\kappa)
\tanh\Bigl(\frac{\theta_{12}+i\eta_1+i\eta_2}{2}\Bigr)
\coth\Bigl(\frac{\th_{12}+\kappa+i\eta_1}{2}\Bigr)
\coth\Bigl(\frac{\kappa-i\eta_2}{2}\Bigr)\nn
\fl &&\qquad\qquad+i\bs\delta(\th_{32})
\tanh\Bigl(\frac{\theta_{12}+i\eta_1'+i\eta_2'}{2}\Bigr)
\coth\Bigl(\frac{\th_{12}-\kappa+i\eta_1'}{2}\Bigr)
\coth\Bigl(\frac{\kappa+i\eta_2'}{2}\Bigr).
\eea
Using that
\bea
&&\tanh\Bigl(\frac{\theta_{12}+i\eta_1+i\eta_2}{2}\Bigr)
\coth\Bigl(\frac{\th_{12}+\kappa+i\eta_1}{2}\Bigr)\nn
&&\simeq
1-(\kappa-i\eta_2)\left[\frac{1}{\sinh\th_{12}}-\frac{1}{\th_{12}}
+\frac{1}{\th_{12}+\kappa+i\eta_1}\right],
\eea
this can be simplified to
\bea
\Gamma_{\rm cross,1}&=&-2i\bs\delta(\th_{32}+\kappa)\Bigl[
\frac{1}{\kappa-i\eta_2}
-\frac{1}{\sinh\th_{12}}+\frac{1}{\th_{12}}
-\frac{1}{\th_{12}+\kappa+i\eta_1}\Bigr]\nn
&&+2i\bs\delta(\th_{32})\Bigl[
\frac{1}{\kappa+i\eta_2}
+\frac{1}{\sinh\th_{12}}-\frac{1}{\th_{12}}
+\frac{1}{\th_{12}-\kappa+i\eta_1}\Bigr]\nn
&=&2\pi\bs\delta(\kappa)\left[\delta(\th_{32}+\kappa)+\delta(\th_{32})\right]
+i\bs\frac{2\kappa}{\kappa^2+\eta_2^2}\left[\delta(\th_{32})-
\delta(\th_{32}+\kappa)\right]\nn
&&+2i\bs\left[\delta(\th_{32}+\kappa)+\delta(\th_{32})\right]\left[\frac{1}{\sinh\th_{12}}
-\frac{1}{\th_{12}}\right]\nn
&&+2i\bs\left[\frac{\delta(\th_{32})}{\th_{12}-\kappa+i\eta_1}
+\frac{\delta(\th_{32}+\kappa)}{\th_{12}+\kappa+i\eta_1}\right].
\eea
In the limit $\kappa\to 0$ this becomes
\bea
\Gamma_{\rm
  cross,1}&\to&4\pi\bs\delta(\th_{32})\delta(\kappa)
+2i\bs\left[\frac{\delta(\th_{32})}{\th_{12}-\kappa+i0}
+\frac{\delta(\th_{32}+\kappa)}{\th_{12}+\kappa+i0}\right]\nn
&-&2i\bs\delta'(\th_{32})
+4i\bs\delta(\th_{32})\left[\frac{1}{\sinh\th_{12}}
-\frac{1}{\th_{12}}\right].
\eea
Similarly we find
\bea
\Gamma_{\rm cross,2}&\to&4\pi\bs\delta(\th_{31})\delta(\kappa)
+2i\bs\left[\frac{\delta(\th_{31})}{\th_{12}+\kappa+i0}
+\frac{\delta(\th_{31}+\kappa)}{\th_{12}-\kappa+i0}\right]\nn
&+&2i\bs\delta'(\th_{31})
+4i\bs\delta(\th_{31})\left[\frac{1}{\sinh\th_{12}}
-\frac{1}{\th_{12}}\right].
\eea
We note the emergence of a term involving the derivative of the delta
function. 
Our final result for $\Gamma^{\rm dis,2}$ is then
\bea
\lim_{\kappa\to 0}\Gamma_{\rm dis,2}&=&\lim_{\kappa\to 0}\Bigl\{
8\pi^2\bs^2\left[\delta(\th_{31})+\delta(\th_{32})\right]
\delta(\kappa)\nn
&+&4\pi i\bs^2\left[\delta'(\th_{31})-\delta'(\th_{32})\right]\nn
&+&8\pi i\bs^2\left[\delta(\th_{31})+\delta(\th_{32})\right]
\left[\frac{1}{\sinh\th_{12}}-\frac{1}{\th_{12}}\right]\nn
&+&4\pi i\bs^2\left[\frac{\delta(\th_{31})+\delta(\th_{32}+\kappa)}
{\th_{12}+\kappa+i0}
+\frac{\delta(\th_{31}+\kappa)+\delta(\th_{32})}{\th_{12}-\kappa+i0}
\right]\Bigr\}.
\label{Gcross}
\eea
The infinite volume regularization for the form factor squared is then
\bea
|\langle\theta_3|\sigma(0,0)|\theta_2\theta_1\rangle|^2&=&
\bs^2\tanh^2\Bigl(\frac{\th_{12}}{2}\Bigr)
\coth^2\Bigl(\frac{\th_{13}+i0}{2}\Bigr)
\coth^2\Bigl(\frac{\th_{23}-i0}{2}\Bigr)\nn
&&+\lim_{\kappa\to 0} 
\left[\Gamma^{\rm dis,1}+\Gamma^{\rm dis,2}\right],
\label{FF2ising}
\eea
where $\Gamma^{\rm dis,1}$ and $\Gamma^{\rm dis,2}$
are given by \fr{Gdis} and \fr{Gcross} respectively.
For our purposes we only need the part of $\Gamma^{\rm dis}=\Gamma^{\rm
  dis,1}+\Gamma^{\rm dis,2}$ symmetric under
$\th_1\leftrightarrow\th_2$ as the remainder of the integrands we
consider all have this symmetry. The symmetric part is
\bea
\lim_{\kappa\to 0}\frac{\Gamma^{\rm dis}(\th_1,\th_2,\th_3)
+\Gamma^{\rm dis}(\th_2,\th_1,\th_3)}{2}&=&
4\pi^2\bs^2\delta(\kappa)\left[\delta(\th_{31})+\delta(\th_{32})\right]\nn
&&+8\pi^2\bs^2\ \delta(\th_{31})\ \delta(\th_{32}).
\eea

\section{Products of Form Factors for the O(3) nonlinear $\sigma$-Model}
\label{app:B}
In this Appendix we determine the form factor squared
\bea
\Gamma(\th_1,\th_2,\th_3)=\sum_{b,b_1,b_2}
|{}_{b_1b_2}\langle\theta_1,\theta_2|\Phi^a(0)|\theta_3\rangle_b|^2.
\label{SMffsquared}
\eea
Using \fr{SMFFsq}, which is a consequence of the crossing relations
discussed in \ref{app:xing}, we obtain the decomposition
\be
\fl
\Gamma(\th_1,\th_2,\th_3)=\lim_{\kappa,\eta_{1,2}\to 0}\left[
\Gamma^{\rm conn}(\th_1,\th_2,\th_3)+
\Gamma^{\rm dis,1}(\th_1,\th_2,\th_3)+
\Gamma^{\rm dis,2}(\th_1,\th_2,\th_3)\right].
\label{FFdecomp}
\ee
Here
\bea
\Gamma^{\rm conn}&=&
\sum_{b,b_1,b_2}
{}_{b_1b_2}\langle\theta_1-i\eta_1,\theta_2+i\eta_2|\Phi^a(0)|
\theta_3\rangle_b^*\ \nn
&&\qquad\times\
{}_{b_1b_2}\langle\theta_1+i\eta_1,\theta_2-i\eta_2|\Phi^a(0)|\theta_3
+\kappa\rangle_b\ ,
\label{Gconn}
\eea
\bea
\Gamma^{\rm dis,1}&=&\sum_b
(2\pi)^2\ S_{ab}^{ab}(\th_{21})\left[
\delta(\th_{32})\delta(\th_{32}+\kappa)
+\delta(\th_{31})\delta(\th_{31}+\kappa)\right]\nn
&+&\sum_b(2\pi)^2\  S_{ba}^{ab}(\th_{21})\left[
\delta(\th_{31})\delta(\th_{32}+\kappa)
+\delta(\th_{32})\delta(\th_{31}+\kappa)\right],
\label{Gdis1}
\eea
\bea
\Gamma^{\rm dis,2}&=&
\sum_b2\pi\delta(\th_{32}+\kappa) \
{}_{ab}\langle\theta_1-i\eta_1,\theta_2+i\eta_2|\Phi^a(0)|\theta_2-\kappa\rangle_b^*\nn
&+&\sum_b2\pi\delta(\th_{31}+\kappa)\ 
{}_{ba}\langle\theta_1-i\eta_1,\theta_2+i\eta_2|\Phi^a(0)|\theta_1-\kappa\rangle_b^*\nn
&+&\sum_b2\pi\delta(\th_{32}) \
{}_{ba}\langle\theta_2-i\eta_2,\theta_1+i\eta_1|\Phi^a(0)|\theta_2+\kappa\rangle_b\nn
&+&\sum_b2\pi\delta(\th_{31}) \ 
{}_{ab}\langle\theta_2-i\eta_2,\theta_1+i\eta_1|\Phi^a(0)|\theta_1+\kappa\rangle_b\ .
\label{Gdis2}
\eea
Using crossing the connected contribution can be written as
\bea
\Gamma^{\rm conn}&=&\sum_{b,b_1,b_2}
\langle0|\Phi^a(0)|\th_1+i\pi^-,\th_2+i\pi^+,\theta_3\rangle_{b_1b_2b}^*\nn
&&\qquad\times
\langle 0|\Phi^a(0)|\th_1+i\pi^+,\th_2+i\pi^-,\theta_3+\kappa\rangle_{b_1b_2b}.
\eea
Using the explicit expression \fr{SM3part} for the three-particle form
factor this can be brought in the form
\bea
\fl
\lim_{\kappa,\eta_{1,2}\to 0}\Gamma^{\rm conn}
&=&\frac{\pi^6}{2}\left[\th_{12}^2+\th_{13}^2+\th_{23}^2+4\pi^2\right]
\frac{\th_{12}^2+\pi^2}{\th_{12}^2(\th_{12}^2+4\pi^2)}
\tanh^4\Bigl(\frac{\th_{12}}{2}\Bigr) \nn
&&\times \frac{(\th_{13}+i0)^2}{(\th_{13}^2+\pi^2)^2}
\coth^4\Bigl(\frac{\th_{13}+i0}{2}\Bigr) 
\frac{(\th_{23}-i0)^2}{(\th_{23}^2+\pi^2)^2}
\coth^4\Bigl(\frac{\th_{23}-i0}{2}\Bigr).
\label{SMconnFF}
\eea
The connected $\Gamma^{\rm conn}$ contribution is now in a form
suitable for further analysis. For our purposes it is 
sufficient to determine the parts of $\Gamma^{\rm dis,1}$,
$\Gamma^{\rm dis,2}$ symmetric in $\th_1$ and $\th_2$
\bea
\Gamma^{\rm dis,j}_+(\th_1,\th_2,\th_3)=
\lim_{\kappa,\eta_{1,2}\to 0}\frac{
\Gamma^{\rm dis,j}(\th_1,\th_2,\th_3)+\Gamma^{\rm
  dis,j}(\th_2,\th_1,\th_3)}{2}\ ,\ j=1,2.
\eea
We therefore concentrate on these parts only in the following. 
We have
\bea
\lim_{\kappa\to 0}\Gamma^{\rm dis,1}_+&=&
\lim_{\kappa\to 0}\sum_b
4\pi^2\ \delta(\kappa)\ {\rm Re} S_{ab}^{ab}(\th_{21})\left[
\delta(\th_{32})+\delta(\th_{31})\right]\nn
&&-24\pi^2\ \delta(\th_{31})\delta(\th_{32})\ .
\label{gdis1+}
\eea
Using
the explicit forms of the three-particle form factors and proceeding
along the same lines as for the Ising model, we obtain after some lengthy
calculations 
\bea
2\Gamma^{\rm dis,2}_+(\th_1,\th_2,\th_3)&=&
\sum_b
(2\pi)^2\delta(\th_{32})\delta(\kappa)\
{\rm Re}\left[1-S^{ab}_{ab}(\th_{21})\right]\nn
&+&\sum_b(2\pi)^2\delta(\th_{31})\delta(\kappa)\
{\rm Re}\left[1-S^{ab}_{ab}(\th_{21})\right]\nn
&+&\sum_b4\pi\delta'(\th_{32})\ {\rm Im} S^{ab}_{ab}(\th_{12})\nn
&+&\sum_b(4\pi)^2\delta(\th_{32})(1-\delta_{ab})\ {\rm Re}\left[\frac{1}{\th_{21}+\pi i}
\frac{1}{\th_{21}-2\pi i}\right]\nn
&+&\sum_b8\pi\delta(\th_{32})\ {\rm Im}S^{ab}_{ab}(\th_{12})\left[
\frac{2}{\sinh\th_{21}}-\frac{2}{\th_{21}}+\frac{2\th_{21}}{\th_{21}^2+\pi^2}
\right]\nn
&+&\sum_b(2\pi)^2\delta(\th_{32}+\kappa)\delta(\th_{31})\ 
{\rm Re}\left[1-S^{ab}_{ab}(\th_{21})\right]\nn
&+&\sum_b(2\pi)^2\delta(\th_{31}+\kappa)\delta(\th_{32})\ 
{\rm Re}\left[1-S^{ab}_{ab}(\th_{21})\right]\nn
&+&\sum_b4\pi\delta(\th_{32}+\kappa)\frac{\th_{31}}{\th_{31}^2+\eta^2}\
{\rm Im} S^{ab}_{ab}(\th_{12})\nn
&+&\sum_b4\pi\delta(\th_{32})\frac{\th_{31}+\kappa}{(\th_{31}+\kappa)^2+\eta^2}\
{\rm Im} S^{ab}_{ab}(\th_{12})\nn
&+&\th_1\leftrightarrow\th_2.
\eea
This is simplified further to
\bea
\lim_{\kappa\to 0}\Gamma^{\rm dis,2}_+&=&4\pi^2\left[
\delta(\th_{32})+\delta(\th_{31})\right]
\delta(\kappa)\left[3-\sum_b {\rm Re} S^{ab}_{ab}(\th_{21})\right]
\nn
&-&16\pi^4\left[\delta'(\th_{32})-\delta'(\th_{31})\right]
\frac{\th_{12}}{(\th_{12}^2+\pi^2)(\th_{12}^2+4\pi^2)}\nn
&+&32\pi^2\delta(\th_{32})\delta(\th_{31})\ \nn
&+&16\pi^2\frac{\delta(\th_{32})+\delta(\th_{31})}
{(\th_{12}^2+\pi^2)(\th_{12}^2+4\pi^2)}
\left[\th_{12}^2+\frac{4\pi^2\th_{12}}{\sinh\th_{12}}
+\frac{4\pi^2\th_{12}^2}{\th_{12}^2+\pi^2}
\right].
\label{gdis2+}
\eea
Our final result for the symmetrized (in $\th_1$ and $\th_2$)
disconnected parts of the form factor squared in the infinite volume
scheme is  
\bea
\fl
\lim_{\kappa\to 0}
\left[\Gamma^{\rm dis,1}_++\Gamma^{\rm dis,2}_+\right]
&=&
12\pi^2\left[
\delta(\th_{32})+\delta(\th_{31})\right]
\delta(\kappa)\nn
&&+8\pi^2\delta(\th_{32})\delta(\th_{31})\ \nn
&&-16\pi^4\left[\delta'(\th_{32})-\delta'(\th_{31})\right]
\frac{\th_{12}}{(\th_{12}^2+\pi^2)(\th_{12}^2+4\pi^2)}\nn
&&+16\pi^2\frac{\delta(\th_{32})+\delta(\th_{31})}
{(\th_{12}^2+\pi^2)(\th_{12}^2+4\pi^2)}
\left[\th_{12}^2+\frac{4\pi^2\th_{12}}{\sinh\th_{12}}
+\frac{4\pi^2\th_{12}^2}{\th_{12}^2+\pi^2}
\right].
\label{regfinal}
\eea


\section*{References}
\begin{thebibliography}{99}

\bibitem{Smirnov92book}
F.~A. Smirnov, {\em Form factors in completely integrable models of quantum
  field theory} (World Scientific, Singapore, 1992).

\bibitem{Lukyanov95}
S. Lukyanov, Commun. Math. Phys. {\bf 167},  183  (1995).

\bibitem{DelMuss}
G. Delfino and G. Mussardo, Nucl. Phys B {\bf 455} 724 (1995);

\bibitem{FF}
G. Delfino and G. Mussardo, Nucl. Phys B {\bf 455} 724 (1995);
S. Lukyanov, Mod. Phys. Lett. A {\bf 12}, 2911 (1997);
H. Babujian, A. Fring, M. Karowski and A. Zapletal, Nucl. Phys. B {\bf
538}, 535 (1999);
S. Lukyanov and A. B. Zamolodchikov, Nucl. Phys. B {\bf 607}, 437 (2001);
H. M. Babujian and M. Karowski, Nucl. Phys. B {\bf 620}, 407 (2002).

\bibitem{Delfino04}
G. Delfino, J.~Phys.~A: Math. Gen. {\bf 37},  R45  (2004).

\bibitem{BH}
J. Balog and T. Hauer, Phys. Lett. {\bf 337}, 115 (1994).

\bibitem{BN} 
J. Balog and M. Niedermaier, Nucl. Phys. B {\bf 500}, 421 (1997).

\bibitem{ks}
A.N. Kirillov and F.A. Smirnov, Int. Jour. Mod. Phys. {\bf A3}, 731 (1998).

\bibitem{mussardobook}
G. Mussardo, ``Statistical Field Theory, An Introduction to Exactly
Solved Models in Statistical Physics'' (Oxford University Press,
Oxford 2009).

\bibitem{karowski}
M. Karowski and P. Weisz, Nucl. Phys. B {\bf 139}, 455 (1978).

\bibitem{review}
F.H.L. Essler and R.M. Konik, in Ian Kogan Memorial Collection ``From
Fields to Strings: Circumnavigating Theoretical Physics'', eds
M. Shifman, A. Vainshtein and J. Wheater, World Scientific Singapore 2005;
cond-mat/0412421; 


\bibitem{QM}
F. H. L. Essler, A. M. Tsvelik and G. Delfino, Phys. Rev. B {\bf 56},
11001 (1997); 
F. H. L. Essler and A. M. Tsvelik, Phys. Rev. B {\bf 57}, 10592 (1998);
F. H. L. Essler, A. Furusaki and T. Hikihara, Phys. Rev. B {\bf 68},
064410 (2003).

\bibitem{MI}
F. H. L. Essler, F. Gebhard and E. Jeckelmann, Phys. Rev. B {\bf 64},
125119 (2001); 
F. H. L. Essler and A. M. Tsvelik, Phys. Rev. B {\bf 65}, 115117 (2002);
F. H. L. Essler and A. M. Tsvelik, Phys. Rev. Lett. {\bf 88}, 096403 (2002);
F. H. L. Essler and A. M. Tsvelik, Phys. Rev. Lett. {\bf 90}, 126401 (2003);
M. J. Bhaseen and A. M. Tsvelik, Phys. Rev. B {\bf 68}, 094405 (2003).

\bibitem{2leg}
E. Orignac and D. Poilblanc, Phys. Rev. B {\bf 68}, 052504 (2003);
D. Poilblanc, E. Orignac, S. R. White and S. Capponi, Phys. Rev. B
{\bf 69}, 220406 (2004); 
D. Controzzi and A. M. Tsvelik, Phys. Rev. B {\bf 72}, 035110 (2005);
F. H. L. Essler and R. M. Konik, Phys. Rev. B {\bf 75}, 144403 (2007).

\bibitem{nano}
R. M. Konik and A. W. W. Ludwig, Phys. Rev. B {\bf 64}, 155112 (2001).

\bibitem{gas}
V. Gritsev, A. Polkovnikov and E. Demler, Phys. Rev. B {\bf 75},
174511 (2007).  

\bibitem{oldff1}
J. Cardy and G. Mussardo, Nucl. Phys. B {\bf 410}, 451 (1993).

\bibitem{deltwo} 
G. Delfino and J. Cardy, Nucl.Phys. B {\bf 519}, 551 (1998).

\bibitem{CET}
D. Controzzi, F.H.L. Essler and A.M. Tsvelik, Phys. Rev. Lett. {\bf
86}, 680 (2001).

\bibitem{ising3}
A. Leclair, F. Lesage, S. Sachdev and H. Saleur, Nucl. Phys. {\bf
B482}, 579 (1996).

\bibitem{muss}  
A. LeClair, G. Mussardo, Nucl. Phys. B {\bf 552}, 624 (1999).

\bibitem{finiteTFF}
H. Saleur, Nucl. Phys. {\bf B567} 602 (2000);
G. Delfino, J. Phys. {\bf A34}, L161 (2001);
G. Mussardo, J. Phys. {\bf A34}, 7399 (2001);
S. Lukyanov, Nucl. Phys. {\bf B612}, 391 (2001);
O.A. Castro Alvaredo and A. Fring, Nucl. Phys. {\bf B636} 611 (2002).

\bibitem{rmk} 
R.M. Konik, Phys. Rev. B {\bf 68}, 104435 (2003).

\bibitem{AKT} 
B.L. Altshuler, R.M. Konik and A.M. Tsvelik, Nucl. Phys. {\bf B739}, 311
(2006). 

\bibitem{takacs} 
B. Pozsgay and G. Takacs, Nucl. Phys. B. {\bf 788}, 167 (2008);
B. Pozsgay and G. Takacs, Nucl. Phys. B. {\bf 788}, 209 (2008).

\bibitem{bugrij} 
A. Bugrij, Theo. and Math. Phys. {\bf 127}, 528 (2001);
A. Bugrij, O. Lisovyy, Phys. Lett. A {\bf 319}, 390 (2003).

\bibitem{doyon}
B. Doyon, J. Stat. Mech., P11006 (2005);
B. Doyon, Sigma {\bf 3}, 11 (2007);

\bibitem{young}
S. Sachdev, A. P. Young, Phys. Rev. Lett. {\bf 78}, 2220 (1997).

\bibitem{Reyes06}
S.A. Reyes, A. Tsvelik, Phys. Rev. B {\bf 73}, 220405(R) (2006).

\bibitem{Reyes06b}
S.A. Reyes, A. Tsvelik, Nucl. Phys. {\bf B744}, 330 (2006).

\bibitem{doyongamsa}
B. Doyon and A. Gamsa, J. Stat. Mech., P03012 (2008).

\bibitem{PDE}
J.H.H. Perk, Phys. Rev. A{\bf 79},1 (1980);
A. Leclair and D. Bernard, Nucl. Phys. {\bf B426}, 534 (1994);
Erratum-ibid. B {\bf 498}, 619 (1997);
R. Konik, A. LeClair, and G. Mussardo, Int. J. Mod. Phys. {\bf A11},
2765 (1996);
O. Lisovyy, Adv. Theor. Math. Phys. {\bf 5}, 909 (2002);
P. Fonseca and A.B. Zamolodchikov, hep-th/0309228.

\bibitem{perk}
J.H.H. Perk and H. Au-Yang, J. Stat. Phys. {\bf 135}, 599 (2009).

\bibitem{kenz}
M. Kenzelmann, R. A. Cowley, W. J. Buyers, R. Coldea, J. S. Gardner,
M. Enderle, D. F. McMorrow and S. M. Bennington, Phys. Rev. Lett. {\bf
  87}, 017201 (2001);  
M. Kenzelmann, R. A. Cowley, W. J. Buyers and D. F. McMorrow,
Phys. Rev. B {\bf 63}, 134417 (2001);
M. Kenzelmann, R.A. Cowley, W.J.L. Buyers, Z. Tun, R. Coldea and
M. Enderle, Phys. Rev. B {\bf 66}, 024407 (2002);
M. Kenzelmann, R. A. Cowley, W. J. Buyers, R. Coldea, M. Enderle and
D. F. McMorrow , Phys. Rev. B {\bf 66}, 174412 (2002).

\bibitem{xu}
G. Xu et al. Science {\bf 317}, 1049 (2007).

\bibitem{zheludev}
A. Zheludev et al, Phys. Rev. Lett. {\bf 100}, 157204, (2008).

\bibitem{EK08}
F.H.L. Essler and R.M. Konik, Phys. Rev. B{\bf 78}, 100403(R) (2008).

\bibitem{sachdevbook}
S. Sachdev, {\sl Quantum Phase Transitions}, Cambridge University
Press, Cambridge 1999.

\bibitem{damle}
K. Damle and S. Sachdev, Phys. Rev. B {\bf 57}, 8307 (1998).

\bibitem{semiclassicsSG}
K. Damle, S. Sachdev, Phys. Rev. Lett. 95, 187201 (2005).

\bibitem{zarand}
A. Rapp, G. Zarand, Phys. Rev. B {\bf 74}, 014433 (2006);
A. Rapp, G. Zarand, Eur. Phys. Jour. {\bf B67}, 7 (2009).

\bibitem{kw}
H. A. Kramers and G. H. Wannier, {\em Physical Review} {\bf 60}, 252 (1941).

\bibitem{kogut}
J.~B. Kogut, {\em Rev. Mod. Phys.} {\bf 51}, 659 (1979).

\bibitem{fradkin}
E. Fradkin and L. Susskind, Phys. Rev. D{\bf 17}, 2637 (1978).

\bibitem{McCoyWu73}
B.~M. McCoy and T.~T. Wu, {\em The two-dimensional {I}sing model} 
(Havard University Press, Cambridge, 1973).

\bibitem{barouch}
T.T. Wu, B.M. McCoy, C.A. Tracy and E. Barouch, Phys. Rev. {\bf B13},
316 (1976);
H.G. Vaidya and C.A. Tracy, Physica 92A, 1 (1978).

\bibitem{Berg-79}
B. Berg, M. Karowski and P. Weisz, Phys. Rev.~D {\bf 19},  2477
(1979).

\bibitem{CardyMuss}
J.L. Cardy and G. Mussardo, Nucl. Phys. {\bf B340}, 387 (1990).

\bibitem{yurov}
V.P. Yurov and Al.B. Zamolodchikov, Int. Jour. Mod. Phys. {\bf A6}, 
3419 (1991).

\bibitem{zamo} 
P. Fonseca, A. Zamolodchikov, J. Stat. Phys. {\bf 110} 527 (2003).

\bibitem{haldane} 
F.D.M. Haldane, Phys. Lett. A {\bf 93}, 464  (1983).

\bibitem{o3}
I. Affleck {\rm in} {\em Fields, Strings and Critical Phenomena},
{\rm eds E. Br\'ezin and J. Zinn-Justin}, (Elsevier, Amsterdam, 1989);
I. Affleck, J. Phys. Cond. Mat. {\bf 1}, 3047 (1989).

\bibitem{luescher}
M. L\"uscher, Nucl. Phys. {\bf B135}, 1 (1978).

\bibitem{o3smat}
A.B. Zamolodchikov and Al.B. Zamolodchikov, Annals of Physics {\bf
120},253 (1979).

\bibitem{O3integrability}
P. B. Wiegmann, Phys. Lett. B{\bf 152},209 (1985);
JETP Lett. {\bf 41}, 95 (1985).

\bibitem{affwes} 
I. Affleck and R. Weston, Phys. Rev. B {\bf 45}, 4667 (1992).

\bibitem{3particle}
M.D.P. Horton and I. Affleck, Phys. Rev. B{\bf 60}, 9864 (1999);
F.H.L. Essler, Phys. Rev. B{\bf 62}, 3264 (2000). 

\bibitem{HWS}
L.~D. Faddeev and L.~A. Takhtajan, J. Soviet Math. {\bf 24}, 241
(1984);
H.J. deVega and M. Karowski, Nucl. Phys. {\bf B280}, 225 (1987);
F.H.L. Essler, V.E. Korepin and K. Schoutens, Phys. Rev. Lett. {\bf 67},
3848 (1991);
F.H.L. Essler, V.E. Korepin and K. Schoutens, Nucl. Phys. {\bf B372},
559 (1992);
A. F\"orster and M. Karowski, Nucl. Phys. {\bf B396}, 611 (1993);
H.J. deVega and A. Gonzales-Ruiz, Phys. Lett. {\bf B332}, 123 (1994);
M. Karowski and A. Zapletal, J. Phys. {\bf A27}, 7419 (1994);
J. Links and A. F\"orster, J. Phys. {\bf A32}, 147 (1999);
T. Deguchi, J. Phys. {\bf A46}, 9755 (2001).

\bibitem{igor}
I.A. Zaliznyak, S.-H. Lee and S.V. Petrov, Phys. Rev. Lett. {\bf 87},
017202 (2001). 

\bibitem{huang}
H. Huang, Phys. Lett. {\bf A 360}, 731 (2007).

\bibitem{mikeska} 
H. J. Mikeska and C. Luckmann, Phys. Rev. B {\bf 73}, 184426 (2006);
A.J.A. James, F.H.L. Essler and R.M. Konik, Phys. Rev. B {\bf 78},
094411 (2008).

\bibitem{Alan}
D.A. Tennant et al, unpublished.

\bibitem{ruegg}
N. Cavadini, Ch. R\"uegg, W. Henggeler, A. Furrer, H.-U. G\"udel,
K. Kr\"amer and H. Mutka, Eur. Phys. J. B{\bf 18}, 565 (2000);
C. R\"uegg et al, Phys. Rev. Lett. {\bf 95}, 267201 (2005).

\bibitem{XXZ}
H.~Yoshizawa, K.~Hirakawa, S.~K. Satija, and G.~Shirane, {\em
  Phys. Rev. B} {\bf 23}, 2298 (1981);
S.~E. Nagler, W.~J.~L. Buyers, R.~L. Armstrong, and B.~Briat, {\em
  Phys.\ Rev. Lett.} {\bf 49}, 590 (1982); 
S.~E. Nagler, W.~J.~L. Buyers, R.~L. Armstrong, and B.~Briat, {\em
  Phys.\ Rev. B} {\bf 28}, 3873 (1983); 
A.~Oosawa, K.~Kakurai, Y.~Nishiwaki, and T.~Kato, {\em
  J. Phys. Soc. Jpn.} {\bf 75}, 074719 (2006);
H.-B. Braun, J. Kulda, B. Roessli, D. Visser, K. W. Kr{\"{a}}mer,
H.-U. G{\"{u}}del, and P. B{\"{o}}ni, {\em Nat. Phys.} {\bf 1}, 159
(2005). 

\bibitem{Andrew2}
A.J.A. James, W.D. Goetze and F.H.L. Essler, Phys. Rev. B{\bf 79},
214408 (2009). 

\bibitem{Villain}
J.~Villain, {\em Physica} {\bf 79B}, 1 (1975).

\bibitem{decay}
G. Delfino, G. Mussardo and P. Simonetti,
Nucl. Phys. B {\bf 473}, 469 (1996);
G. Delfino, P. Grinza and G. Mussardo, Nucl.Phys. {\bf B737}, 291 (2006).
\end{thebibliography}
\end{document}